\documentclass[review,1p]{elsarticle}

\usepackage{amssymb}
\usepackage{amsmath}
\usepackage{amsbsy}
\usepackage{multirow,multicol}
\usepackage{epsfig}
\usepackage[utf8]{inputenc}
\usepackage{multirow}
\usepackage{todonotes}
\usepackage{wrapfig}
\usepackage{subfig}
\usepackage{graphicx}
\usepackage{hyperref}

\usepackage[draft]{fixme}
\fxusetheme{color}
\FXRegisterAuthor{ak}{apek}{\color{blue}AK}
\FXRegisterAuthor{db}{adb}{\color{red}DB}

\begin{document}


\begin{frontmatter}
\title{A Stabilised Nodal Spectral Element Method for \\ Fully Nonlinear Water Waves \\ Part 2: Wave-body interaction}
\author[compute,cere]{A.~P.~Engsig-Karup}
\ead{apek@dtu.dk, Office phone: +45 45 25 30 73}

\author[compute]{C.~Monteserin}

\address[compute]{Department of Applied Mathematics and Computer Science \\ Technical University of Denmark, 2800 Kgs. Lyngby, Denmark. }

\address[cere]{Center for Energy Resources Engineering (CERE) \\ Technical University of Denmark, 2800 Kgs. Lyngby, Denmark. }

\author[cles]{C.~Eskilsson}
\address[cles]{Department of Shipping and Marine Technology \\ Chalmers University of Technology, SE-412 96 Gothenburg, Sweden.}

\date{\today.}

\begin{abstract}
We present a new stabilised and efficient high-order nodal spectral element method based on the Mixed Eulerian Lagrangian (MEL) method for general-purpose simulation of fully nonlinear water waves and wave-body interactions. In this MEL formulation a standard Laplace formulation is used to handle arbitrary body shapes using unstructured -- possibly hybrid -- meshes consisting of high-order curvilinear iso-parametric quadrilateral/triangular elements to represent the body surfaces and for the evolving free surface. Importantly, our numerical analysis highlights that a single top layer of quadrilaterals elements resolves temporal instabilities in the numerical MEL scheme that are known to be associated with mesh topology containing asymmetric element orderings. The 'surface variable only' free surface formulation based on introducing a particle-following (Lagrangian) reference frame contains quartic nonlinear terms that require proper treatment by numerical discretisation due to the possibility of strong aliasing effects. We demonstrate how to stabilise this nonlinear MEL scheme using an efficient combination of (i) global $L^2$ projection without quadrature errors, (ii) mild nonlinear spectral filtering and (iii) re-meshing techniques. 
Numerical experiments revisiting known benchmarks are presented, and highlights that modelling using a high-order spectral element method provides excellent accuracy in prediction of nonlinear and dispersive wave propagation, and of nonlinear wave-induced loads on fixed submerged and surface-piercing bodies.
\end{abstract}


\end{frontmatter}

\section{Introduction}

In the last decades, numerical simulation of hydrodynamics for free surface flows and the design of marine structures has become an indispensable tool in engineering analysis. The numerical tools need to accurately and efficiently account for nonlinear wave-wave and wave-body interactions in large and representative marine regions in order to give reliable estimates of environmental loads on offshore structures such as floating production systems, ship-wave hydrodynamics, offshore wind turbine installations and wave energy converters.

The development of numerical models for time-domain simulation of fully nonlinear and dispersive water wave propagation have been a research topic since the 1960s and with real engineering applications since the 1970s \cite{BeckReed2001}. Models based on fully nonlinear potential flow (FNPF) have been considered relative mature for about two decades, see the review papers \cite{Yeung1982,TsaiYue1996} and the references therein. More recent research focus on improved modelling by incorporating more physics and finally enabling large-scale simulation of nonlinear waves \cite{CavaleriEtAl2007,Ma10}. In many applications, computational speed is far more important than the cost of hardware and therefore algorithms that provide speed and scalability of work effort is of key interest.

The Boundary Element Method (BEM) is a widely used method in applications for free surface flow \cite{Yeung1975} and flow around complex bodies due to the ease in handling complex geometry \cite{KimEtAll1999,HarrisEtAl2014}. However, scalability has been shown to favour Finite Element Methods \cite{WuTaylor1995,ShaoEtAl2014}. The use of Finite Element Methods (FEM) for fully nonlinear potential flow has also received significant attention starting with the original work of Wu \& Eatock Taylor (1994) \cite{WuTaylor1994}. Studies of free surface solvers based on second-order FEM are, e.g. \cite{WuTaylor1995,GreavesBorthwickTaylor1997,MaWuTaylor2001,MaWuTaylor2001b,TurnbullEtAl2003,WangWu2011,Zienkiewicz2014}. For wave-wave interaction and wave propagation with no bodies, the $\sigma$-transformed methods can be used to map the physical domain into a fixed computational domain \cite{CaiEtAl1998,WesthuisAndonowati1998,ClaussSteinhagen:1999tr,TurnbullEtAl2003} is maybe the most relevant approach due to the numerical efficiency. However, most numerical models for free surface flows use the Mixed Eulerian Lagrangian method (MEL) \cite{LonguetHigginsCokelet1976} for updating the free surface variables, especially for solving wave-body interaction problems. The MEL approach requires re-meshing of and development of techniques that improves the efficiency, e.g. QALE-FEM \cite{MaYan2006}, have been put forward. It is well-known that high-order discretisation methods can give significant reduction in computational effort compared to use of conventional low-order methods, especially for long-time simulations \cite{KO72}. Higher-order schemes that has been proposed for FNPF models include high-order Finite Difference Methods (FDM) \cite{EBL08,DucrozetEtAl2010,ShaoEtAl2014}, high-order BEM \cite{WangYaoTulin1995,Ferrant1996,CelebiEtAl1998,XueEtAl2001,FerrantEtAl2003,MaYan2009,YanLiu2011,HarrisEtAl2014b}, the High-Order Spectral (HOS) Method \cite{DOMYOU1987,WestEtAl1987,GouinEtAl2017}  and other pseudo-spectral methods \cite{ChernEtA1999,ChristiansenEtAl2012}  based on global basis functions in a single element (domain). Indeed, it has been shown that efficient algorithms \cite{LiFleming1997,KorsmeyerEtal1999,EngsigKarup2014} together with other means of acceleration (such as multi-domain approaches and massively parallel computing \cite{EngsigKarupEtAl2011,NimmalaEtAl2013} via software implementations on modern many-core hardware \cite{EGNL13}) render FNPF models practically feasible for analysis of wave propagation on standard work stations - even with a real-time perspective within reach for applications with appreciable domain sizes as discussed in \cite{EngsigKarupEtAl2011}. 

Even though high-order FDM have shown to be very efficient, second-order FEM models remain popular because of the geometrical flexibility and the sparse matrix patterns of the discretisation. In contrast, a high-order finite element method, such as the Spectral Element Method (SEM) due to Patera (1984) \cite{PAT84}, has historically received the least attention for FNPF equations. However, SEM is thought to be highly attractive as it combines the high-order accuracy of spectral methods for problems with sufficiently smooth solutions with the geometric flexibility via adaptive meshing capability of finite element methods. A previous study employing SEM for the FNPF model include \cite{RobertsonSherwin1999} where an Arbitrary Lagrangian Eulerian (ALE) technique was used to track the free-surface motion identified that numerical instabilities were caused by asymmetry in the mesh. In order to stabilise the model, they added a diffusive term proportional to the mesh skewness to the kinematic free surface condition. Recently, a SEM model allowing for the computation of very steep waves has been proposed \cite{EngsigKarupEtAl2016}. The use of $\sigma$-transformed domains partitioned into a single vertical layer of elements is shown to avoid the fundamental problem of instabilities caused by mesh asymmetry. It was also found that the quartic nonlinear terms present in the Zakharov form \cite{ZAK1968} of the free surface conditions could cause severe aliasing problems and consequently numerical instability for marginally resolved or very steep nonlinear waves. This problem was mitigated through over-integration of the free surface equations and application of a gentle spectral filtering using a cap of 1\% of the highest modal coefficient. Finally, in the context of SEM for FNPF models we also mention the HOSE method \cite{ZhuEtAl2003}, however, it is only the boundary domains that are discretised with SEM, not the interior domain. 

Thus, the challenge we seek to address in this work is the development of a robust SEM based FNPF solver with support for geometric flexibility for handling wave-body interaction problems. This is achieved by extending the results of \cite{EngsigKarupEtAl2016}, keeping the same de-aliasing techniques. In order to include arbitrary shaped bodies, we discard the $\sigma$-transform in favour for the MEL method. To avoid the temporal stability issues associated with asymmetric mesh configurations near the free surface boundary \cite{Westhuis2001,RobertsonSherwin1999} we propose to use hybrid meshes consisting of a layer of quadrilaterals at the surface and unstructured triangles below. Eigenvalue analysis of the semi-discrete formulation for small-amplitude waves is used to illustrate that this mesh configuration is linearly stable and during the simulation the vertical interfaces of the quadrilaterals are kept vertical. Global re-meshing is employed where the vertices of the mesh topology are repositioned similar to the technique used in the QALE-FEM method \cite{MaYan2006} to improve general stability of the model for nonlinear waves. This is combined with local re-meshing of the node distributions of the free surface elements to keep the discrete operators well-conditioned for all times, and is shown to increase the temporal stability significantly. In the following, several test cases are used to show the robustness and accuracy of the proposed SEM model based on MEL.

\subsection{On high-order methods for fully nonlinear potential flow models}

High-order methods allow for convergence rates faster than quadratic with mesh refinement, and is attractive for improving efficiency of numerical methods - especially for long-time integration \cite{KO72}. Indeed, high-order discretisation methods are needed for realistic large-scale applications. To understand why, we need to first to understand that the key criterion for deciding between practical numerical tools is based on identifying the tool which is the most numerically efficient one, e.g. measured in terms of CPU time. In general, we can express the minimum work effort for a numerical scheme in terms of the discretization parameter as
\begin{align*}
W \leq c_1 h^{-d}
\end{align*}
where $d$ is the spatial dimension of the problem and $h$ a characteristic mesh size. At the same time, for a numerical method the spatial error is bound by 
\begin{align*}
||e||\leq c_2 h^{q}
\end{align*}
where $q$ characterise the order of accuracy (rate of convergence). For SEM the optimal order is $q=p+1$ with $p$ the order of the local polynomial expansions. So, to be efficient, we desire that the error can be reduced much faster than the work increases, i.e. that $||e||\to0$ faster than $W\to\infty$ when resolution is increased ($h\to0$). Thus,
\begin{align*}
\lim_{h\to0} W||e|| = c_3h^{q-d} 
\end{align*}
indicating that for larger dimensions, it is necessary that a higher order spatial discretisation is used or that the work is balanced or minimised, e.g. through adapting the mesh resolution. So in three spatial dimensions ($d=3$) it is not possible to be efficient for large-scale applications with low-order methods since the convergence rate is $q\leq 2$ for a low-order method. Another key requirement is the development of methods which leads to scalable work effort, i.e. linear scaling with the problem size when subject to mesh refinement pursuing better accuracy. We see that this is only possible if the order of accuracy at least matches the dimension of the problem, i.e. $q>d$.

\subsection{Paper contribution}

The key challenge we seek to address in this work is the development of a FNPF solver with high-order convergence rate that also has support for geometric flexibility for handling arbitrary body shapes and other structures. To this end we exploit and extend the results of the recent work of Engsig-Karup, Eskilsson \& Bigoni (2016) \cite{EngsigKarupEtAl2016} on a stabilised $\sigma$-transformed spectral element method for efficient and accurate marine hydrodynamics applications. We consider in this work a Mixed Eulerian Lagrangian (MEL) method \cite{LonguetHigginsCokelet1976} based on explicit time-stepping, we discard the $\sigma$-transformation to be able to include arbitrary shaped bodies in the fluid domain, and propose a remedy to the temporal stability issues associated with the use of asymmetric mesh configurations near the free surface boundary as identified \cite{Westhuis2001} for a classical low-order Galerkin finite element method and \cite{RobertsonSherwin1999} using a high-order spectral element method for a fully nonlinear potential flow solver. These new developments constitutes a new robust SEM methodology for nonlinear wave-body interactions.

\section{Mixed Eulerian-Lagrangian (MEL) formulation}

The governing equations for Fully Nonlinear Potential Flow (FNPF) is given in the following. Let the fluid domain $\Omega\subset \mathbb{R}^d$ ($d=2$) be a bounded, connected domain with piece-wise smooth boundary $\Gamma$ and introduce restrictions to the free surface $\Gamma^{\textrm{FS}}\subset \mathbb{R}^{d-1}$ and the bathymetry $\Gamma^{b}\subset \mathbb{R}^{d-1}$. Let $T:t\geq 0$ be the time domain. 
\begin{figure}[!htb]
\begin{center}
\includegraphics[height=2.75cm]{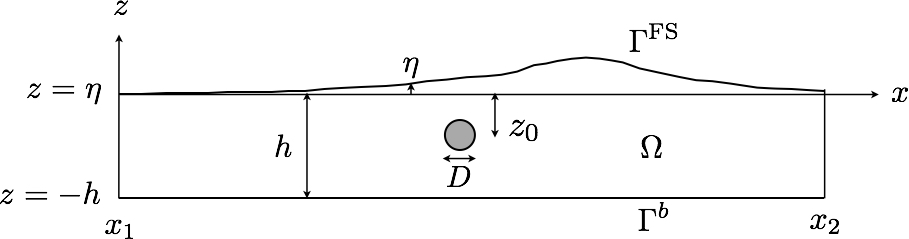}
\end{center}
\caption{Notations for physical domain ($\Omega$) with a cylinder introduced.}
\label{fig:domainorig}
\end{figure}
We seek a scalar velocity potential function  $\phi(x,z,t):\Omega\times T \to \mathbb{R}$ satisfying the Laplace problem
\begin{subequations}
\begin{align}
\phi  =  \tilde{\phi}, \quad z = \eta \quad &\textrm{on} \quad \Gamma^{\textrm{\textrm{FS}}} &\\
\nabla^2 \phi =  0, \quad -h(x)<z<\eta \quad &\textrm{in}\quad \Omega \label{Laplace} \\
\frac{\partial \phi}{\partial z}  + \frac{\partial h}{\partial x}\frac{\partial \phi}{\partial x} = 0, \quad z=-h(x) \quad &\textrm{on} \quad \Gamma^b \label{KB}
\end{align}
\label{eq:laplaceproblem}
\end{subequations}
where $h(x):\Gamma^{\textrm{FS}}\mapsto \mathbb{R}$ describes variation in the still water depth. The evolution of the free surface boundary is described by $\eta(x,t):\Gamma^{\textrm{FS}} \times T \to \mathbb{R}$. The notations are illustrated in Figure \ref{fig:domainorig}.

The MEL time marching technique assumes a particle-following (Lagrangian) reference frame for the free surface particles with changes in position in time given by 
\begin{align}
\frac{D{\bf x}}{Dt}=\boldsymbol{\nabla}\phi, \quad {\bf x}\in\Omega 
\label{LagkinbcFS}
\end{align}
where $\boldsymbol{\nabla}=\left(\tfrac{\partial}{\partial x},\tfrac{\partial}{\partial z}\right)$ having assumed only two space dimensions.

Thus, at the free surface boundary this implies kinematic conditions of the form
\begin{subequations}
\begin{align}
\frac{Dx}{Dt} &= \frac{\partial \phi}{\partial x}, \quad \frac{D\eta}{Dt} = \frac{\partial \phi}{\partial z} = w, \quad z=\eta,
\end{align}
that must be satisfied together with the dynamic boundary condition stated in terms of Bernoulli's equation
\begin{align}
\frac{D\phi}{Dt} = \frac{1}{2}\boldsymbol{\nabla}\phi\cdot\boldsymbol{\nabla}\phi - gz-p, \quad z=\eta,
\end{align}
\end{subequations}
where the material derivative connects the Eulerian and Lagrangian reference frame through the relation
\begin{align}
\frac{D}{Dt}\equiv \frac{\partial }{\partial t} + \boldsymbol{V} \cdot \boldsymbol{\nabla},
\label{eq:materialderivative}
\end{align}
where $\boldsymbol{V}$ is a velocity vector for the moving frame of reference that can be chosen arbitrarily. If $\boldsymbol{V}=0$ we obtain an Eulerian frame of reference and if $\boldsymbol{V}=\boldsymbol{\nabla} \phi$ we have a Lagrangian frame of reference. 

Spatial and temporal differentiation of free surface variables are given by the chain rules
\begin{align}
\frac{\partial \tilde{\phi}}{\partial x} &= \frac{\partial \phi}{\partial x} \Big|_{z=\eta} + \frac{\partial \phi}{\partial z}\Big|_{z=\eta}\frac{\partial \eta}{\partial x}, \quad \frac{\partial \tilde{\phi}}{\partial t} = \frac{\partial \phi}{\partial t} \Big|_{z=\eta} + \frac{\partial \phi}{\partial z}\Big|_{z=\eta}\frac{\partial \eta}{\partial t},
\label{eq:subexp}
\end{align}
and is useful for expressing the free surface equations valid at $z=\eta(x,t)$ in terms of free surface variables only, cf. \cite{EGNL13}, in the form
\begin{subequations}
\begin{align}
\frac{Dx}{Dt} &= \widetilde{\frac{\partial\phi}{\partial x}} = \frac{\partial \tilde{\phi}}{\partial x} - \tilde{w}\frac{\partial \eta}{\partial x}, \\
\frac{D\eta}{Dt} &= \tilde{w}, \\
\frac{D\tilde{\phi}}{Dt} &= 
 \frac{1}{2}\left(  \left(\frac{\partial\tilde{\phi}}{\partial x}\right)^2 - 2 \tilde{w}\frac{\partial\eta}{\partial x}\frac{\partial\tilde{\phi}}{\partial x} + \tilde{w}^2\left(\frac{\partial\eta}{\partial x}\right)^2 + \tilde{w}^2 \right) - g\eta,
\end{align}
\label{eq:MELForm}
\end{subequations}
having assumed a zero reference pressure at the free surface. The tilde '$\sim$' is used to denote that a variable is evaluated at the free surface, e.g.
$\tilde{w} = \widetilde{\tfrac{\partial\phi}{\partial z} } = \tfrac{\partial \phi}{\partial z} \big |_{z=\eta}$. Note, this formulation contains quartic nonlinear terms as in the Zakharov form \cite{ZAK1968} for the Eulerian formulation of FNPF equations. These nonlinear terms needs proper treatment to deal with aliasing effects. 

\subsection{Boundary conditions}

For the solution of the Laplace problem, the following free surface boundary condition is specified
\begin{align}
\label{eq:FSeqs}
\Phi=\tilde{\phi} \quad \textrm{on} \quad \Gamma^{\textrm{\textrm{FS}}},
\end{align}
while at fixed vertical boundaries, impermeable wall boundary conditions are assumed
\begin{align}
{\bf n}\cdot {\boldsymbol u} =0 \quad \textrm{on} \quad \Gamma\backslash (\Gamma^\textrm{\textrm{FS}}\cup\Gamma^b),
\end{align}
where ${\bf n}=(n_x,n_z)$ denotes an outward pointing unit normal vector to $\Gamma$. At rigid vertical wall boundaries the domain boundary conditions at free surface variables imposed are
\begin{align}
\partial_n \eta = 0, \quad \partial_n \phi =0 \quad \textrm{on} \quad \Gamma \cap \Gamma^\textrm{\textrm{FS}}.
\end{align}

Wave generation and absorption zones are included using the embedded penalty forcing technique described in \cite{EGNL13}. 

\section{Numerical discretisation}

Following \cite{EngsigKarupEtAl2016}, we present the discretization of the governing equations in a general computational framework based on the method of lines, where first a semi-discrete system of ordinary differential equations is formed by spatial discretisation in two space dimensions using a nodal SEM. The semi-discrete system is subject to temporal integration performed using an explicit fourth-order Runge-Kutta method.

\subsection{Weak Galerkin formulation and discretisation }

We form a partition of the domain $\Gamma_h^{\textrm{FS}}\subseteq\Gamma^{\textrm{FS}}$ to obtain a tessellation $\mathcal{T}_h^{\textrm{FS}}$ of $\Gamma^{\textrm{FS}}$ consisting of $N_{el}$ non-overlapping shape-regular elements $\mathcal{T}_h^{\textrm{FS},k}$ such that $\cup_{k=1}^{N_{el}}\mathcal{T}_h^{\textrm{FS},k}=\mathcal{T}_h^{\textrm{FS}}$ with $k$ denoting the $k$'th element. We introduce for any tessellation $\mathcal{T}_h$ the spectral element approximation space of continuous, piece-wise polynomial functions of degree at most $P$, 
\begin{align*}
V=\{ v_h\in C^0(\mathcal{T}_h); \forall k \in \{ 1,...,N_{el} \}, v_{h|\mathcal{T}_h^k}\in\mathbb{P}^P \}. 
\end{align*}
which is used to form finite-dimensional nodal spectral element approximations
\begin{align}
f_h(x,t) = \sum_{i=1}^{N^{\textrm{FS}}} f_i(t) N_i(x),
\label{eq:representation}
\end{align}
where $\{N_i\}_{i=1}^{N^{\textrm{FS}}}\in V$ is the $N^{\textrm{FS}}$ global finite element basis functions with cardinal property $N_i(x_j)=\delta_{ij}$ at mesh nodes with $\delta_{ij}$ the Kronecker Symbol. 

\subsubsection{Unsteady free surface equations}
The weak formulation of the free surface equations \eqref{eq:MELForm} is derived in the following form. Find $f\in V$ where $f\in\{x,\eta, \tilde{\phi}\}$ such that 
\begin{subequations}
\begin{align}
\int_{\mathcal{T}_h^{\textrm{FS}}}\frac{Dx}{Dt}v dx &= \int_{\mathcal{T}_h^{\textrm{FS}}}\left[ \frac{\partial \tilde{\phi}}{\partial x} - \tilde{w}\frac{\partial \eta}{\partial x}   \right] v dx, \\
\int_{\mathcal{T}_h^{\textrm{FS}}} \frac{D\eta}{Dt} v dx &= \int_{\mathcal{T}_h^{\textrm{FS}}}\left[\tilde{w}\right]vdx, \\
\int_{\mathcal{T}_h^{\textrm{FS}}}\frac{D\tilde{\phi}}{Dt} v dx &= \int_{\mathcal{T}_h^{\textrm{FS}}}\left[ \frac{1}{2}\left(  \left(\frac{\partial\tilde{\phi}}{\partial x}\right)^2 - 2 \tilde{w}\frac{\partial\eta}{\partial x}\frac{\partial\tilde{\phi}}{\partial x} + \tilde{w}^2\left(\frac{\partial\eta}{\partial x}\right)^2 + \tilde{w}^2 \right) - g\eta \right]vdx,
\end{align}
\label{FSproblem}
\end{subequations}
for all $v\in V$.
Substitute the expressions in \eqref{eq:representation} into \eqref{FSproblem} and choose $v(x)\in V$. The discretisation in one spatial dimension becomes 
\begin{subequations}
\label{eq:semidiscrete}
\begin{align}
\mathcal{M} \frac{Dx_h}{Dt} &= \mathcal{A}_x \tilde{\phi}_h - \mathcal{A}_x^{\tilde{w}_h}\eta_h, \\
 \frac{D\eta_h}{Dt} &=  \tilde{w}_h, 
\\
\mathcal{M}\frac{D\tilde{\phi}_h}{Dt} &= -\mathcal{M}^g \eta_h + \frac{1}{2} \left[ \mathcal{A}_x^{\left(\frac{\partial\tilde{\phi}}{\partial x}\right)_h} \tilde{\phi}_h +  \mathcal{M}^{\tilde{w}_h} \tilde{w}_h + \mathcal{A}_x^{\tilde{w}_h^2\left(\frac{\partial\eta}{\partial x}\right)_h} \eta_h\right ] -\mathcal{A}_x^{\tilde{w}_h\left(\frac{\partial\eta}{\partial x}\right)_h} \tilde{\phi}_h,
\end{align}
\end{subequations}
where the following global matrices have been introduced
\begin{align}
\mathcal{M}_{ij} \equiv \int_{\mathcal{T}_h^{\textrm{FS}}}N_iN_j dx, \quad \mathcal{M}^b_{ij} \equiv \int_{\mathcal{T}_h^{\textrm{FS}}}b(x) N_iN_j dx, \quad (\mathcal{A}_x^{b})_{ij} \equiv \int_{\mathcal{T}_h^{\textrm{FS}}}b(x) N_i \frac{dN_j }{d x}dx.
\label{eq:globalsem}
\end{align}
where $f_h\in\mathbb{R}^{N^{\textrm{FS}}}$ is a vector containing the set of discrete nodal values.

Following \cite{EngsigKarupEtAl2016}, the gradients of the free surface state variables are recovered via a global gradient recovery technique based on global Galerkin $L^2(\Gamma^{\textrm{FS}}_h)$ projections that work for arbitrary unstructured meshes in the SEM framework as described in \cite{EngsigKarupEtAl2016}. Aliasing effects are effectively handled using exact quadrature for nonlinear terms combined with a mild spectral filtering technique \cite{HW08} that gently removes high-frequency noise that may arise as a result of marginal resolution.

{\bf Remark}: The free surface node positions are changing in time, and this implies that the mesh must change accordingly. Thus, the scheme needs to recompute the global spectral element matrices \eqref{eq:globalsem} at every time step which impacts the computational efficiency of the scheme.

\subsubsection{Curvilinear iso-parametric elements}

To handle arbitrary body shapes we partition the fluid domain $\Omega_h$ to obtain another tessellation $\mathcal{H}_h$ consisting of $N_{el}^{2D}=N_{el}^Q+N_{el}^T$ non-overlapping shape-regular elements such that the tessellation can be formed by combining $N_{el}^Q$ quadrilateral and $N_{el}^T$ triangular curvilinear elements into an unstructured hybrid mesh such that $\mathcal{H}_h= \mathcal{Q}_h\cup \mathcal{T}_h = \left(\cup_{k=1}^{N_{el}^Q}\mathcal{Q}_k\right)\cup\left(\cup_{k=1}^{N_{el}^T}\mathcal{T}_k\right)$. In two space dimensions, 
the nodal spectral element approximations takes the form
\begin{align}
f_h(t,{\bf x}) = \sum_{i=1}^{n} f_i(t) N_i({\bf x}).
\label{eq:representation}
\end{align}
where $n$ is the total degrees of freedom in the discretisation.

\begin{figure}[!htb]
\begin{center}
\begin{minipage}{0.45\textwidth}
\begin{center}
(a) Hybrid Mesh 1 \\
\includegraphics[height=2.45cm]{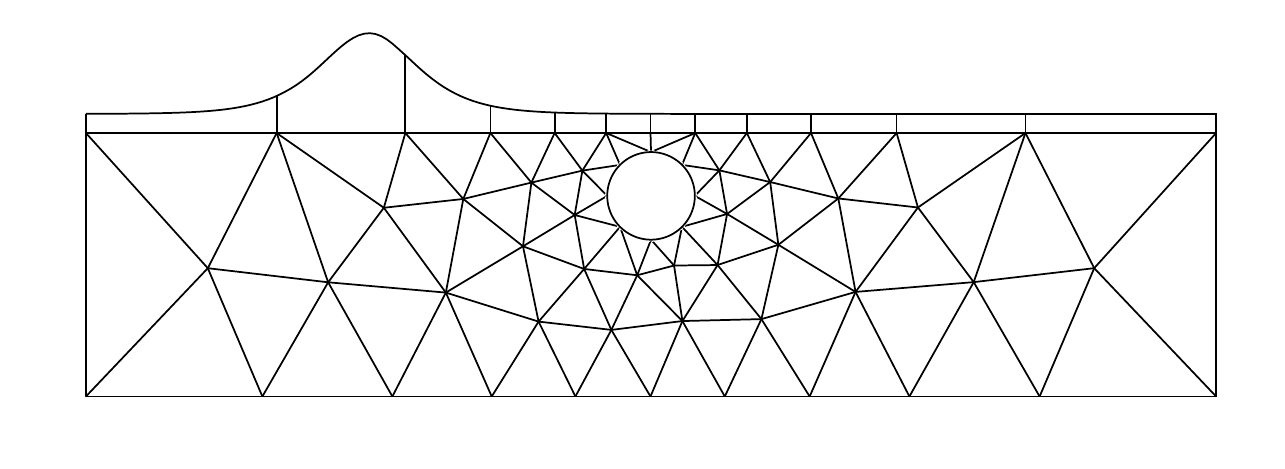} 
\end{center}
\end{minipage}
\;\;
\begin{minipage}{0.45\textwidth}
\begin{center}
(b) Hybrid Mesh 2 \\
\includegraphics[height=2.45cm]{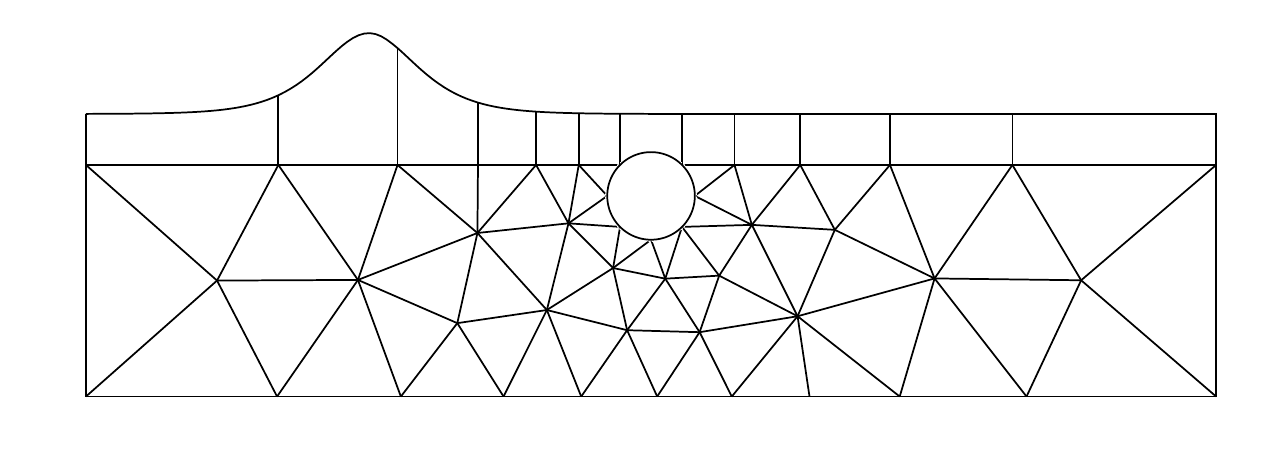} 
\end{center}
\end{minipage}
\vspace{-0.3cm}
\caption{Illustration of two mesh topologies for a fully submerged cylinder. A curvilinear layer of quadrilaterals is used near the free surface. (a) The cylinder is represented using curvilinear triangles, and (b) the cylinder is represented using a hybrid combination of curvilinear quadrilateral and curvilinear triangular elements.}
\label{fig:hybridmeshes}
\end{center}
\end{figure}

\begin{figure}[!htb]
\begin{center}
\subfloat[Triangle]{\includegraphics[height=2.75cm]{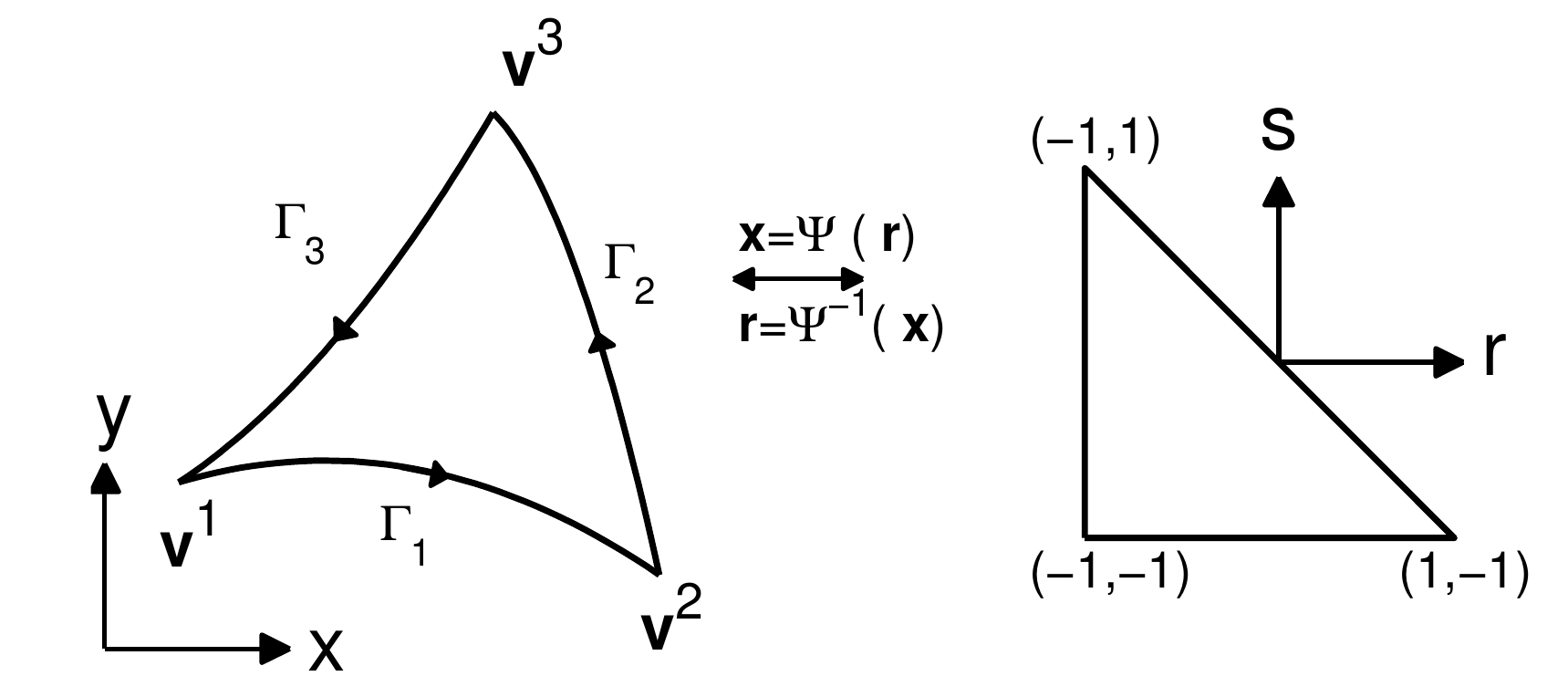}} \quad 
\subfloat[Quadrilateral]{\includegraphics[height=2.75cm]{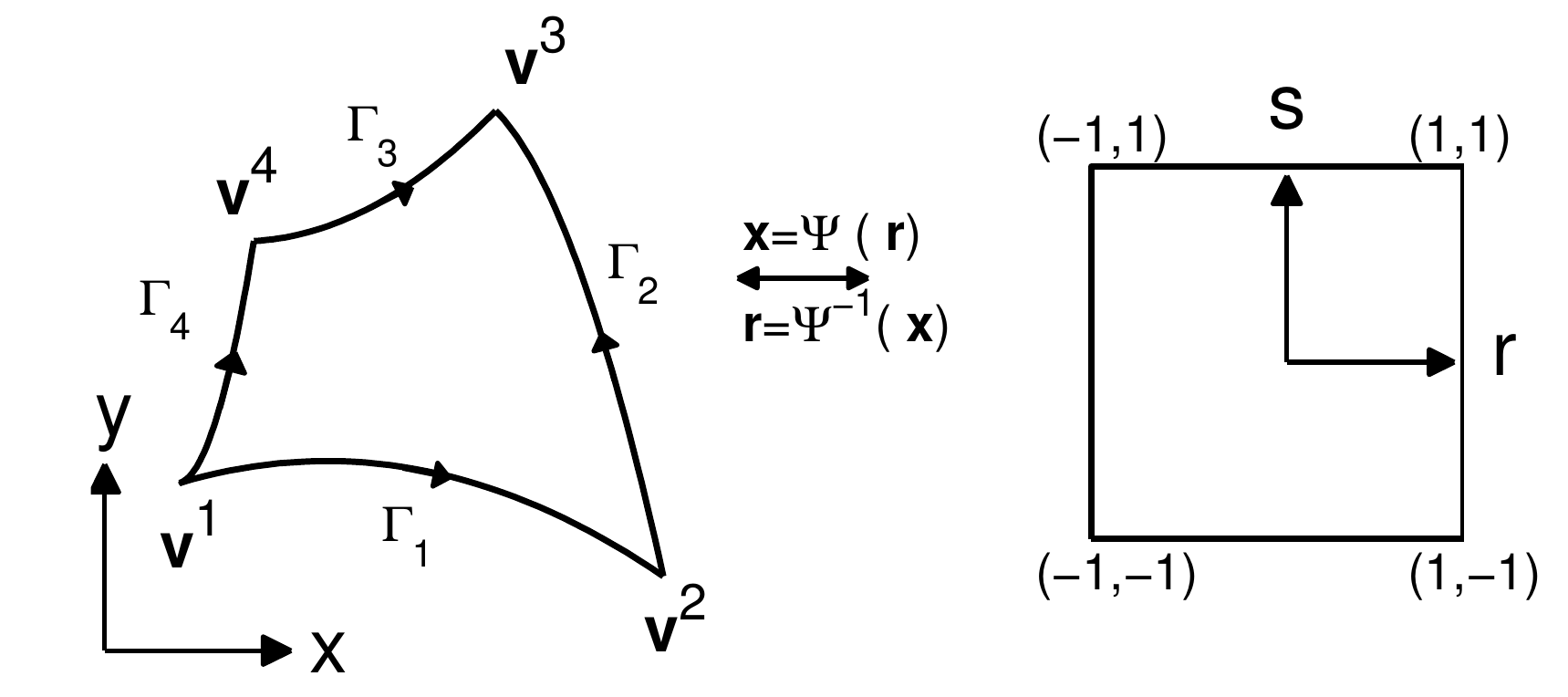}} 
\end{center}
\caption{Conventions for curvilinear elements and their transformation to a reference element.}
\label{fig:elmtrans}
\end{figure}

The curvilinear elements makes it possible to treat the deformations in the free surface and the body surfaces as illustrated in Figure \ref{fig:hybridmeshes} with two different hybrid unstructured meshes. In both cases a single quadrilateral layer is used just below the free surface.
Consider the $k$'th element $\mathcal{H}_{h}^k\subset \mathcal{H}_h$. On this element, we form a local polynomial expansion expressed as 
\begin{align}
f_h^k({\bf x},t) = \sum_{j=1}^{N_P} \hat{f}_j^k(t)\phi_j(\Psi_k^{-1}({\bf x})) = \sum_{j=1}^{N_P} f_j^k(t)L_j(\Psi_k^{-1}({\bf x})),
\end{align}
where both a modal and a nodal expansion in the reference element is given in terms of $N_p$ nodes/modes. We have introduced a map to take nodes from the physical element to a reference element, $\Psi_k:\mathcal{H}_h^k\to\mathcal{H}_{r}$ where $\mathcal{H}_r$ is a single computational reference element. This dual representation can be exploited to form the curvilinear representations, since the coefficient vectors are related through
\begin{align}
 {\bf f}_h = \mathcal{V} \boldsymbol{\hat{f}} , \quad \mathcal{V}_{ij} = \phi_j({\bf r}_i),
\end{align}
where $\phi_j$, $j=1,2...,N_P$ is the set of orthonormal basis functions and with nodes ${\bf r}=\{r_i\}_{i=1}^{N_p}$ in the reference element that defines the Lagrangian basis. For quadrilaterals a tensor product grid formed by Legendre-Gauss-Lobatto nodes in 1D is used. For triangles, the node distribution is determined using the explicit warp \& blend procedure \cite{War06}. The one-to-one mapping from a general curvilinear element to the reference element is highlighted in Figure \ref{fig:elmtrans}. The edges of the physical quadrilateral elements are defined by the functions $\Gamma_j$, $j=1,2,3,4$, and by introducing iso-parametric polynomial interpolants \cite{Zienkewitch1971} of the same order as the spectral approximations of the form
\begin{align}
\mathcal{I}_N\Gamma_j(s) = \sum_{n=0}^N \Gamma(s_n)h_n(s), \quad j=1,...,4,
\end{align}
it is possible to represent curved boundaries.
Using transfinite interpolation with linear blending \cite{GH73} the affine transformation from the square reference quadrilateral to the physical domain is defined in terms of these edge curves in the form
\begin{align}
{\bf \Psi}_k(r,s)&=\tfrac{1}{2}\left[ (1-r)\Gamma_4(s)+(1+r)\Gamma_2(s)+(1-s)\Gamma_1(r)+(1+s)\Gamma_3(r) \right] \nonumber \\
&-\tfrac{1}{4}\left[ (1-r)\{ (1-s)\Gamma_1(-1) + (1+s)\Gamma_3(-1)  \} \right. \nonumber \\
&+\left. (1-r)\{ (1-s)\Gamma_1(1) + (1+s)\Gamma_3(1)  \} \right].
\end{align}
Similarly, using transfinite interpolation with linear blending, the transformation for triangles is given as
\begin{align}
{\bf \Psi}_k(r,s)&=\tfrac{1}{2}\left[ (1-s)\Gamma_1(r)-(1+r)\Gamma_1(-s)+(1-r)\Gamma_2(s)-(1+s)\Gamma_2(-r) \right] \nonumber \\
+& (1+\tfrac{(r+s)}{2})\Gamma_3(r) - \tfrac{(1+r)}{2}\Gamma_2(-(1+r+s)) + \tfrac{1+r}{2} \Gamma_1(1) + \tfrac{r+s}{2} \Gamma_1(-1).
\end{align}
The use of curvilinear elements, implies that the Jacobian of the mapping is no longer constant as for straight-sided triangular elements. Thus, to avoid quadrature errors higher order quadratures are employed in the discrete Galerkin projections and this increases the cost of the scheme proportionally for the elements in question. Note, the introduction of curvilinear elements, can be setup in a pre-processing step and therefore does not add any significant additional complexity to the numerical scheme.

\subsubsection{Spatial discretisation of the Laplace problem}

Consider the discretisation of the governing equations for the Laplace problem \eqref{eq:laplaceproblem}. We seek to construct a linear system of the form
\begin{align}
\label{eq:laplaceproblemdiscrete}
\mathcal{L} \Phi_h = {\bf b}, \quad \mathcal{L}\in\mathbb{R}^{n\times n}, \quad \Phi_h, {\bf b}\in\mathbb{R}^{n}.
\end{align}
The starting point is a weak Galerkin formulation that can be expressed as: find $\Phi\in V$ such that 
\begin{align}
\int_{\mathcal{H}_h} \nabla\cdot( \nabla\Phi)  v d\boldsymbol{x} = \oint_{\partial\mathcal{H}_h} v {\bf n}\cdot (\nabla\Phi) d\boldsymbol{x}  - \int_{\mathcal{H}_h}  ( \nabla\Phi)  \cdot (\nabla v) d\boldsymbol{x} = 0,
\end{align}
for all $v\in V$ where the boundary integrals vanish at domain boundaries where impermeable walls are assumed. 
The discrete system operator is defined as
\begin{align}
\mathcal{L}_{ij} \equiv -\int_{\mathcal{H}_h} (\nabla N_j)  \cdot (\nabla N_i) d\boldsymbol{x}  = -\sum_{k=1}^{N_{el}^{2D}}   \int_{\mathcal{H}_h^k} (\nabla N_j)  \cdot (\nabla N_i) d\boldsymbol{x} .
\end{align}
The elemental integrals are approximated through change of variables as
\begin{align}
&\int_{\mathcal{H}_h^k} (\nabla N_j)  \cdot (\nabla N_i) d\boldsymbol{x}  = \int_{\mathcal{H}_r} |\mathcal{J}^k| (\nabla N_j ) \cdot (\nabla N_i) d\boldsymbol{r},
\end{align}
where $\mathcal{J}^k$ is the Jacobian of the affine mapping $\chi^k:\mathcal{H}_h^k\to\mathcal{H}_{r}$. The global assembly of this operator preserves the symmetry, and the resulting linear system is modified to impose the Dirichlet boundary conditions \eqref{eq:FSeqs} at the free surface. The vertical free surface velocity $\tilde{w}_h$ is recovered from the potential $\Phi_h$ via a global Galerkin $L^2(\mathcal{H}_h)$ projection that involves a global matrix for the vertical derivative.

\section{Numerical properties}
\label{sec:numanal}

We start out by considering the numerical properties of the model related to the temporal stability and convergence of the numerical MEL scheme. Results of comparison with the stabilised Eulerian formulation \cite{EngsigKarupEtAl2016} is included since the two schemes are complementary.

\subsection{Temporal linear stability analysis of semi-discrete system}

We revisit the analysis of temporal instability following the works of \cite{Westhuis2001,RobertsonSherwin1999} by considering the semi-discrete free surface formulation that arise under the assumption of small-amplitude waves
\begin{subequations}
\begin{align}
\frac{Dx}{Dt} &= \frac{\partial \tilde{\phi}}{\partial x}, \\
\frac{D\eta}{Dt} &= \tilde{w}, \\
\frac{D\tilde{\phi}}{Dt} &= - g\eta,
\end{align}
\label{eq:MELFormLinear}
\end{subequations}
The discretization of this semidiscrete systems leads to 
\begin{align}
\frac{D}{Dt} \left[ \begin{array}{c} x \\ \eta \\ \tilde{\phi} \end{array} \right] = \mathcal{J} \left[ \begin{array}{c} x \\ \eta \\ \tilde{\phi} \end{array} \right] 
, \quad \mathcal{J} = \left[ \begin{array}{ccc} \boldsymbol{0} & \boldsymbol{0} & \mathcal{M}^{-1}\mathcal{A}_x \\  \boldsymbol{0} & \boldsymbol{0} & \mathcal{J}_{23} \\ \boldsymbol{0} & -g \mathcal{I} & \boldsymbol{0} \end{array} \right] ,
\end{align} 
where $\mathcal{I}_{ij}=\delta_{ij}$ and
\begin{align}
\mathcal{J}_{23} &= [(\mathcal{D}_z)_{bi} \phi_i + (\mathcal{D}_z)_{bb} \phi_b] = [ (\mathcal{D}_z)_{bb} -  (\mathcal{D}_z)_{bi}\mathcal{L}_{ii}^{-1}\mathcal{L}_{ib}  ]\phi_{b}, 
\end{align}
where 
\begin{align}
\mathcal{D}_z = \hat{\mathcal{M}}^{-1}\hat{\mathcal{A}}_z, \quad (\hat{\mathcal{A}}_z)_{ij} \equiv \iint_{\mathcal{T}_h} N_i \frac{dN_j }{d z}dxdz, \quad \hat{M}_{ij} \equiv \iint_{\mathcal{T}_h}N_iN_j dxdz.
\end{align}
and having introduced matrix decompositions of the global matrices of the form
\begin{align}
\mathcal{L} = \left[ \begin{array}{cc} \mathcal{L}_{bb} & \mathcal{L}_{bi} \\ \mathcal{L}_{ib} & \mathcal{L}_{ii}  \end{array} \right],
\end{align}
where the subscript indices 'b' refers to the free surface nodes and the 'i' refers to all interior nodes. The eigenspectrum of $\lambda(\mathcal{J})$ determines the temporal stability of this system.
\begin{figure}[!htb]
  \centering
  \begin{minipage}[c]{.24\textwidth}
  \subfloat[Mesh]{\includegraphics[width=\textwidth]{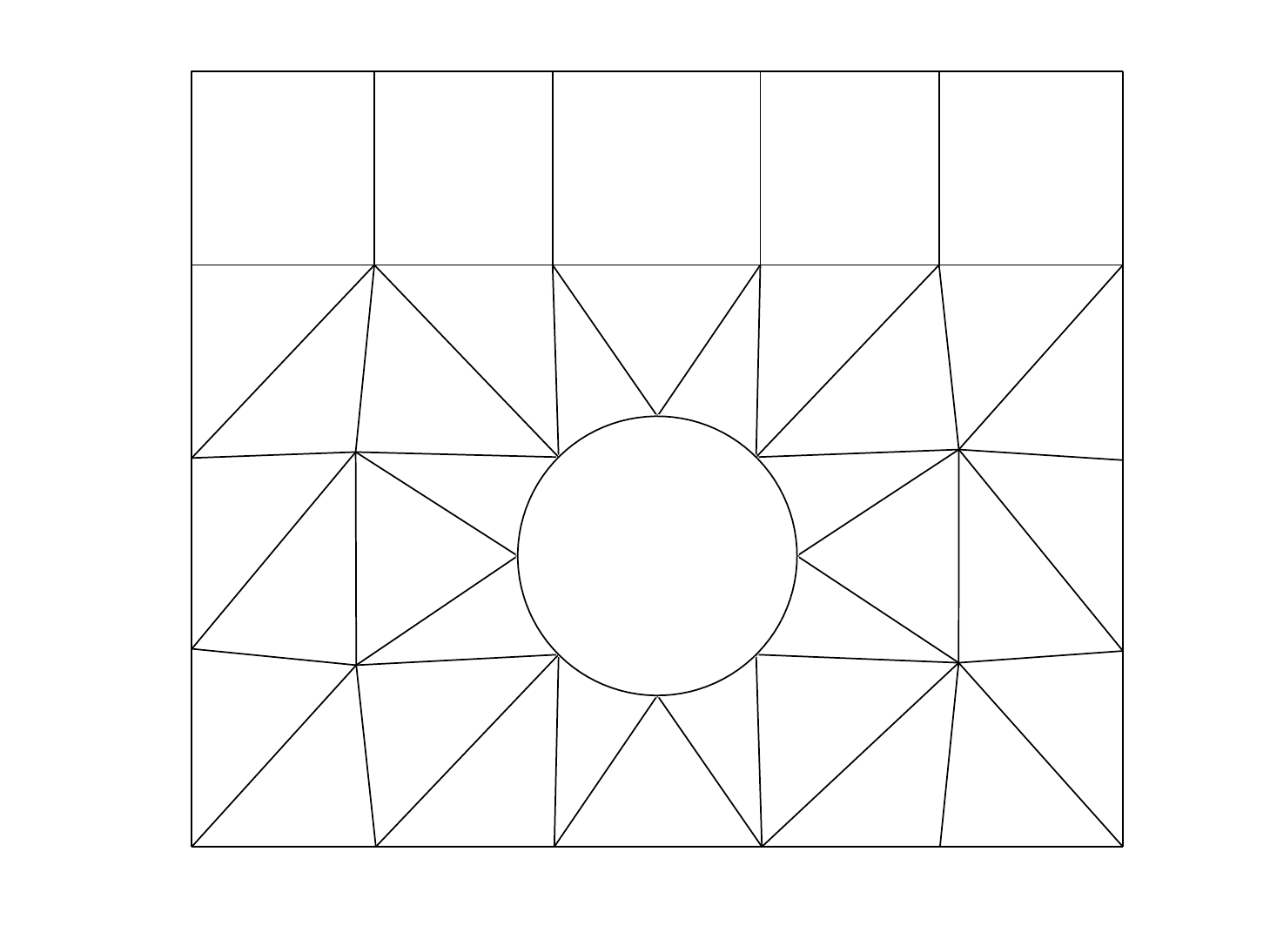}} 
  \end{minipage}
\begin{minipage}[c]{.75\textwidth}
  \subfloat[Eigenvalues]{\includegraphics[width=\textwidth]{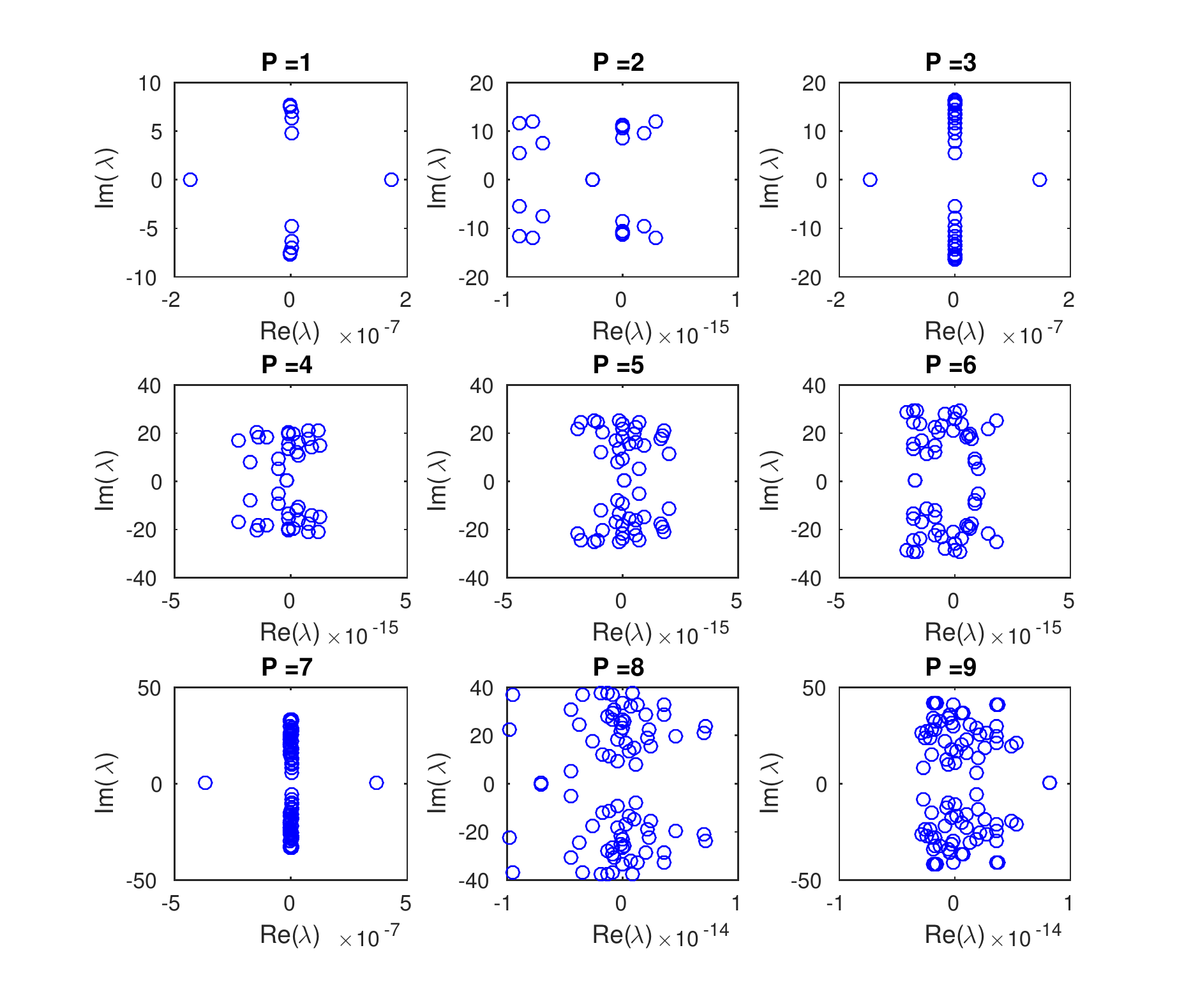}} 
  \end{minipage}
  \caption{Linear stability analysis. Purely imaginary eigenvalues for different polynomial order to within roundoff accuracy corresponding to hybrid unstructured mesh with submerged circular cylinder.} 
\end{figure}

\begin{figure}
  \centering
    \begin{minipage}[c]{.24\textwidth}
  \subfloat[Mesh]{\includegraphics[width=\textwidth]{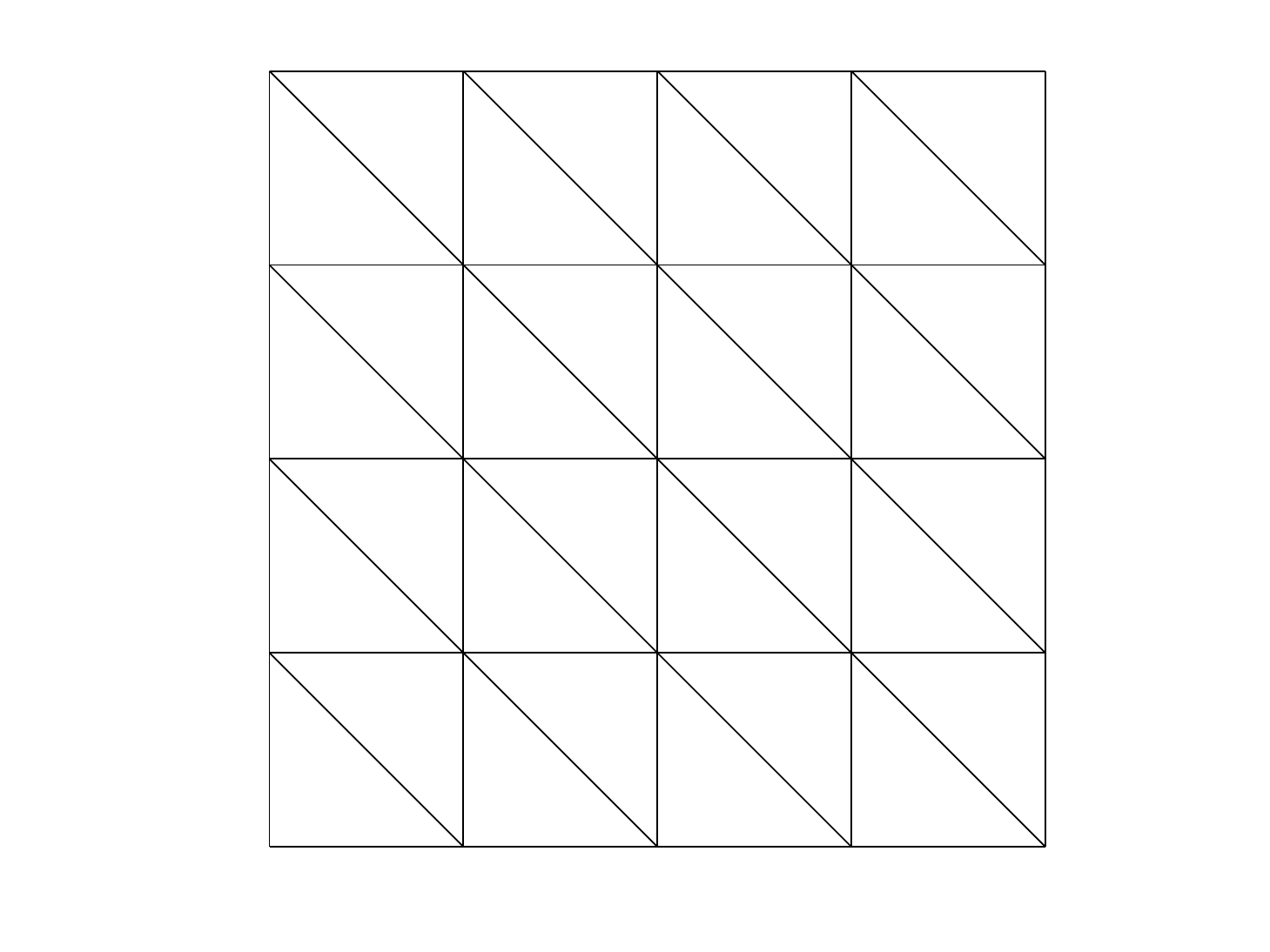}\label{fig:AC_Mesh_asym_4x4}} 
  \end{minipage}
    \begin{minipage}[c]{.75\textwidth}
  \subfloat[Eigenvalues]{\includegraphics[width=\textwidth]{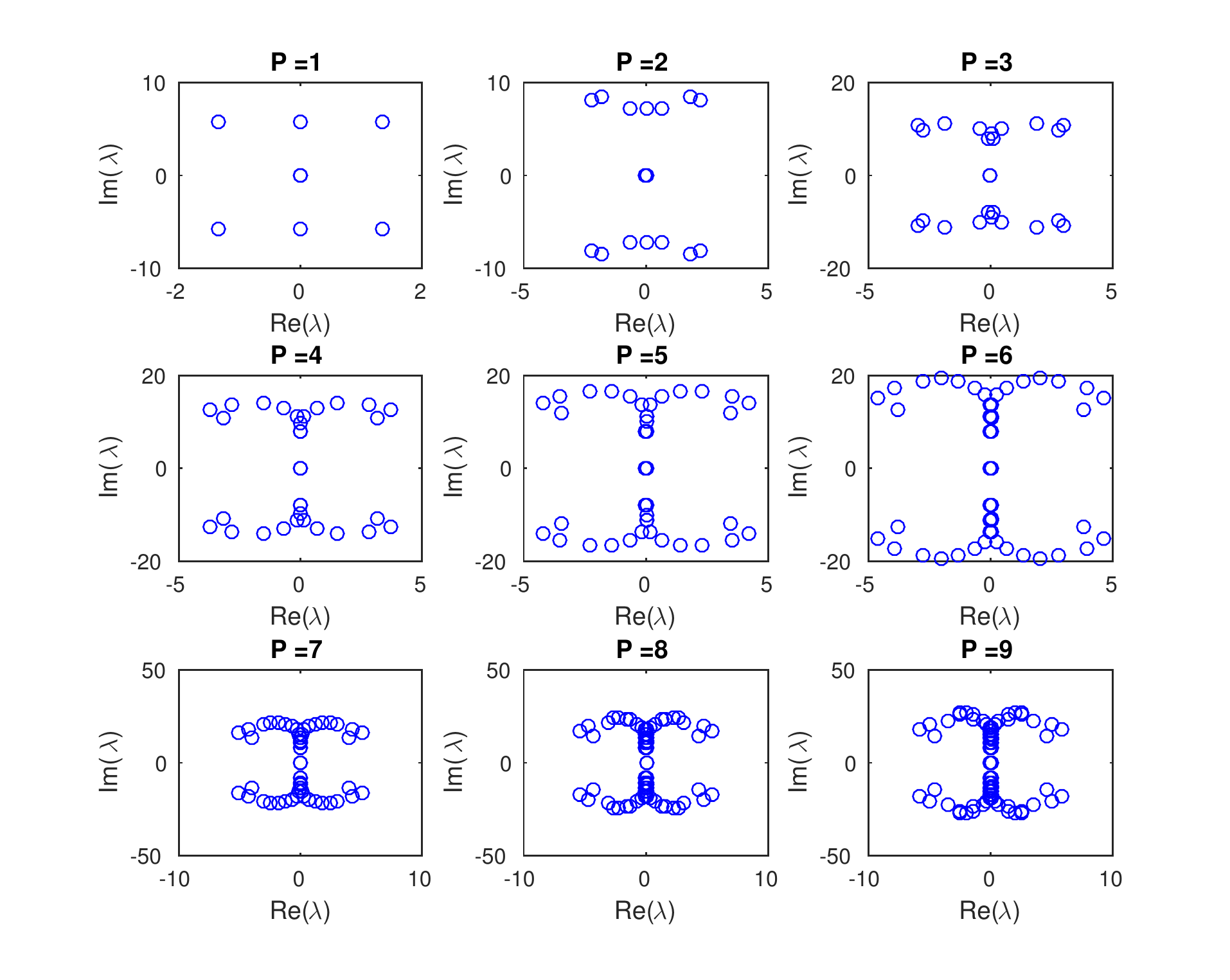}
   \label{fig:AC_Eig_asym_4x4}} 
  \end{minipage}
  \caption{Linear stability analysis. Eigenvalues for different polynomial order corresponding to an asymmetric structured mesh of triangles. Positive real part of eigenvalues causes temporal instability.} \label{fig:AC_asym_4x4}
\end{figure}

\begin{figure}
  \centering
    \begin{minipage}[c]{.24\textwidth}
  \subfloat[Mesh]{\includegraphics[width=\textwidth]{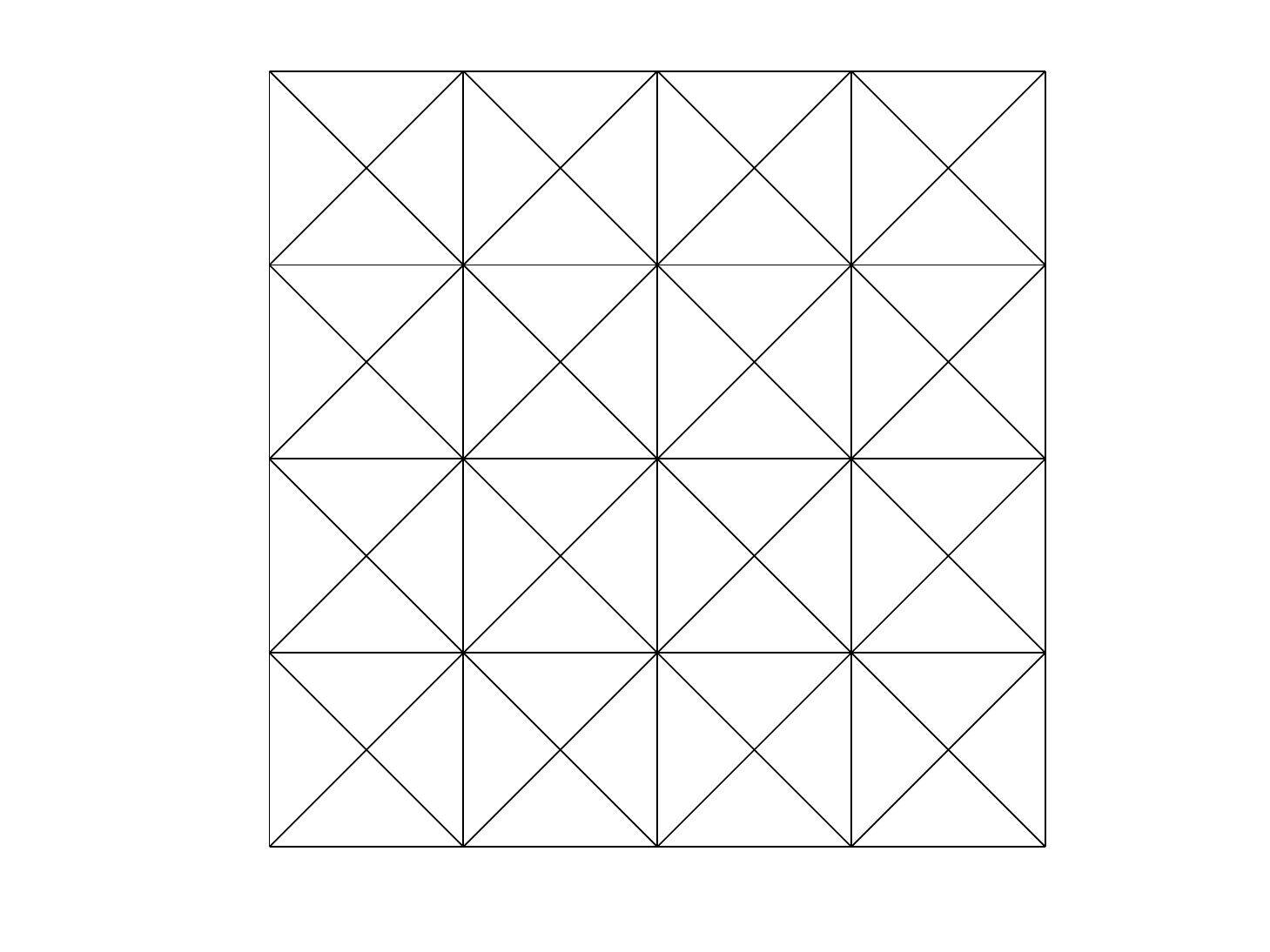}\label{fig:AC_Mesh_sym_4x4}} 
  \end{minipage}
    \begin{minipage}[c]{.75\textwidth}
  \subfloat[Eigenvalues]{\includegraphics[width=\textwidth]{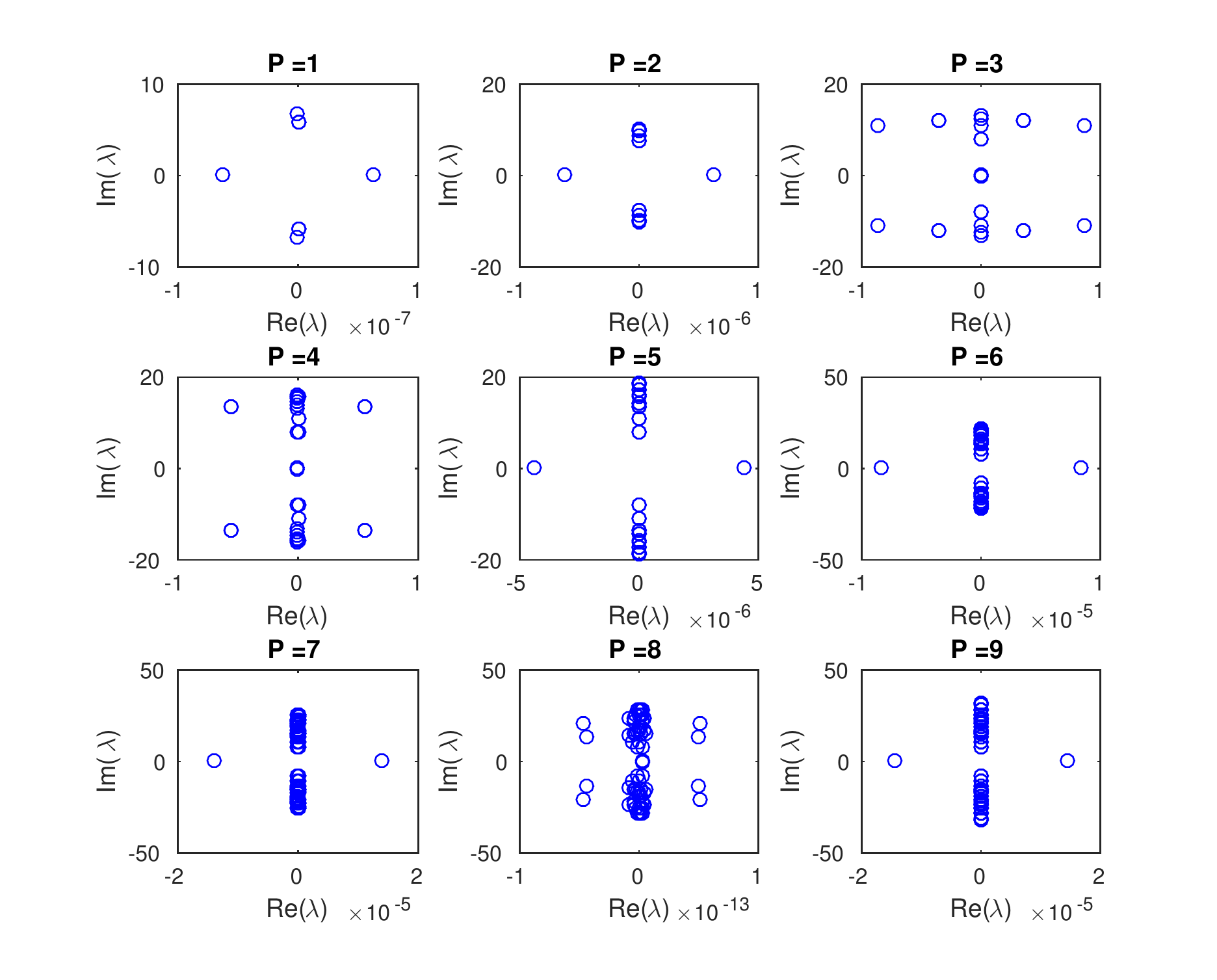}
   \label{fig:AC_Eig_sym_4x4}} 
  \end{minipage}
  \caption{Linear stability analysis. Eigenvalues for different polynomial order corresponding to a symmetric structured mesh of triangles. Positive real part of eigenvalues causes temporal instability for polynomial orders 3 and 4.} \label{fig:AC_sym_4x4}
\end{figure}

\begin{figure}
  \centering
    \begin{minipage}[c]{.24\textwidth}
  \subfloat[Mesh]{\includegraphics[width=\textwidth]{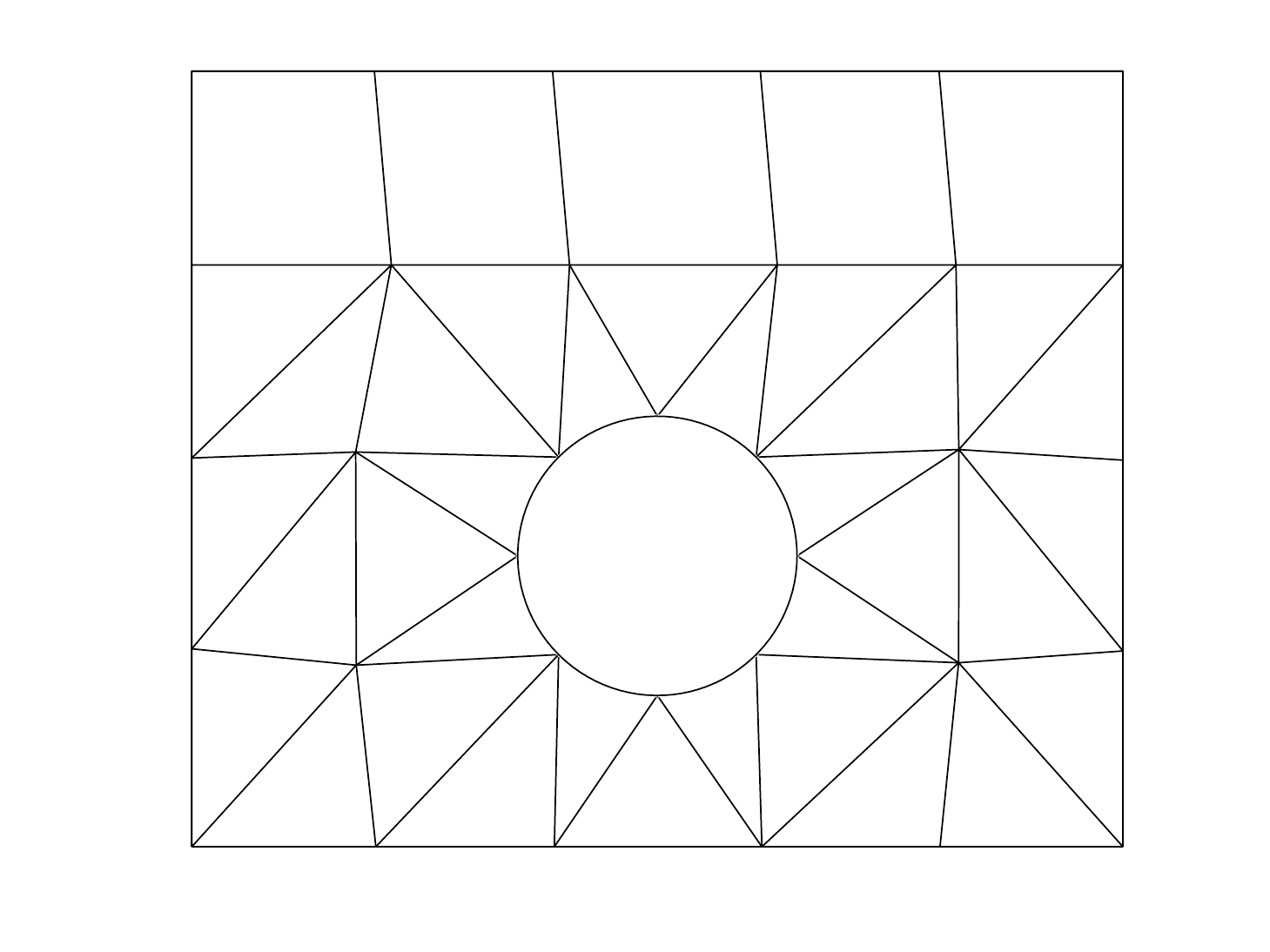}
  \label{fig:AC_Mesh_asym_triquad}} 
  \end{minipage}
    \begin{minipage}[c]{.75\textwidth}
  \subfloat[Eigenvalues]{\includegraphics[width=\textwidth]{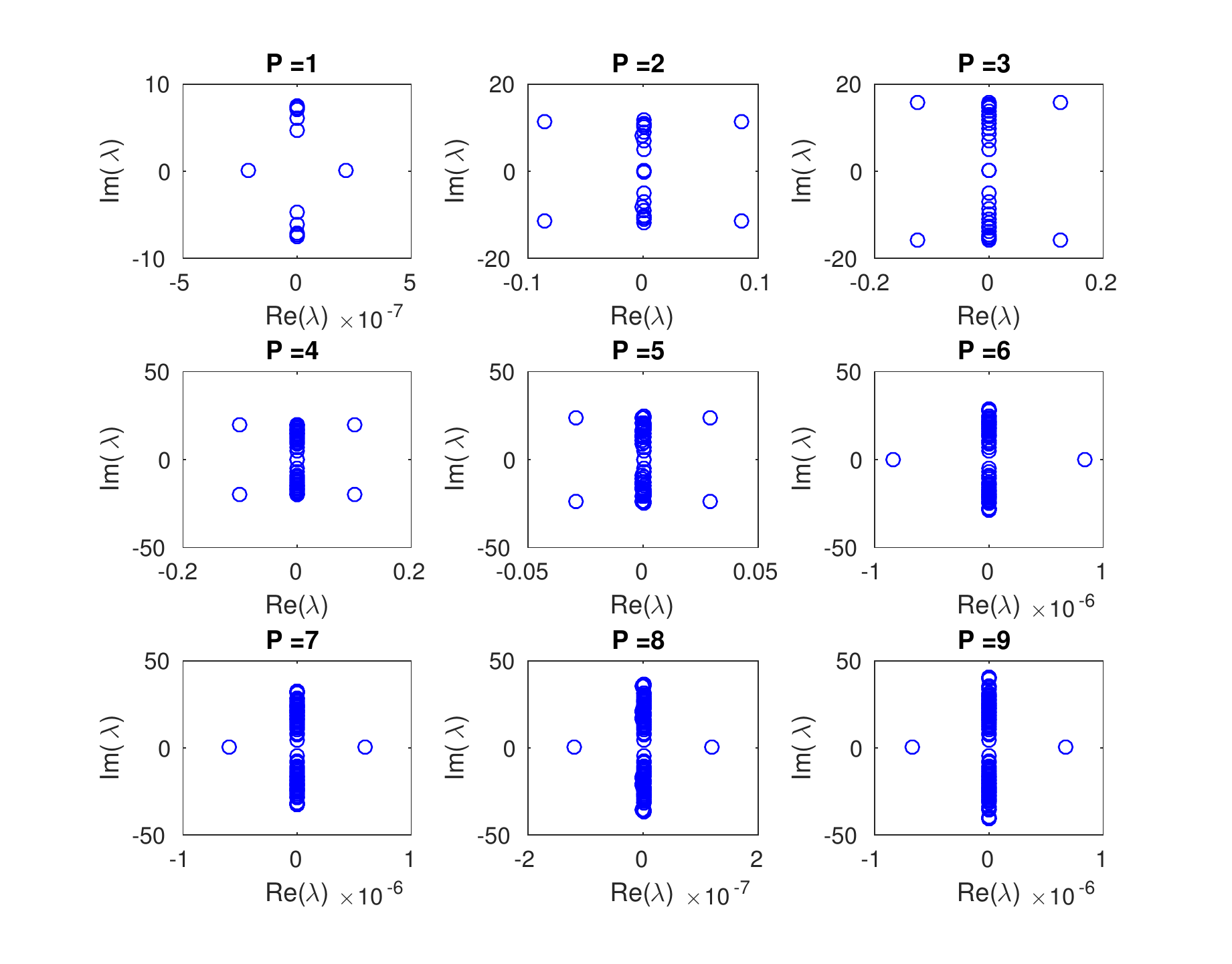} 
   \label{fig:AC_Eig_asym_triquad}} 
  \end{minipage}
  \caption{Linear stability analysis. Eigenvalues for different polynomial order corresponding to an asymmetric structured mesh of quads and curvilinear triangles. Positive real part of eigenvalues causes temporal instability.} \label{fig:AC_asym_triquad}
\end{figure}

In the context of SEM, we seek to first confirm the results given in \cite{RobertsonSherwin1999}. By doing a similar eigenanalysis using a triangulated asymmetric mesh, we confirm that we have temporal instability, e.g. see representative results in Figure \ref{fig:AC_asym_4x4}. To fix this problem, we need to avoid asymmetric meshes near the free surface layer. Following \cite{RobertsonSherwin1999} we can also consider a triangulated symmetric mesh in Figure \ref{fig:AC_Eig_sym_4x4} which turns out to be stable for all polynomial orders except 3 and 4 used in the analysis. This is in line with the results and conclusions presented by Robertson \& Sherwin (1999), but reveals that numerical issues may arise for such triangulated meshes in specific configurations. Furthermore, since our objective is to introduce arbitrary shaped bodies inside the fluid domain, in the next experiment we use a hybrid mesh. This mesh is consisting of a layer of quadrilateral below the free surface of the fluid similar to the meshes used in \cite{EngsigKarupEtAl2016} and combine this layer with an additional triangulated layer used to represent complex geometries such as a submerged cylinder as illustrated in Figure \ref{fig:AC_18}. The eigenanalysis shows that we then reach temporal stability for arbitrary polynomial expansion orders. These results are in line with other experiments we have carried out that are not presented here, confirming that by introducing a quadrilateral layer we can fix the temporal stability problem to obtain purely imaginary eigenspectra (to machine precision). Using instead a similar mesh but with slightly skewed quadrilaterals, reveals again that the mesh asymmetry leads to temporal instability as shown in Figure \ref{fig:AC_asym_triquad}. So, the temporal instability is associated with the accuracy of the vertical gradient approximation that is used to compute $\tilde{w}$ at the free surface and that determines the dispersive properties of the model. So poor accuracy in $\tilde{w}$ destroys the general applicability of the model since the wave propagation cannot be resolved accurately. This makes it clear that a quadrilateral layer with vertical alignment of nodes close to the free surface provides accurate recovery of the vertical free surface velocities and fixes the temporal instability problem.

{\bf Remark: } This new insight paves the road towards also stabilising nonlinear simulations using a high-order accurate unstructured method such as SEM following the recipe for stabilisations recently laid out in \cite{EngsigKarupEtAl2016} for an Eulerian scheme based on quadrilaterals only. 

\begin{figure}[!htb]
 \centering
\subfloat[Steep stream function wave]{
  \includegraphics[width=0.475\textwidth]{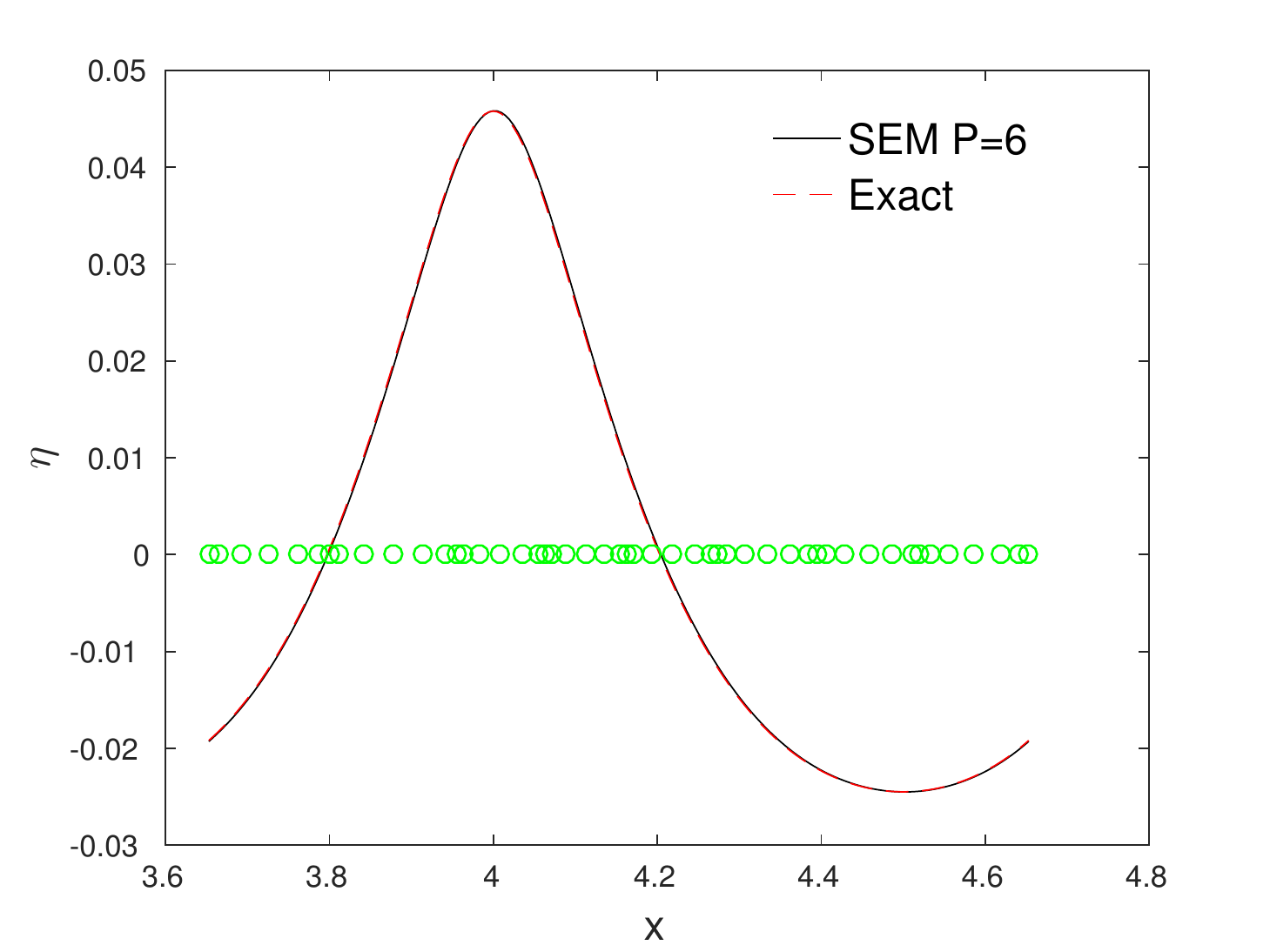} 
  \label{fig:4_Lima}} \,
\subfloat[Mass and energy]{
  \includegraphics[width=0.475\textwidth]{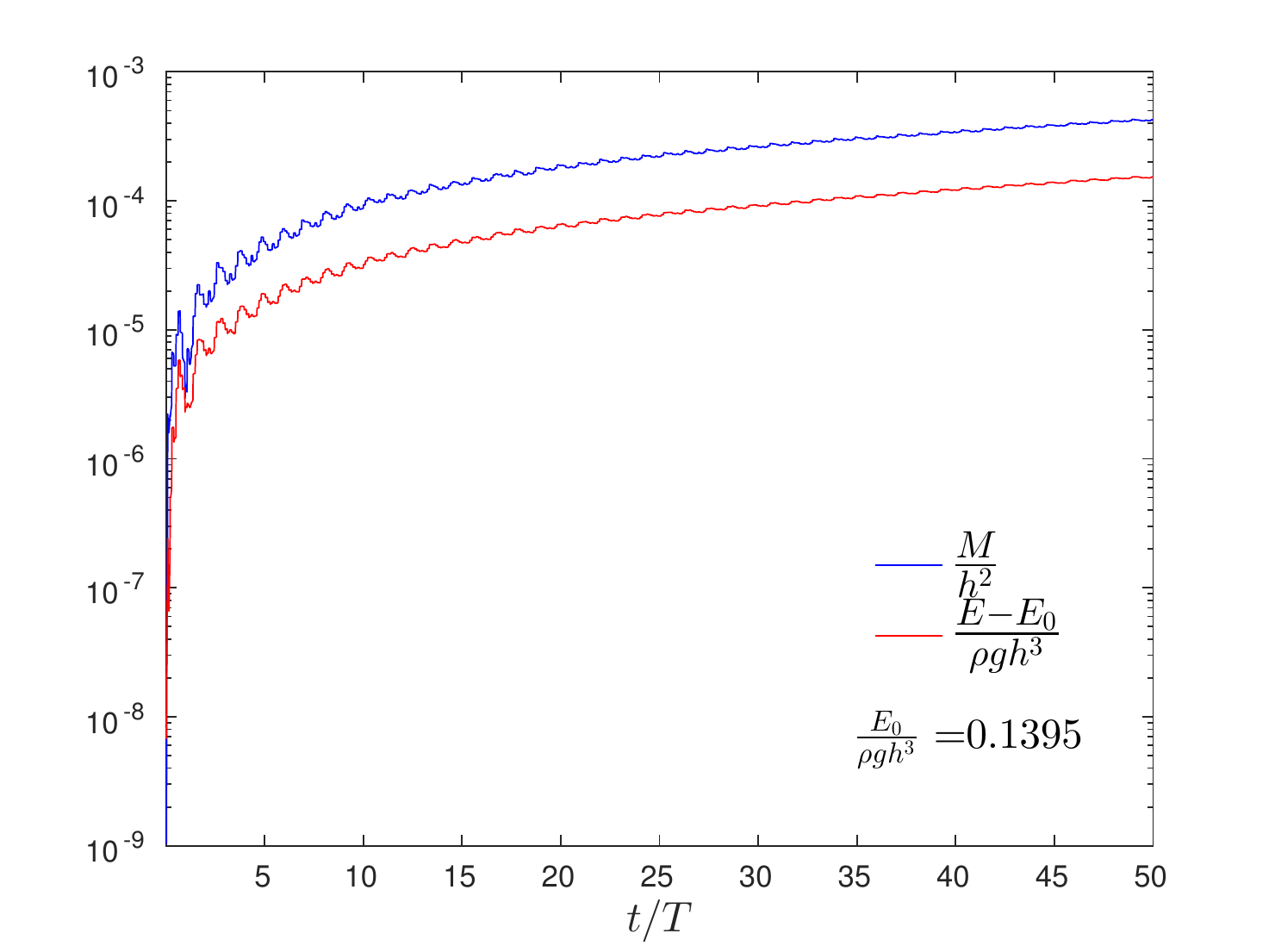} 
  \label{fig:4_Limb}} \\
  \subfloat[Re-meshing off]{
  \includegraphics[width=0.475\textwidth]{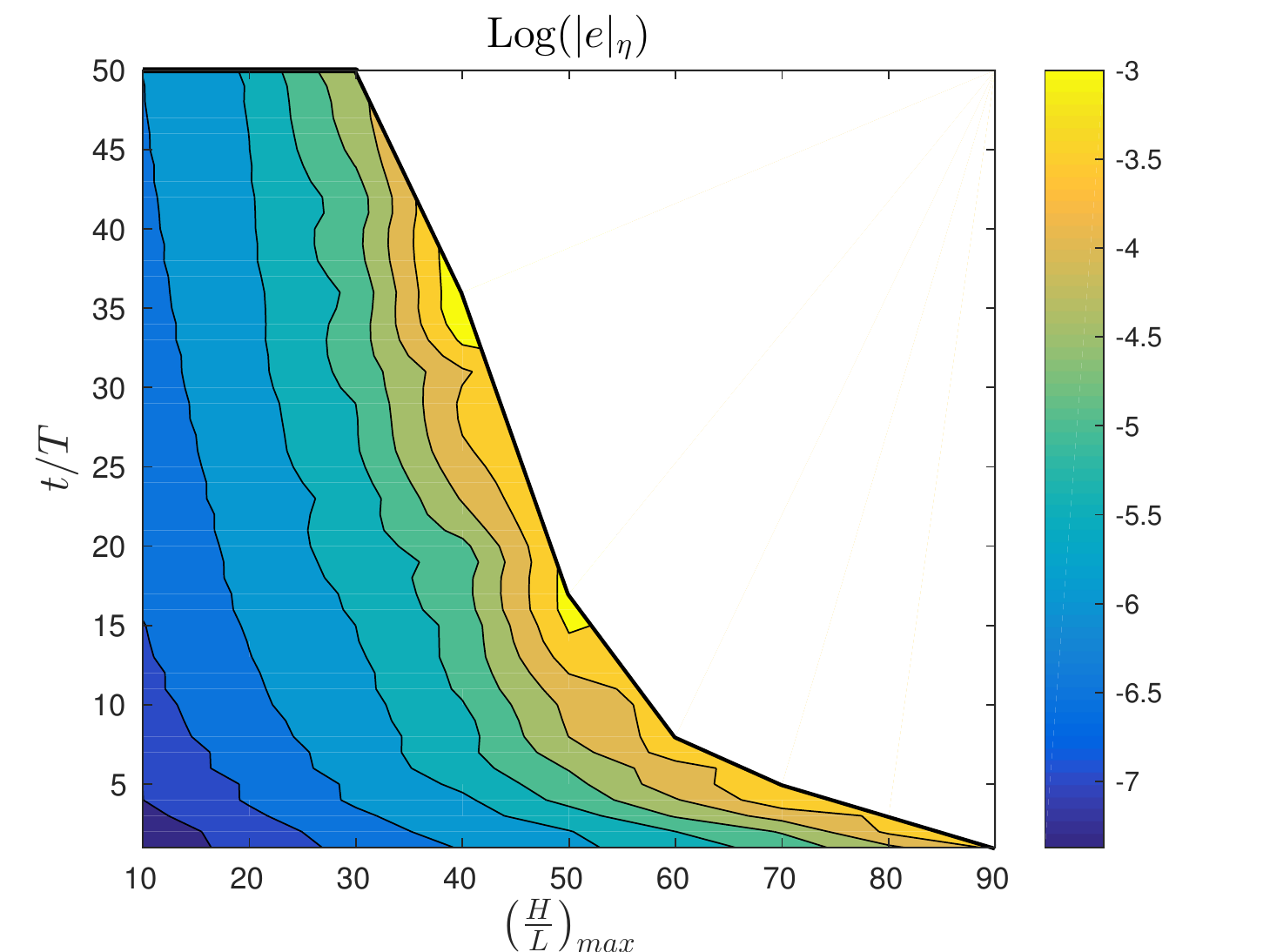} 
  \label{fig:4_Limc}} \,
\subfloat[Re-meshing on]{
  \includegraphics[width=0.475\textwidth]{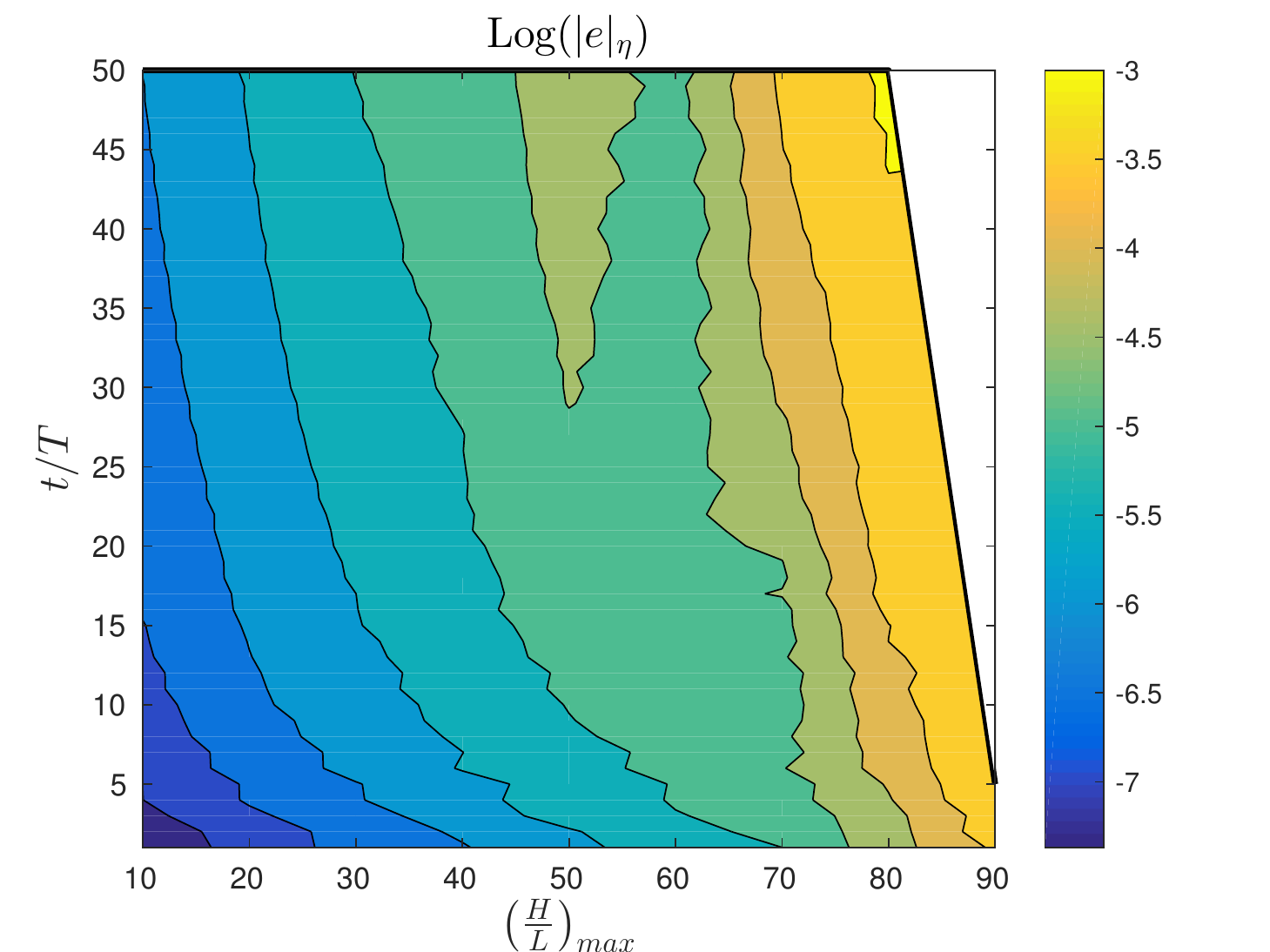} 
  \label{fig:4_Limd}} 
\caption{Stabilisation of $(H/L)_{max}=70 \%$ stream function wave with dispersion parameter $kh=1$ using a re-meshing algorithm. (a) Snapshot at time $t/T=50$. (b) Mass and energy conservation history. Contour plots of $\log(|e|_{\eta})$ (c) without re-meshing and (d) with re-meshing. Experiments corresponding to $T/\Delta t=80$ using a mesh with 8$\times$1 elements of polynomial order $P=6$. Exact integrations and a $1\%$ top mode spectral filter is employed.}
  \label{fig:4_Lim}
\end{figure}

In the following, we consider the strenuous benchmark of nonlinear wave propagation of steep nonlinear stream function waves by setting up a mesh with periodicity conditions imposed to connect the eastern and western boundaries. Then, we examine the conservation of energy and accuracy to assess the stability of the model. This analysis shows that the exact integration used in the discrete global projections has a big effect on stabilising the solution. Following \cite{EngsigKarupEtAl2016}, a test is performed for different amounts of filtering in the last mode and for different maximum steepness ($(H/L)_{max}$). For example, the analysis in Figure \ref{fig:4_Lim} demonstrates that a 1\% filter works well for the MEL Formulation. 

\begin{figure}[!htb]
  \centering
\subfloat[MEL formulation]{
  \includegraphics[width=0.475\textwidth]{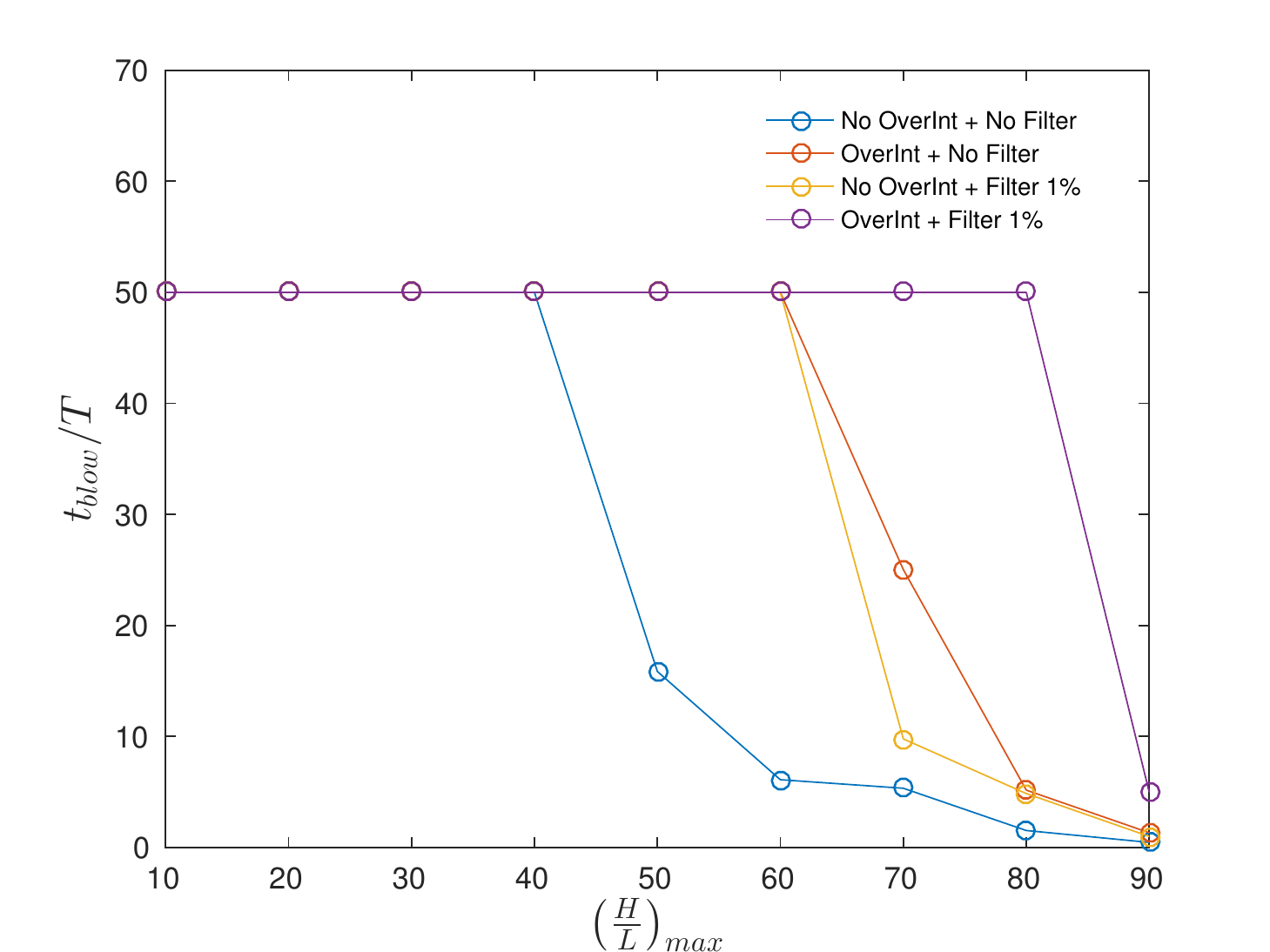} 
  \label{fig:4_5a}}\,   
\subfloat[Eulerian formulation]{
  \includegraphics[width=0.475\textwidth]{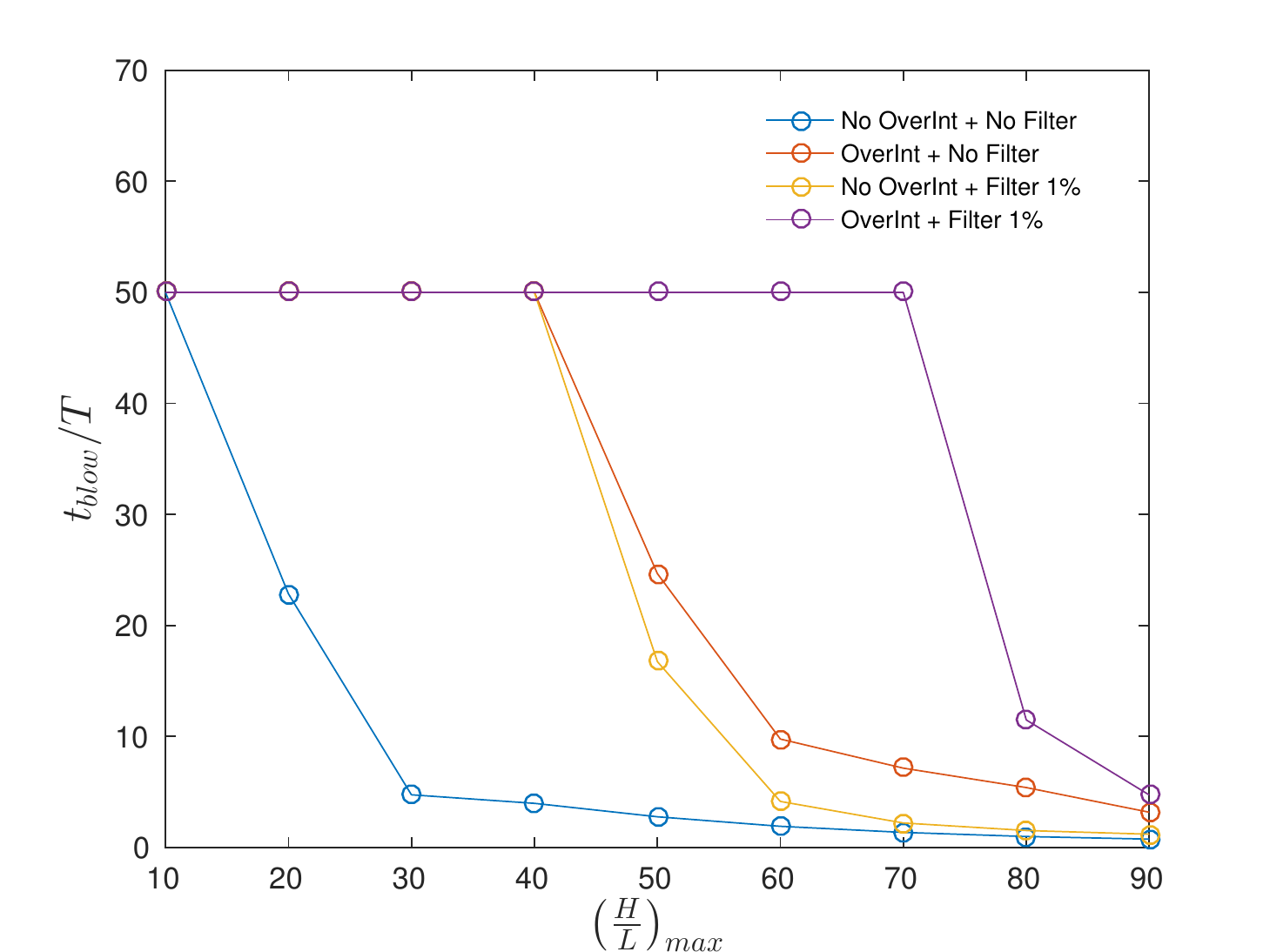} 
  \label{fig:4_5b}} 
\caption{Results of stabilisation of a nonlinear stream function waves ranging from $(H/L)_{max}=10-90 \%$. Time of numerical instability ($t_{blow}$) using different de-aliasing strategies (over-integration, 1\% filter) for (a) the MEL formulation and (b) the Eulerian \cite{EngsigKarupEtAl2016} formulation. Experiment corresponding to $kh=1$, $dt=T/80$ using wave-form mesh 8$\times$1 and polynomial order $P=6$. Calculations are assumed stable and stopped if time reaches $t_{final}/T=50$.}
  \label{fig:4_5}
\end{figure}

To maintain temporal stability when using explicit time-stepping methods, it is necessary to employ re-meshing  \cite{DOMYOU1987}. The objective is to either prevent strict CFL restrictions although this is not severe for FNPF methods where numerical stability is influenced only weakly by spatial resolution in the horizontal \cite{EGNL13,EngsigKarupEtAl2016}, and to retain a proper mesh quality to avoid temporal instabilities induced by the mesh or ill-conditioning in the global operators (discussed in \cite{BjontegaardRonquist2009}). Steep nonlinear waves are associated with larger node movements producing deformations in intra-node positions in the free surface elements. These deformations changes the conditioning of Vandermonde matrices and may impact the accuracy of the numerical scheme. This problem can be prevented by including a local re-mesh operation which changes the interior node distribution while not changing the initial mesh topology. For example, whenever an element goes below 75\% or above 125\% of its original size and with the initial free surface elements assumed of uniform size. The idea is to stop tracking the original material nodes at the free surface, and reposition the intra-element nodes through a local operation via interpolation to the original node distribution used in each element based on Legendre-Gauss-Lobatto nodes. A simple local re-mesh operation of this kind helps to stabilise the solver by improving accuracy in the numerical solutions and is a technique used in similar FEM solvers, e.g. see \cite{WuTaylor1994}. The effects of local re-meshing on temporal stability for a steep nonlinear stream function wave is presented in Figure \ref{fig:4_Lim}, and shows that local re-meshing is essential to maintain temporal stability for longer integration times for steep nonlinear waves. A snapshot of a stream function wave after propagating for $50$ wave periods the wave is still represented very accurately with essentially no amplitude or dispersion errors, cf. Figure \ref{fig:4_Lim} (a). Furthermore, both mass and energy is conserved with high accuracy, cf. Figure \ref{fig:4_Lim} (b). 
Also, we compare the MEL method and the Eulerian method due to \cite{EngsigKarupEtAl2016} in terms of different stabilisation techniques in Figure \ref{fig:4_5}. We find that with our current strategies the stabilised MEL method is more robust than the stabilised Eulerian scheme. 

\subsection{Convergence tests}

\begin{figure}[!htb]
\begin{center}
\begin{minipage}{4.2cm}
\centering (a) $P=3$, $H/L=10\%$ \\
\includegraphics[height=3.6cm]{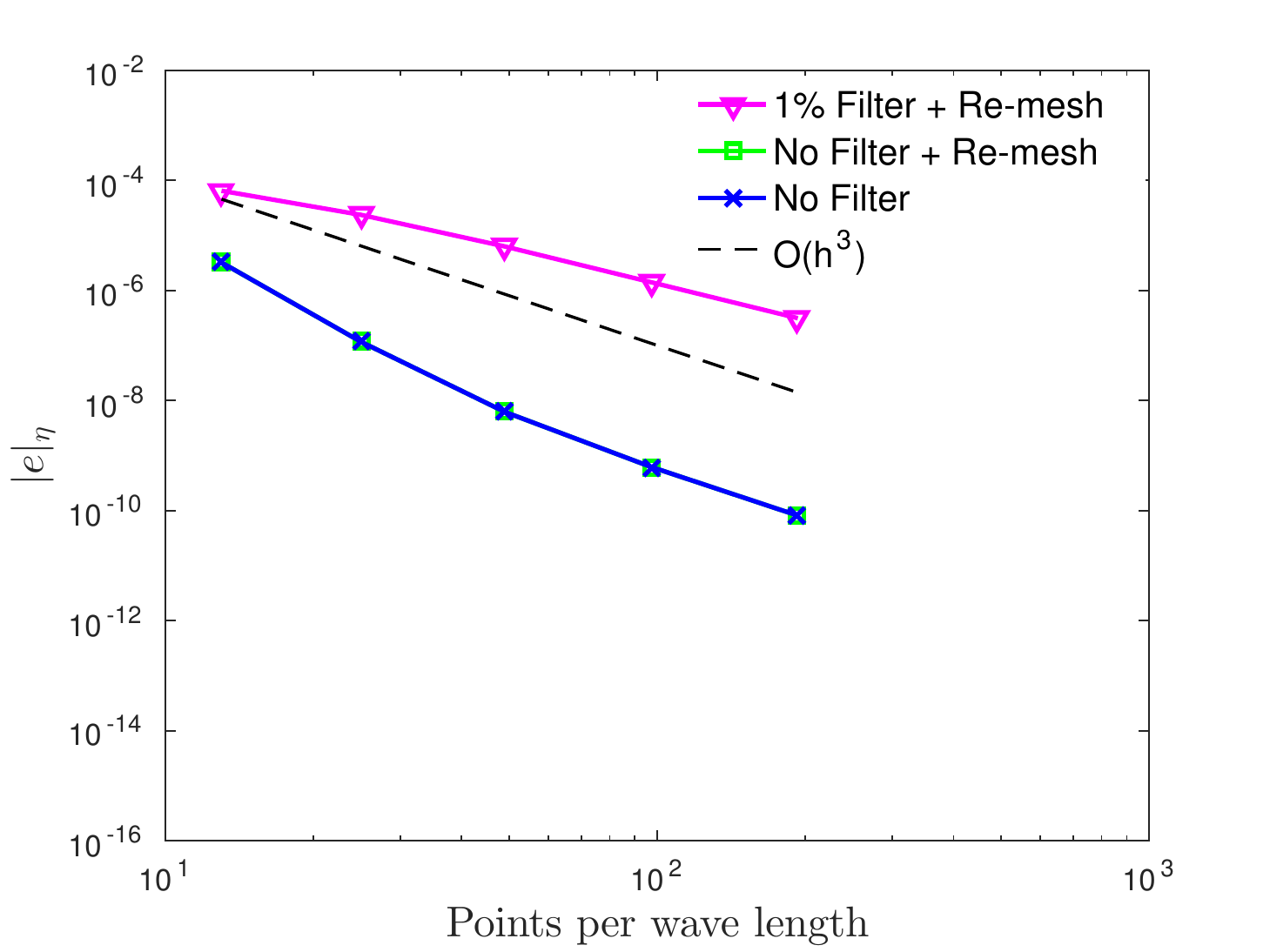} \\
\centering (d) $P=4$, $H/L=10\%$ \\
\includegraphics[height=3.6cm]{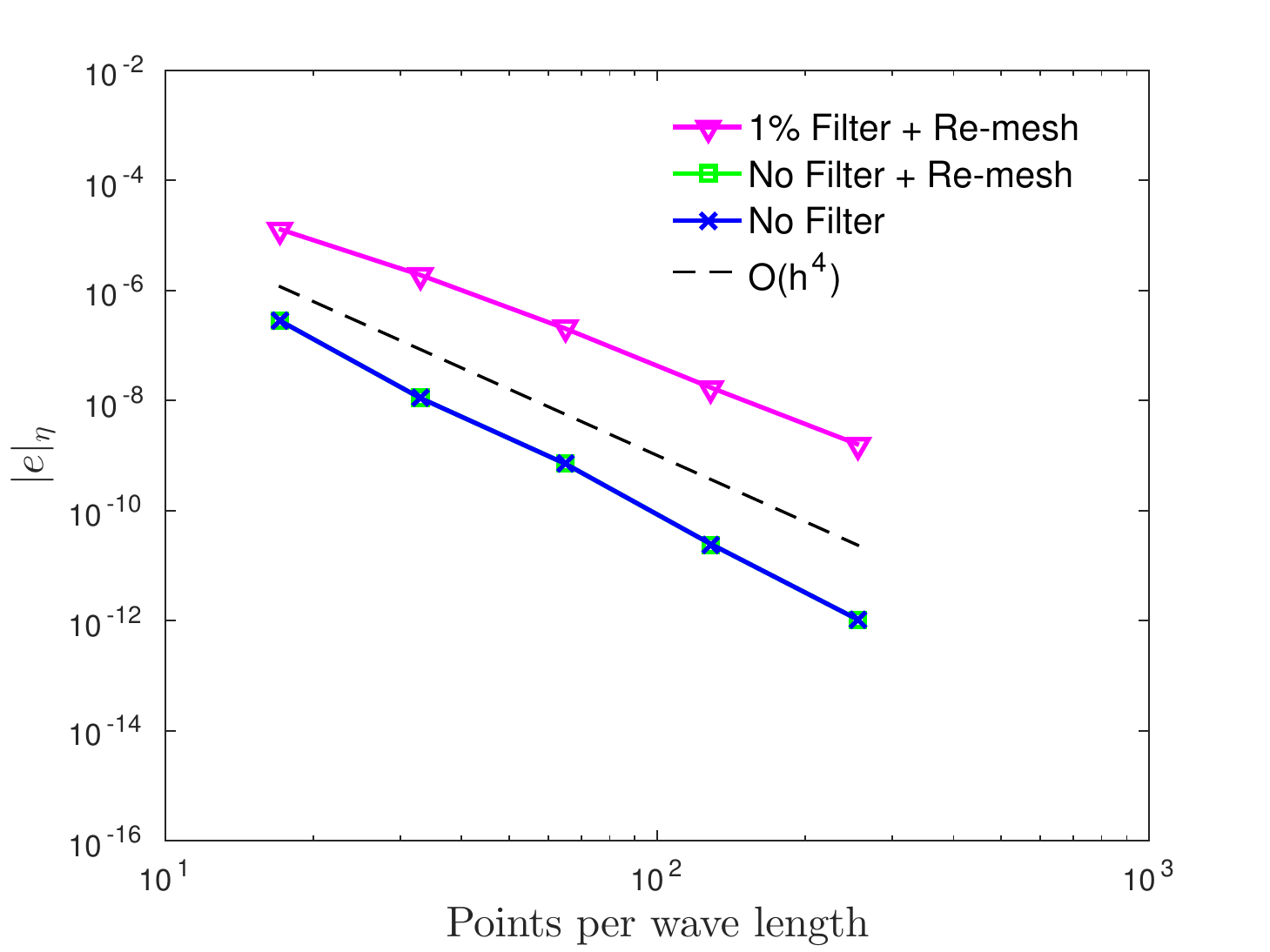} \\
\centering (g) $P=5$, $H/L=10\%$ \\
\includegraphics[height=3.6cm]{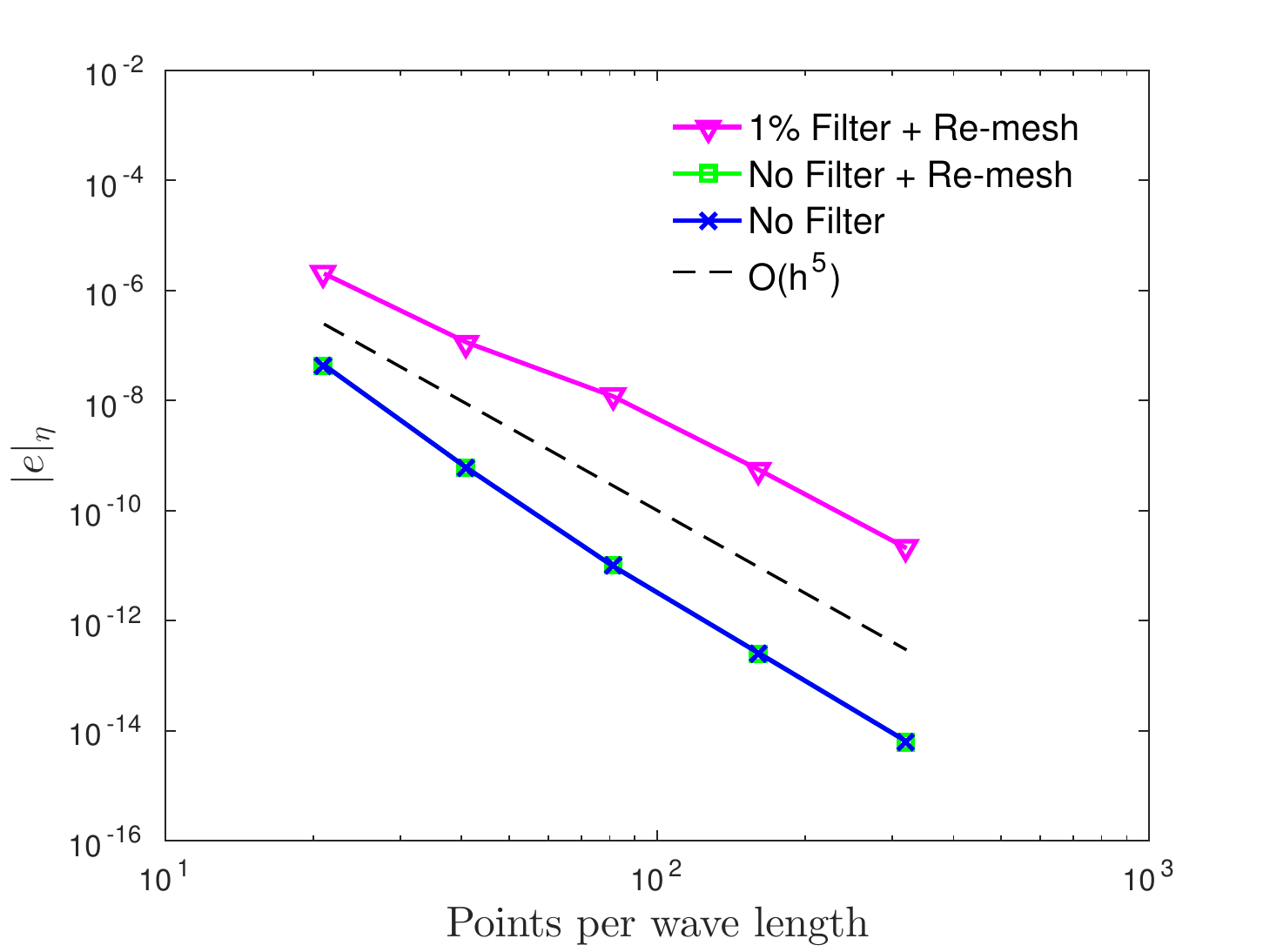} \\
\centering (j) $P=6$, $H/L=10\%$ \\
\includegraphics[height=3.6cm]{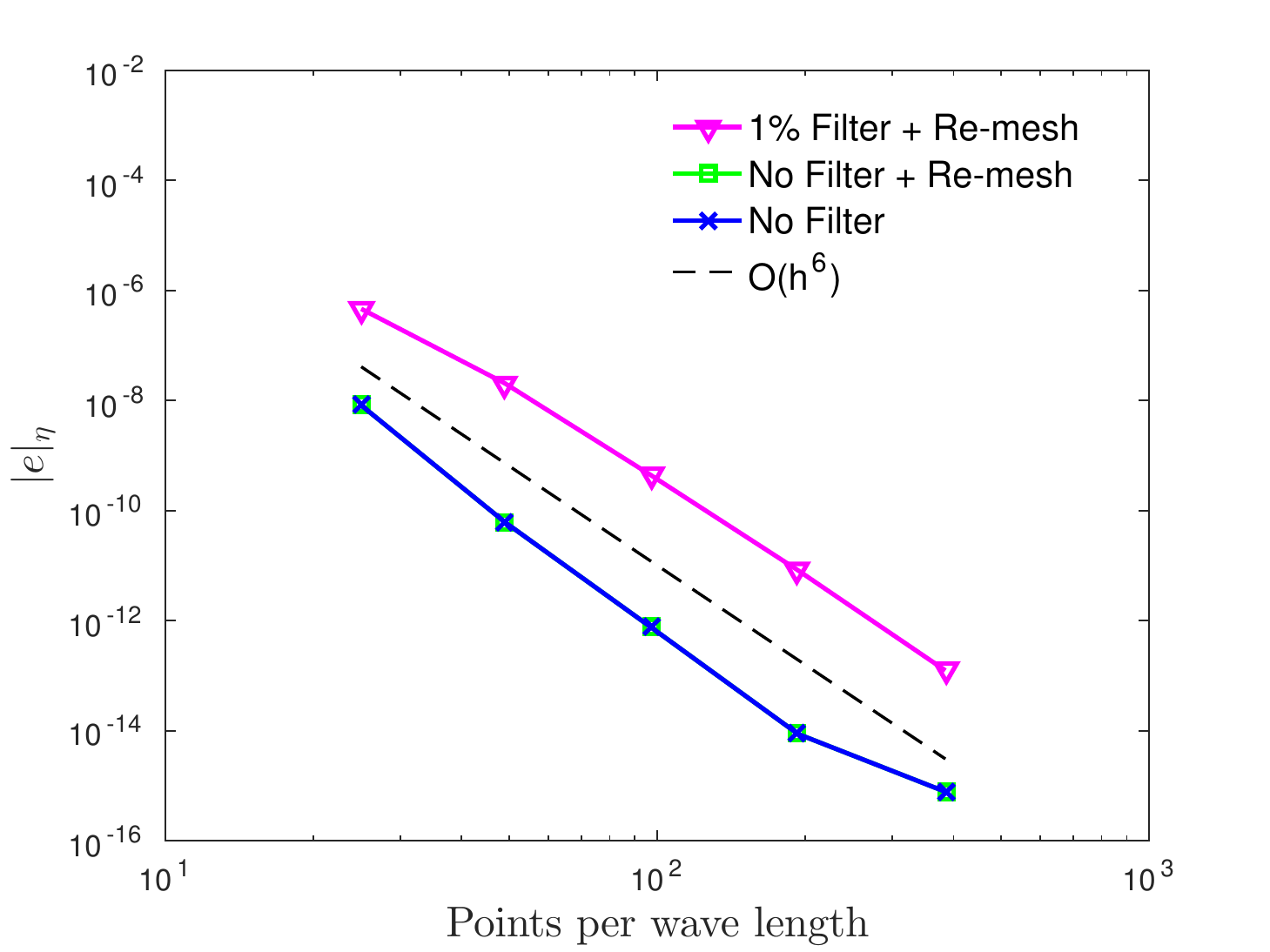}
\end{minipage}
\begin{minipage}{4.2cm}
\centering (b) $P=3$, $H/L=50\%$ \\
\includegraphics[height=3.6cm]{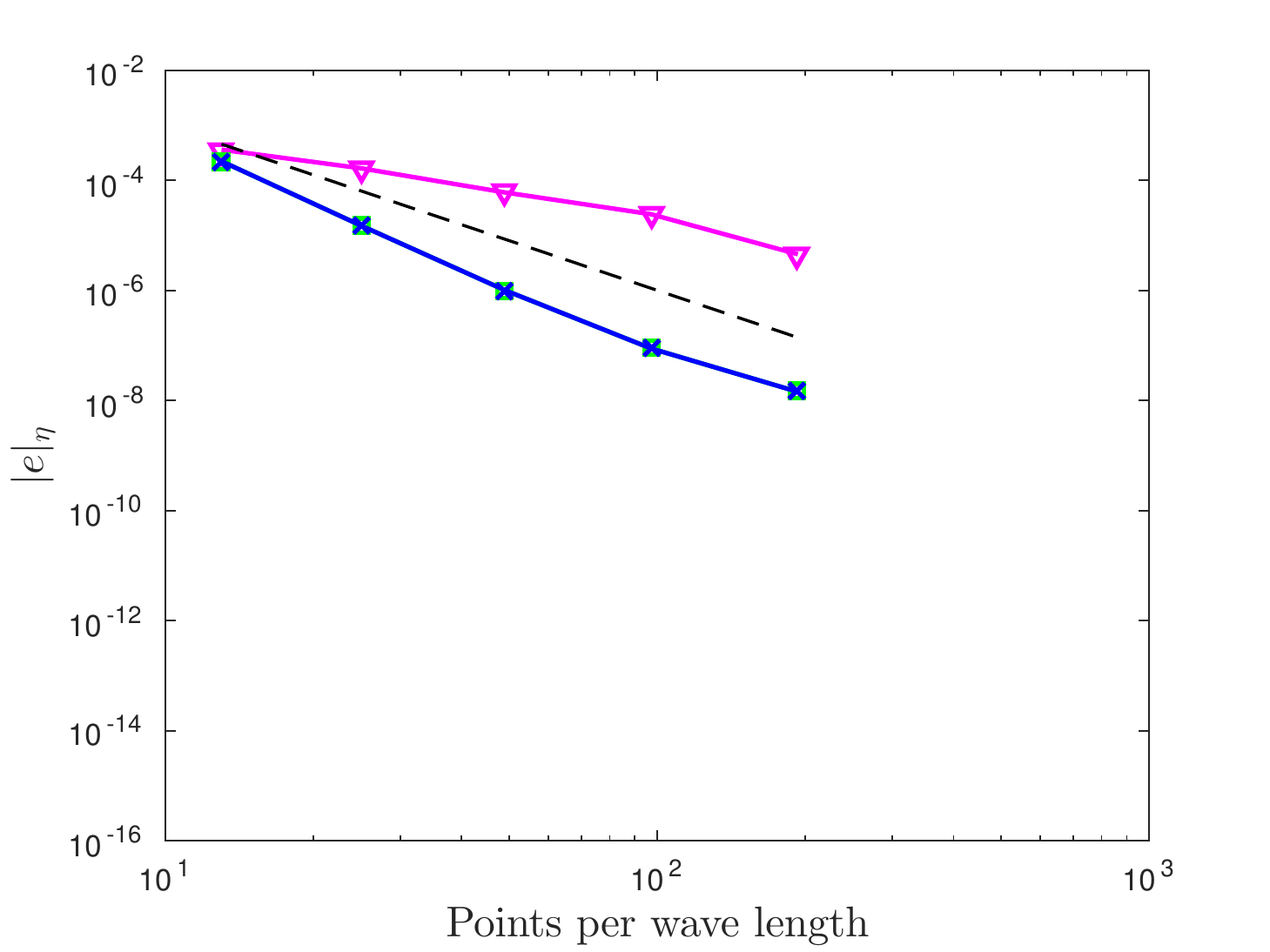} \\
\centering (e) $P=4$, $H/L=50\%$ \\
\includegraphics[height=3.6cm]{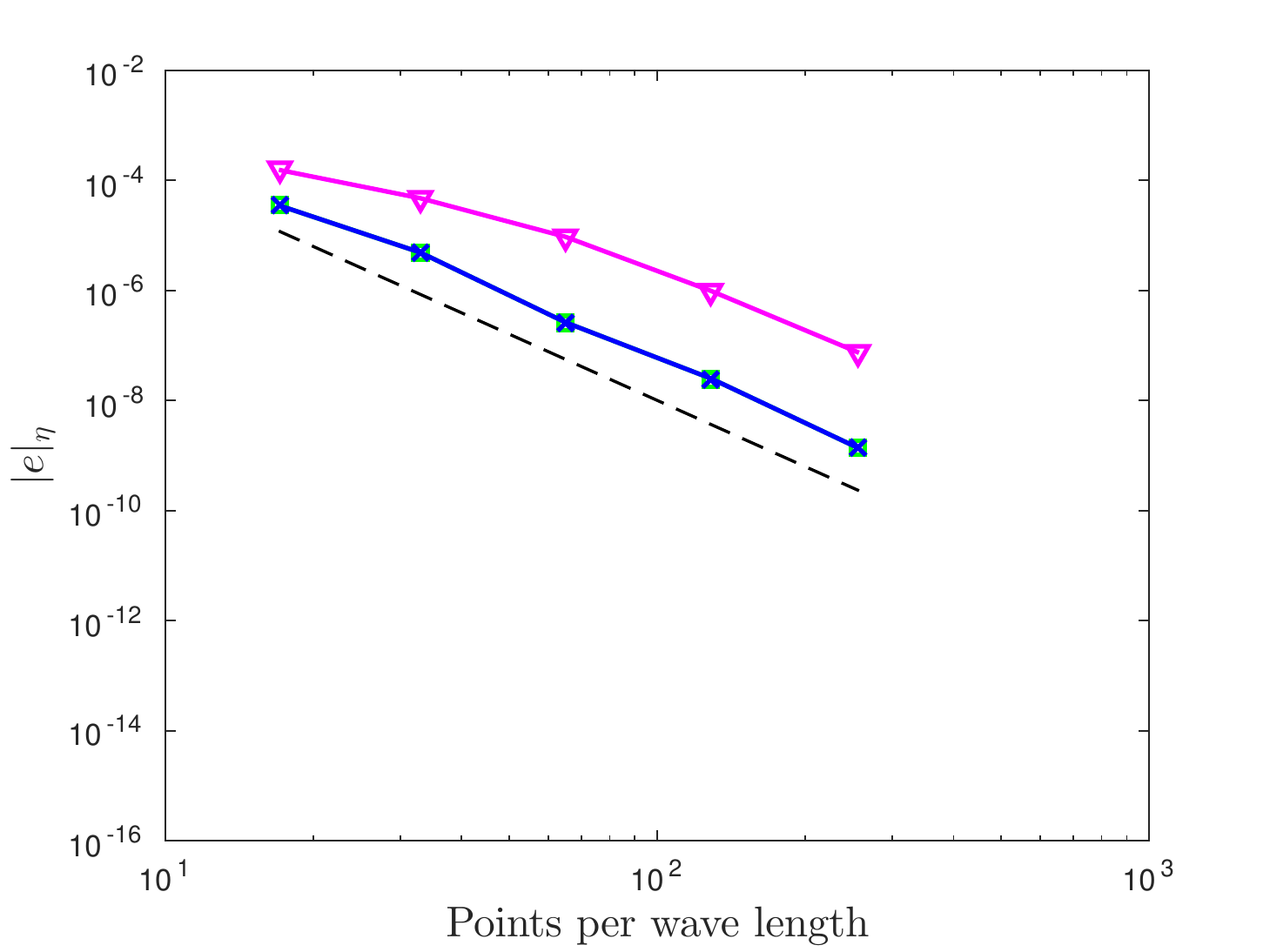} \\
\centering (h) $P=5$, $H/L=50\%$ \\
\includegraphics[height=3.6cm]{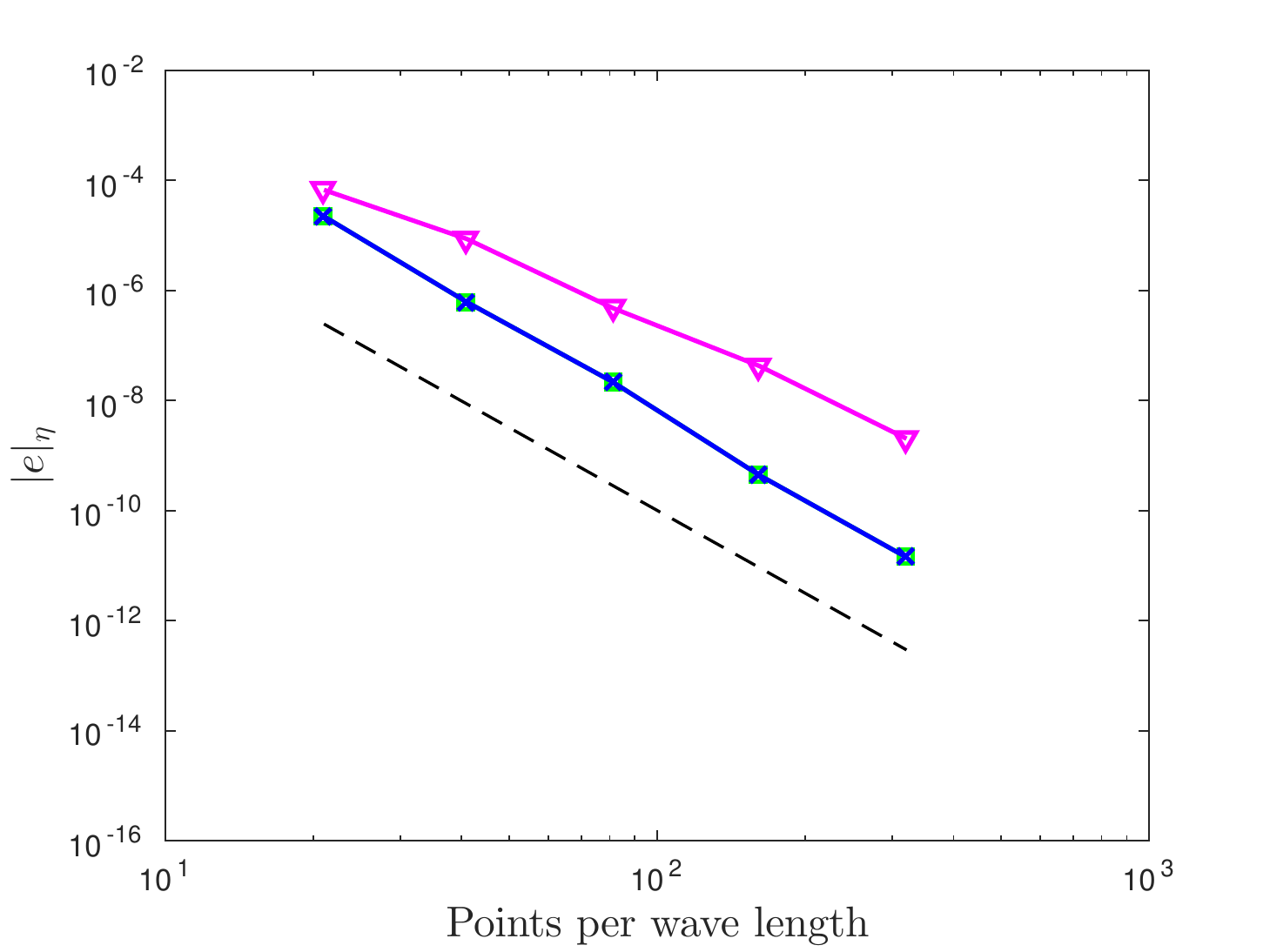} \\
\centering (k) $P=6$, $H/L=50\%$ \\
\includegraphics[height=3.6cm]{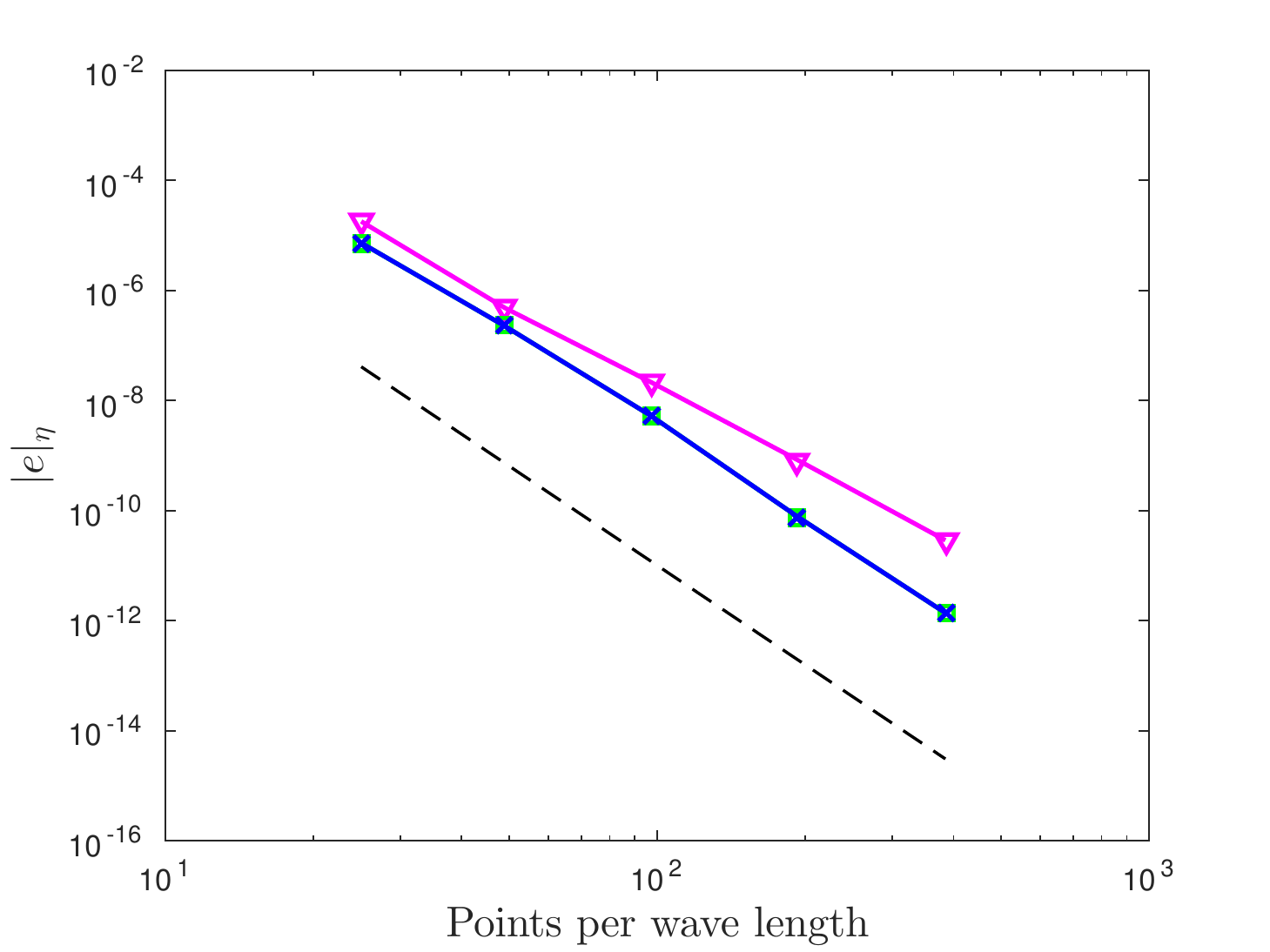}
\end{minipage}
\begin{minipage}{4.2cm}
\centering (c) $P=3$, $H/L=90\%$ \\
\includegraphics[height=3.6cm]{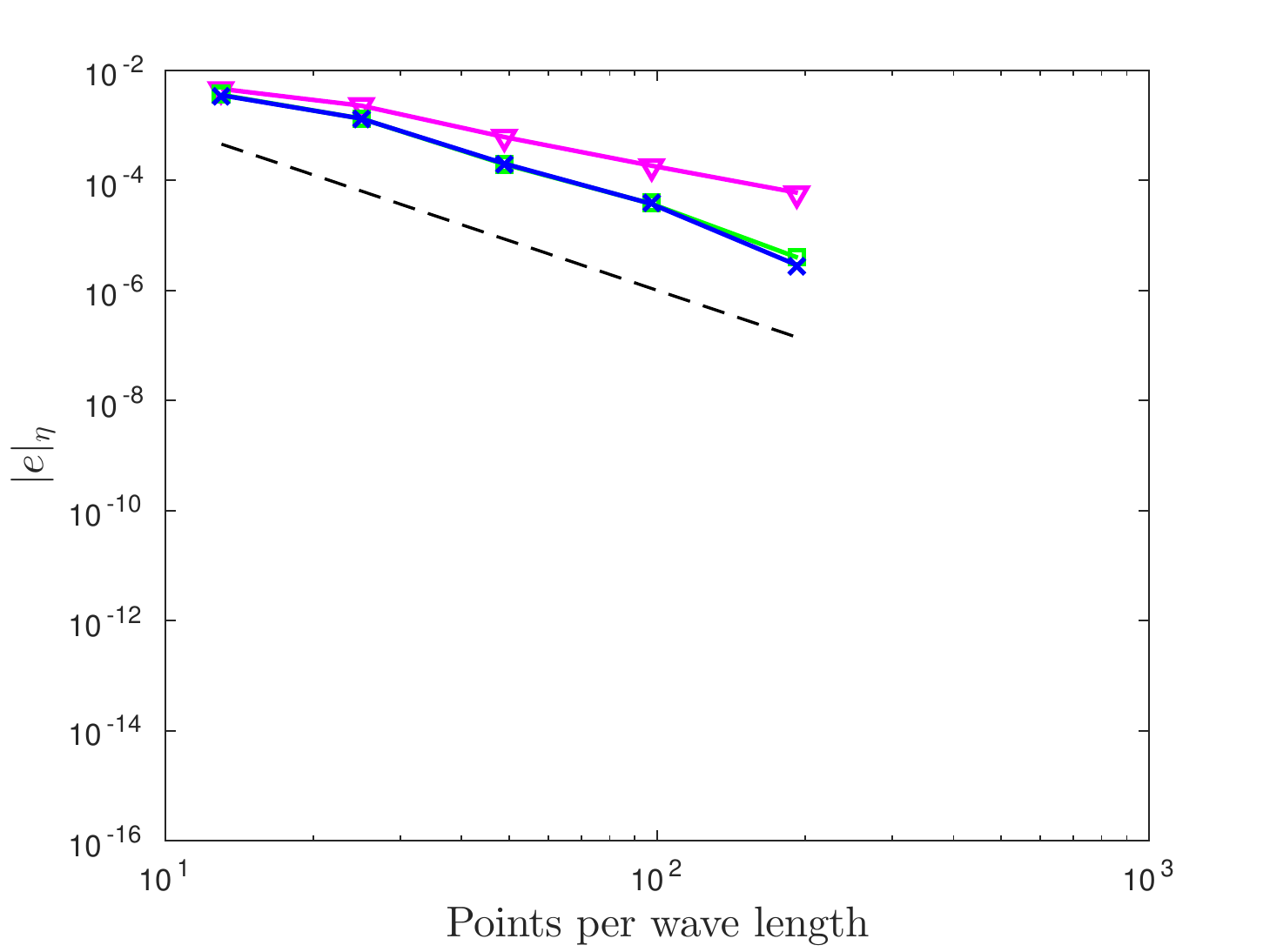} \\
\centering (f) $P=4$, $H/L=90\%$ \\
\includegraphics[height=3.6cm]{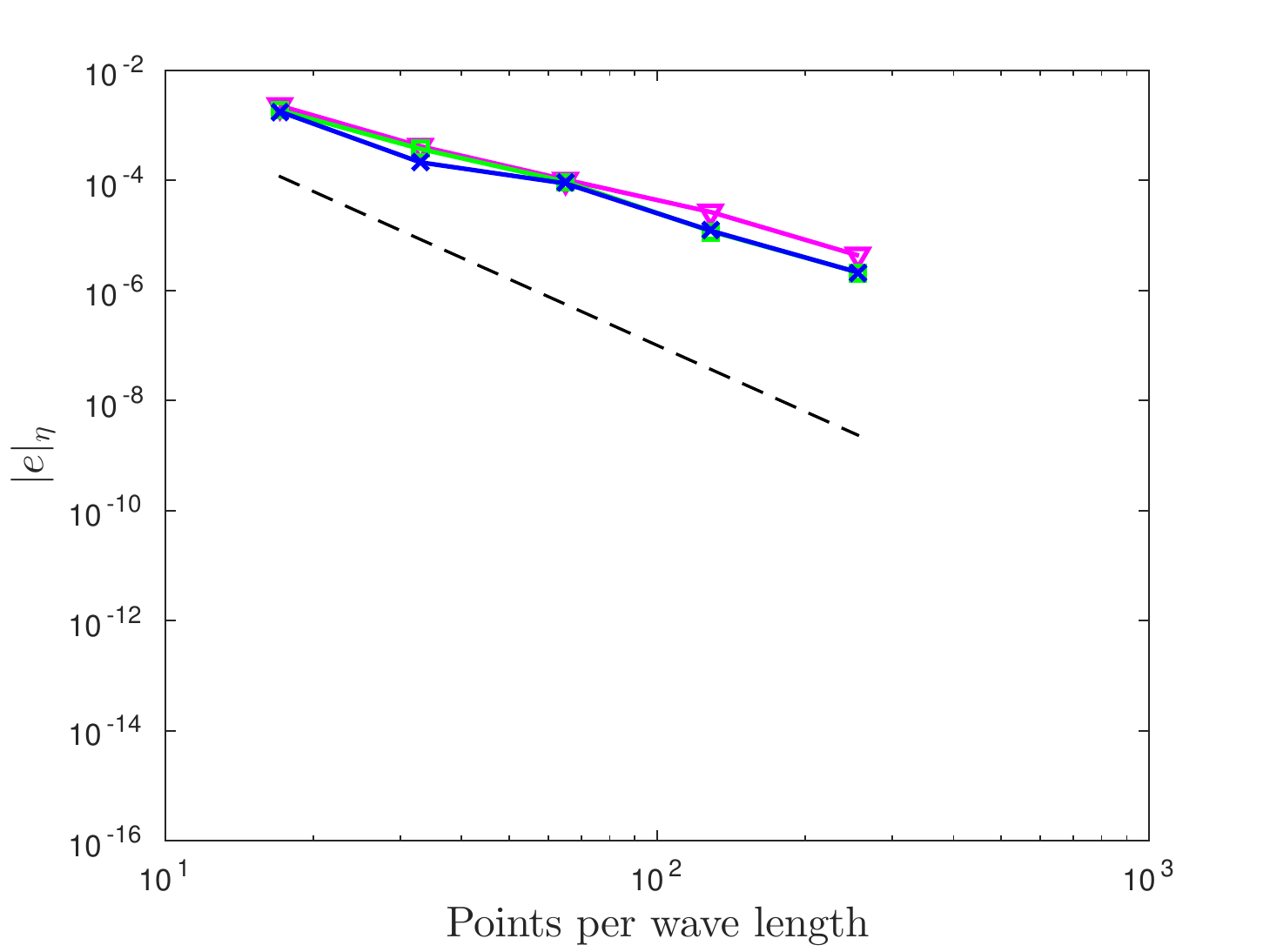} \\
\centering (i) $P=5$, $H/L=90\%$ \\
\includegraphics[height=3.6cm]{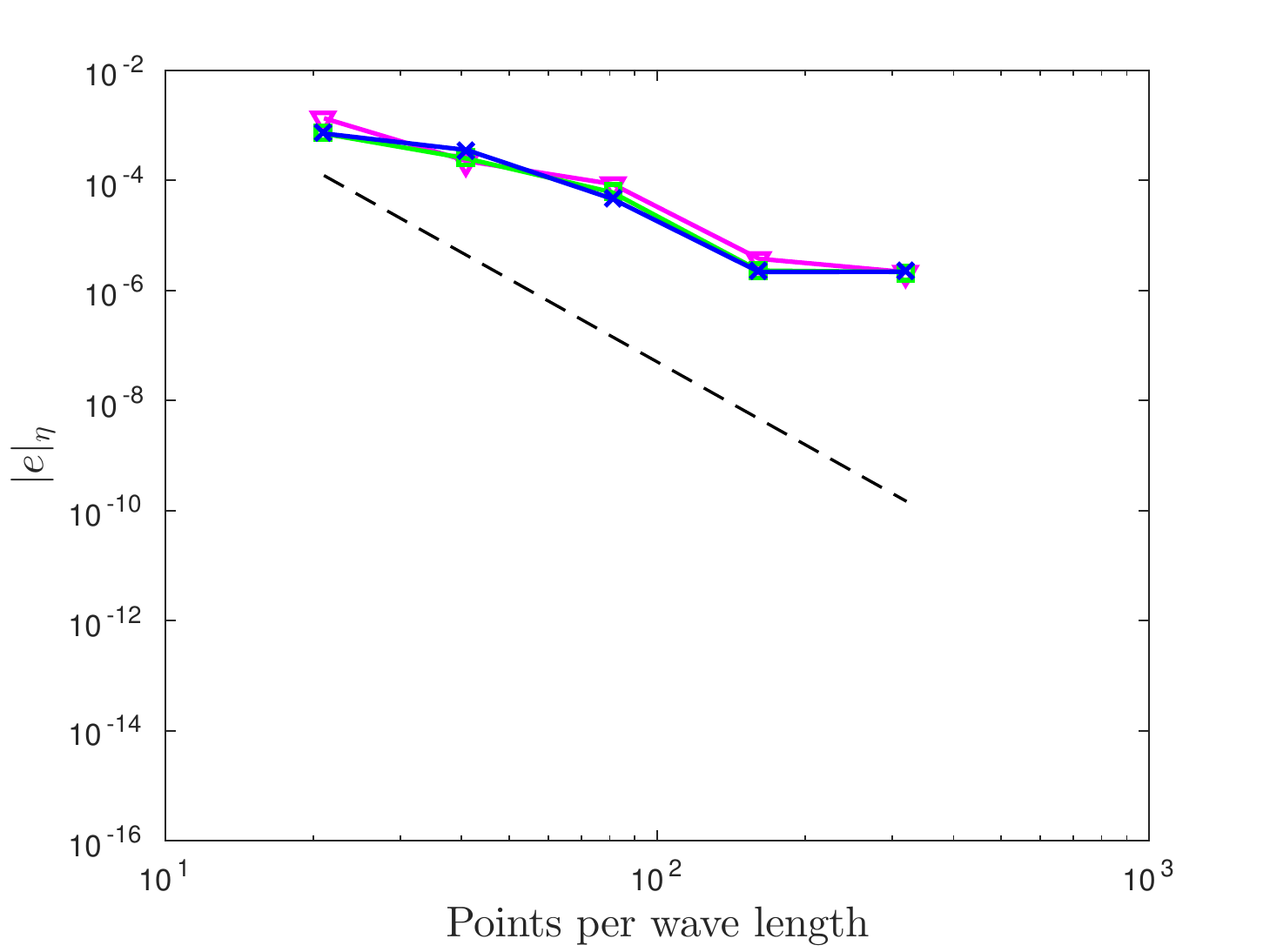} \\
\centering (l) $P=6$, $H/L=90\%$ \\
\includegraphics[height=3.6cm]{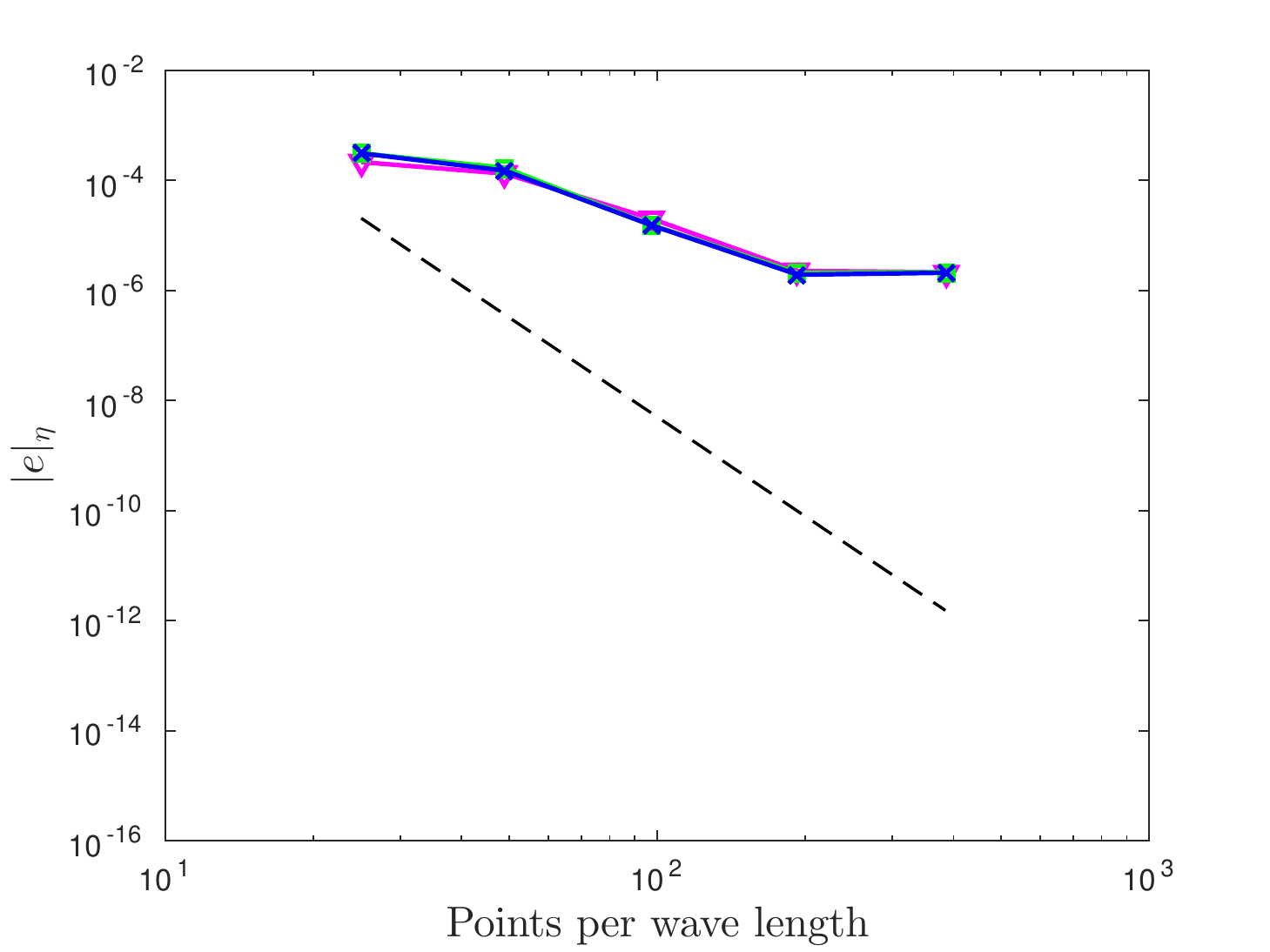}
\end{minipage}
\end{center}
\caption{Convergence tests with different expansion order $P$ in horizontal for nonlinear stream function wave solutions with parameters $kh=1$ and $H/L$ ratios of maximum wave steepness. A Galerkin scheme with over-integration is used with either no filtering, no filtering with re-meshing, or a re-meshing and 1\% filter applied. The time step size in all simulations is set to be small enough for spatial truncation errors to dominate. }
\label{fig:convtests}
\end{figure}

To validate the high-order spectral element method, we demonstrate for the MEL scheme that the high-order convergence rate is $\mathcal{O}(h^P)$ in line with \cite{EngsigKarupEtAl2016} and the accuracy of the method using convergence tests as depicted in Figure \ref{fig:convtests}. These tests, demonstrate that we can exert control over approximations errors in the scheme by adjusting the resolution in terms of choosing the points per wave length. For the most nonlinear waves of 90\% of maximum steepness, the curves find a plateau at the size of truncation error before convergence to machine precision due to insufficient accuracy in our numerical solution of stream function waves used.

\section{Numerical experiments}
\label{sec:numexp}

We examine different strenuous test cases that serve as validation of the numerical spectral element model proposed.

\subsection{Reflection of high-amplitude solitary waves from a vertical wall}
\label{ssec:5_3_3}

We setup a solitary wave as initial condition using the high-order accurate spectral numerical scheme due to \cite{DutCla14} and consider the propagation of solitary waves of different amplitudes above a flat bed that are reflected by a vertical solid wall. In this experiment, a solitary wave approaches the wall with constant speed and starts to accelerate forward when the crest is at a distance of approximately $2h$ from the wall. The water level at the wall position grows leading to the formation of a thin jet shooting up along the wall surface. When the maximum of free surface elevation is reached at the wall position it is said that the wave is attached to the wall, at this moment $t=t_a$ and the height of the crest is $\eta_a$. Thereafter, the jet forms and reaches its maximum run up $\eta_0$ at time $t_0$. After this event, the jet collapses slightly faster than it developed. The detachment time $t_d$ corresponds to the wave crest leaving the wall. The height of the wave at that instant is $\eta_d$, always smaller than the attachment height (i.e. $\eta_d < \eta_a < \eta_0$). Then a reflected wave propagates in the opposite direction with the characteristics of a solitary wave of reduced amplitude. This adjustment in the height produces a dispersive trail behind the wave. The depression becomes more abrupt for increasing wave steepness. 

This test case requires the fully nonlinear free surface boundary conditions to capture the steepest solitary waves and their nonlinear interaction with the wall. The case was first studied using a numerical method by Cooker et al. \cite{Cooker97}, where a Boundary-Integral Method (BIM) was used to solve the Euler equations with fully nonlinear boundary conditions. The results obtained using BIM showed to be in excellent agreement with the experimental data given in \cite{Max76}, however, showed difficulties in the down run phase for the steepest wave of $a/h=0.7$ indicating that for the most nonlinear cases, numerical modelling is difficult. Accurate simulation of the steepest waves  requires sufficient spatial resolution and high accuracy in the kinematics to capture the finest details of the changing kinematics of the fluid near the wall. This problem was recently been addressed using other high-order numerical models, e.g. a high-order Boussinesq model with same fully nonlinear free surface boundary conditions based on a high-order FDM model \cite{MBL02} and a nodal discontinuous Galerkin spectral element method \cite{EHBM06} and both studies showed excellent results for the wave run-up and depth-integrated force histories up to $a/h=0.5$. 

The numerical experiments are carried out using a domain size $x\in[-22.5, 22.5]$ m, with the initial position of the center of the solitary wave at $x_0=0$ m. We employ  a structured mesh consisting of quadrilaterals with $N_x$ elements in a single layer, where $N_x\in[40,120]$ is varied proportional to the wave height to resolve the waves in the range $a/h\in[0.2,0.6]$. Each element is based on polynomial expansion orders $(P_x,P_z)=(6,7)$.  The time step size is chosen in the interval $\Delta t\in[0.01, 0.02]$ s. Stabilisation of the numerical scheme is achieved using exact integration and mild spectral filtering using a $1\%$ top mode spectral filter every time step using the MEL scheme. 
\begin{figure}[!htb]
  \centering
\subfloat[before max runup]{
  \includegraphics[width=1.\textwidth]{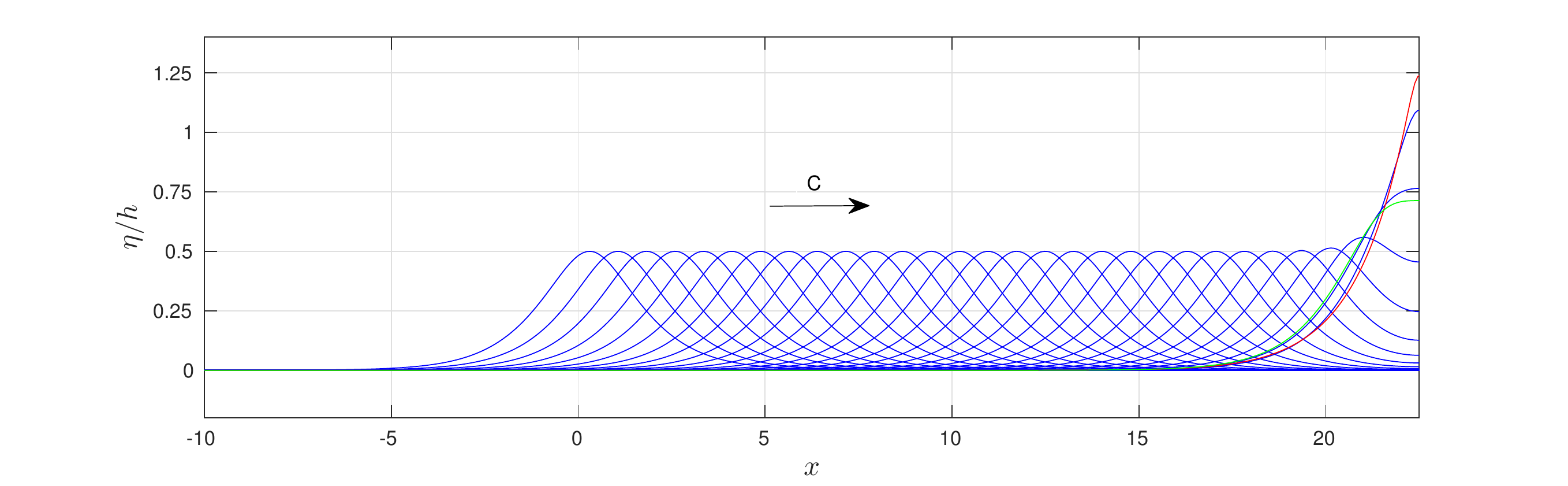} 
  \label{fig:4_S1a}}\\   
\subfloat[after max runup]{
  \includegraphics[width=1.\textwidth]{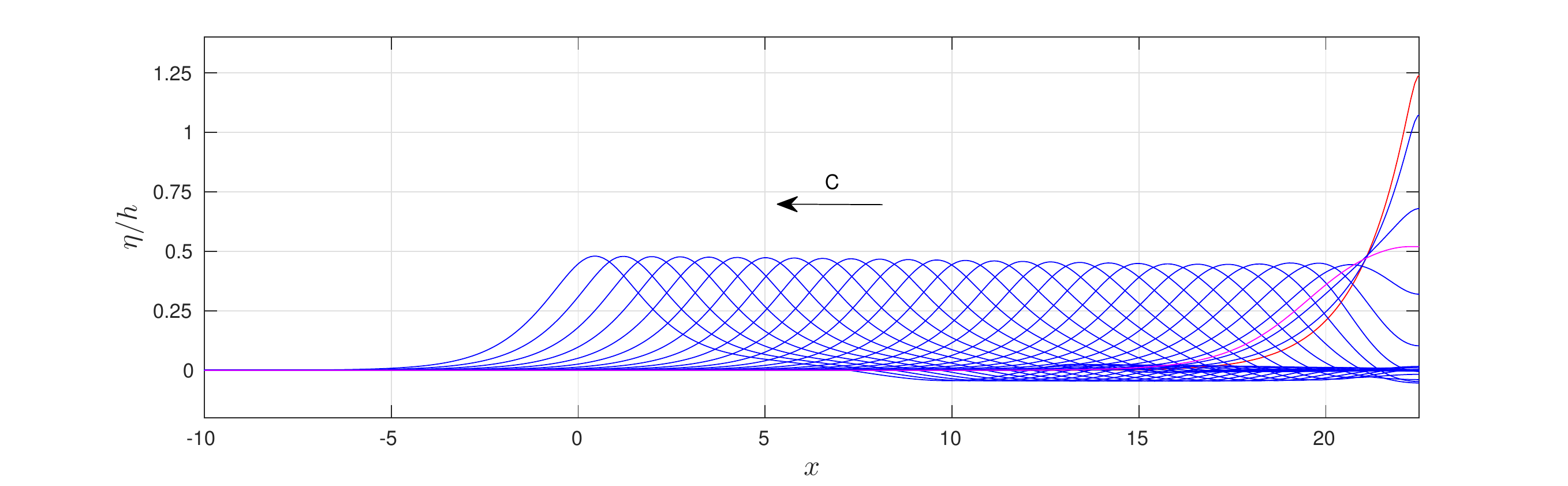}
  \label{fig:4_S1b}} \\
\subfloat[Heigth of the crest]{
  \includegraphics[width=0.42\textwidth]{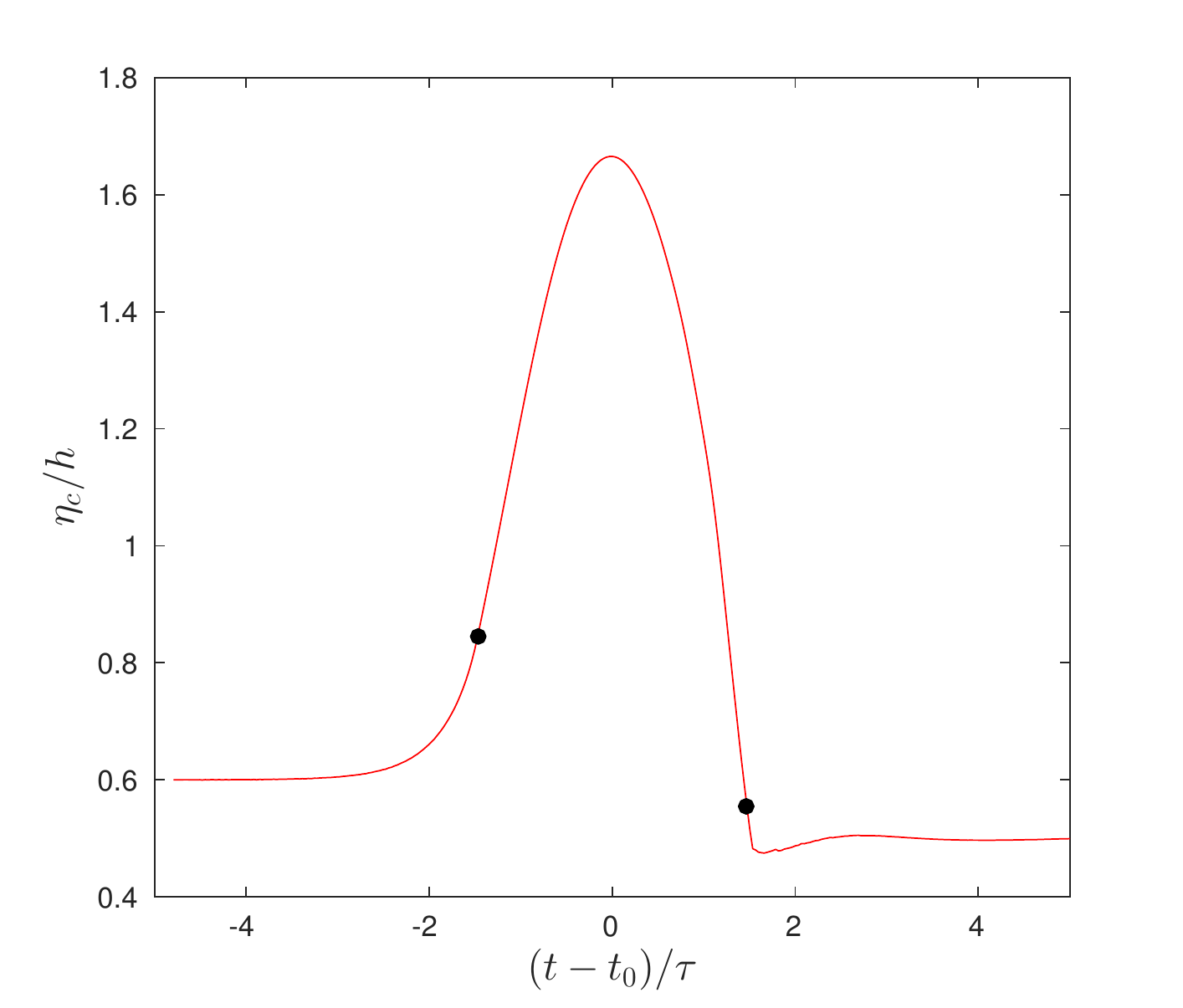} 
  \label{fig:4_S1c}} \,  
\subfloat[Mass and energy]{
  \includegraphics[width=0.42\textwidth]{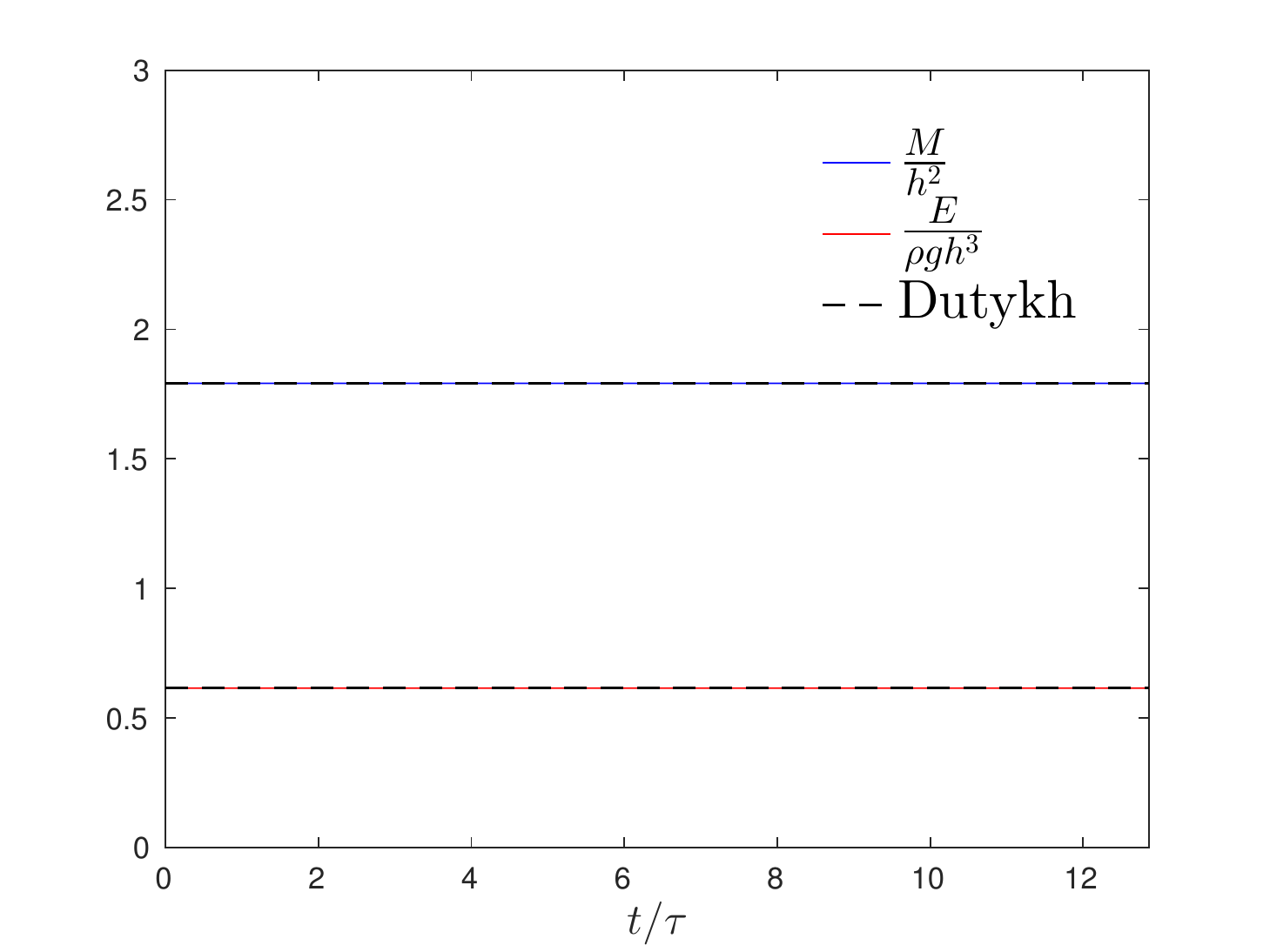} 
  \label{fig:4_S1d}}
  \caption{Simulation of Solitary wave reflection with relative amplitude $a/h=0.5$: Sequence of free surface elevation with time step size $\Delta t=0.2$ s for $t<t_0$ (a) and $t>t_0$ (b), superimposed free surface profiles at attachment, maximum run up and detachment times (green, red and magenta respectively). Time history of crest's height during the collision, attachment/detachment times are marked with black dots (c). Evolution of wave mass and energy during the simulation (d).} 
  \label{fig:4_S1}
\end{figure}

\begin{figure}[!htb]
  \centering
\subfloat[]{
  \includegraphics[width=0.42\textwidth]{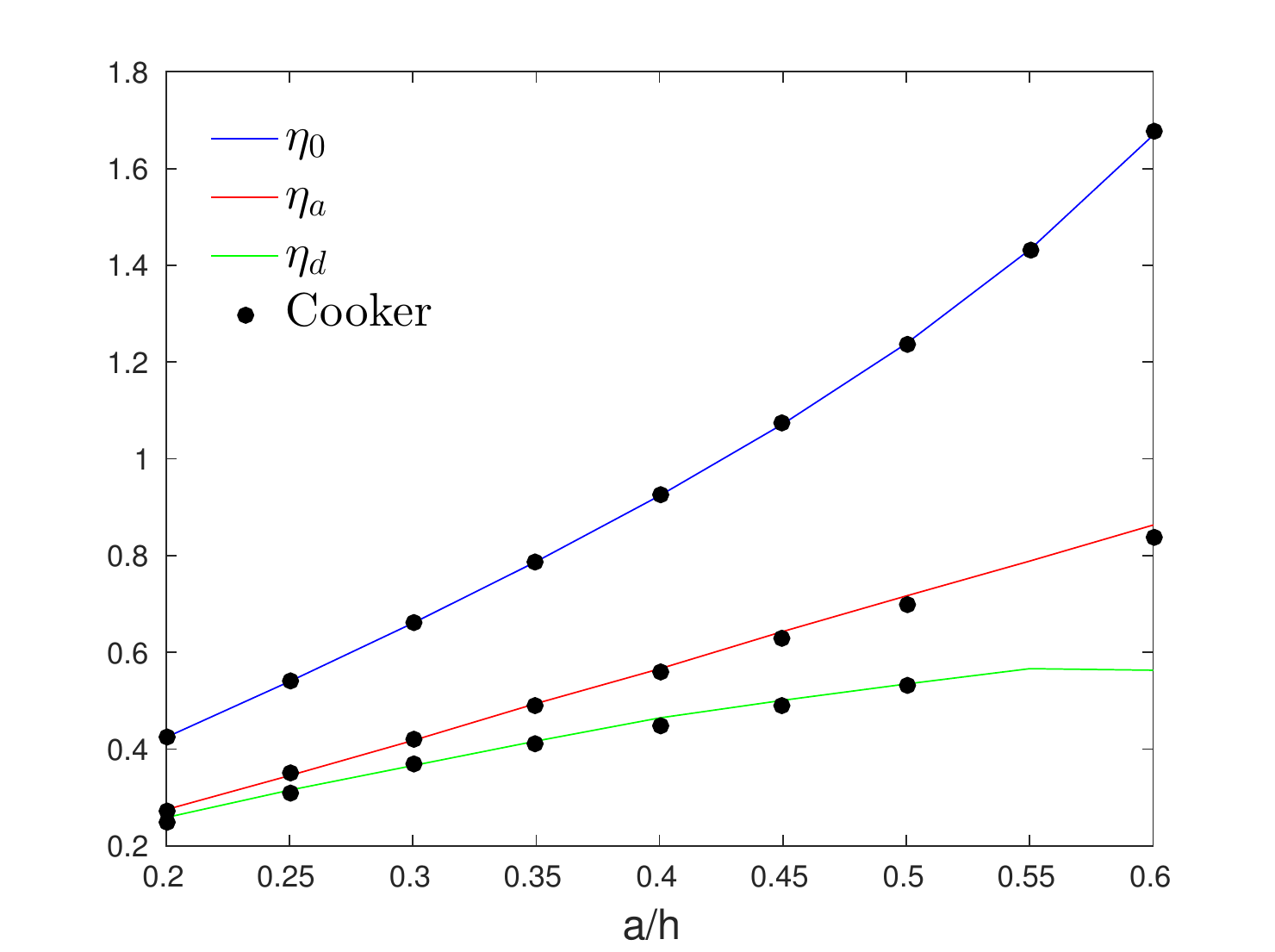} 
  \label{fig:4_S2a}} \,   
\subfloat[]{
  \includegraphics[width=0.42\textwidth]{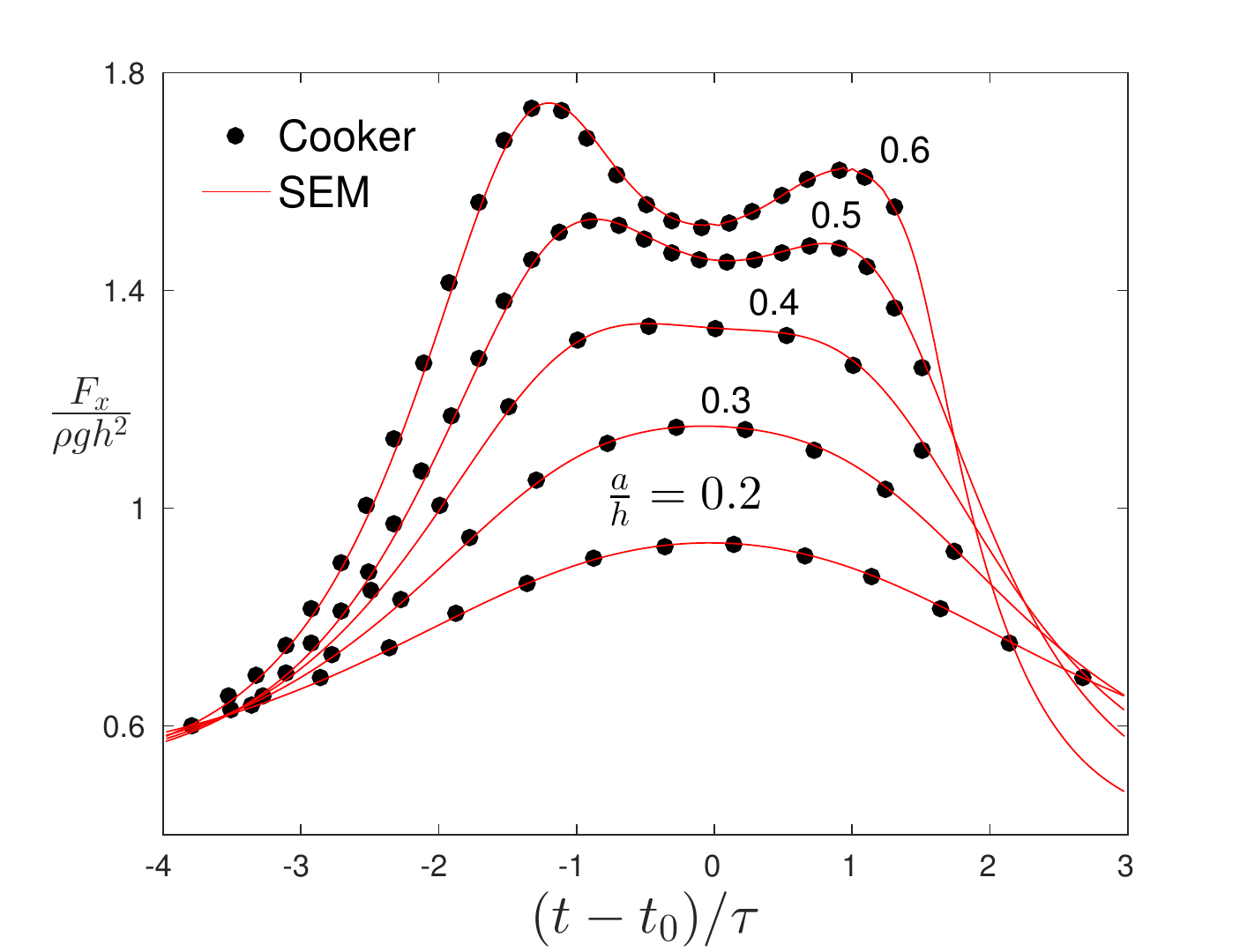}
  \label{fig:4_S2b}} 
  \caption{Results corresponding to MEL formulation for case of solitary wave reflection. (a) Attachment analysis. (b) Dimensionless horizontal forces at right wall normalised using $\tau=\sqrt{h/g}$.} 
  \label{fig:4_S2}
\end{figure}

The attachment analysis of \cite{Cooker97} is reproduced using SEM with the initial profile of the solitary wave \cite{DutCla14} up to $a/h = 0.6$ with excellent agreement, cf. Figures \ref{fig:4_S2a} and \ref{fig:4_S2b}. This time history of the wave propagation is illustrated using the MEL scheme in Figure \ref{fig:4_S1} for a high-amplitude wave with relative amplitude $a/h=0.5$.  
The forces are computed numerically as a post-processing step by integration of the local pressure obtained from Bernoulli's equation
\begin{align}
p = - \rho g z - \rho \frac{\partial \phi}{\partial t} - \rho \frac{1}{2} \nabla\phi \cdot \nabla\phi,
\end{align}
along the wet part of the wall. Approximation of the time derivative $\partial \phi/\partial t$ is approximated by using the acceleration potential method due to \cite{Tanizawa1995}. This method estimates $\phi_t$ by solving an additional boundary value problem based on a Laplace problem defined in terms of $\phi_t$ in the form
\begin{align}
\nabla^2\phi_t = 0, \quad \textrm{in}\quad \Omega
\end{align}
subject to the boundary conditions
\begin{subequations}
\begin{align}
\phi_t&=-g\eta-\frac{1}{2}|\nabla\phi|^2, \quad \textrm{on} \quad \Gamma^{\textrm{FS}} \\
\frac{\partial \phi_t}{\partial n} &= 0, \quad \textrm{on}\quad \Gamma^b
\end{align}
\end{subequations}
The wall force vector is determined numerically from pressure using
\begin{align}
F = \int_{\partial\Omega} p {\bf n} dS.
\end{align}

\subsection{Rectangular obstacle piercing the free surface}

In Lin (2006) \cite{Lin06} a numerical method for Navier-Stokes equations is proposed based on transformation of the fluid domain using a multiple-layer $\sigma$-coordinate model. Wave-structure interactions of a solitary wave of amplitude $a/h=0.1$ (mildly nonlinear) with a rectangular obstacle fixed at different positions (seated, mid-submergence, floating) are investigated in a two dimensional wave flume of depth $h=1$ m. Free surface elevation histories at three different gauge locations are found in excellent agreement with computed VOF solutions. We consider here the experiment corresponding to the obstacle piercing the free surface (width: $5$ m, height: $0.6$ m, draught: $0.4$ m) for validation of the present methodology dealing with semi-submerged bodies.

\begin{figure}[!htb]
  \centering
\subfloat[Experimental setup]{
\includegraphics[width=0.65\textwidth]{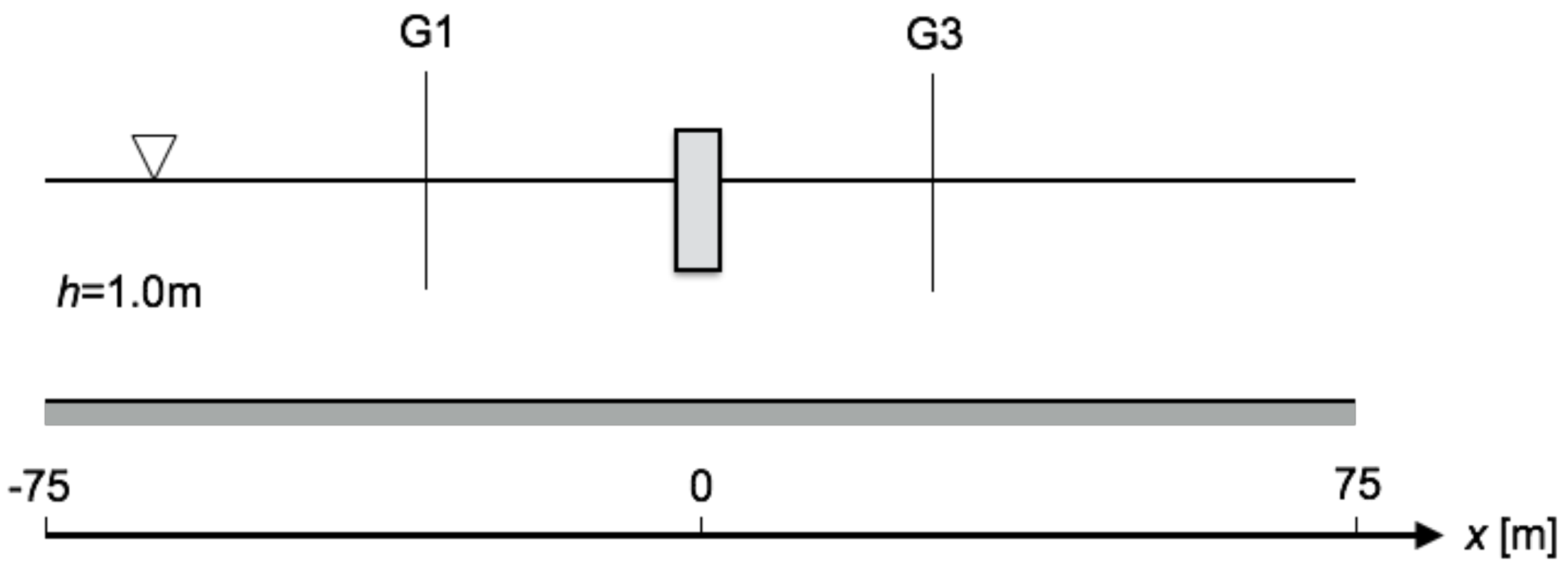}
  \label{fig:HC_23a}} \\
\subfloat[Mesh zoom in at position of structure]{
  \includegraphics[width=0.8\textwidth]{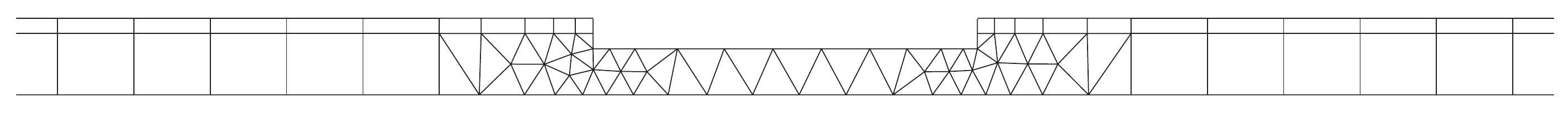} 
  \label{fig:HC_23b}}
   \caption{(a) Sketch of the experiment. (b) Detail of hybrid mesh covering the range $x \in [-10.0; ~10.0] ~m$.}
  \label{fig:HC_23}
\end{figure}

\begin{figure}[!htb]
  \centering
\subfloat[Gauge 1]{
  \includegraphics[width=0.9\textwidth]{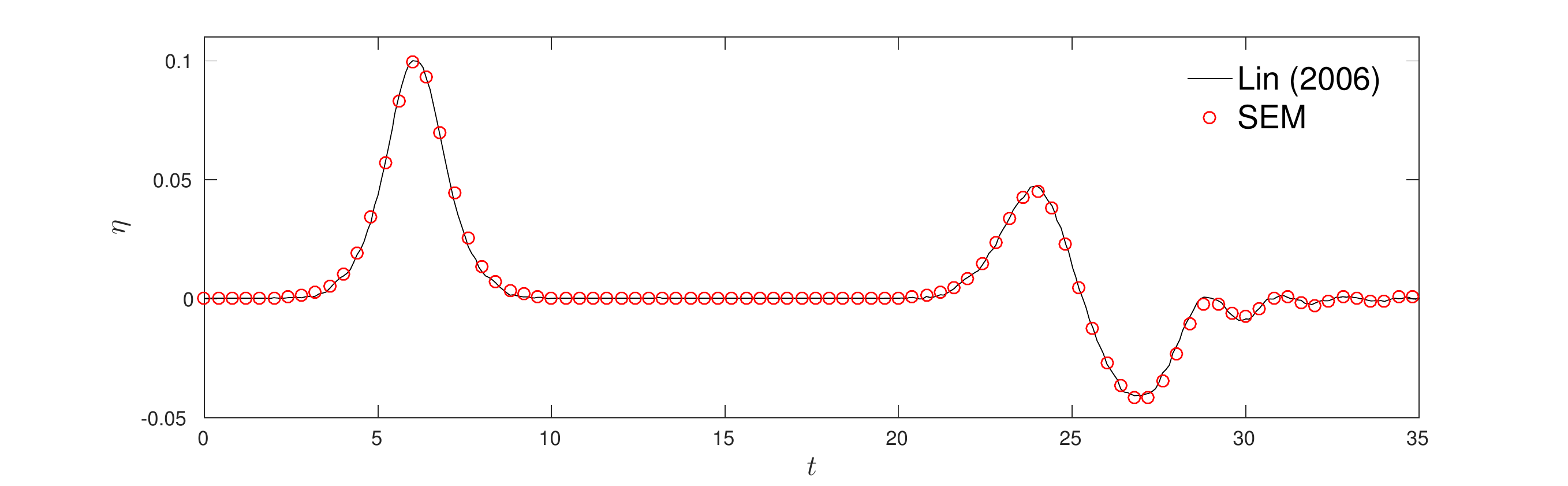}
  \label{fig:HC_24a}} \\
\subfloat[Gauge 3]{
  \includegraphics[width=0.9\textwidth]{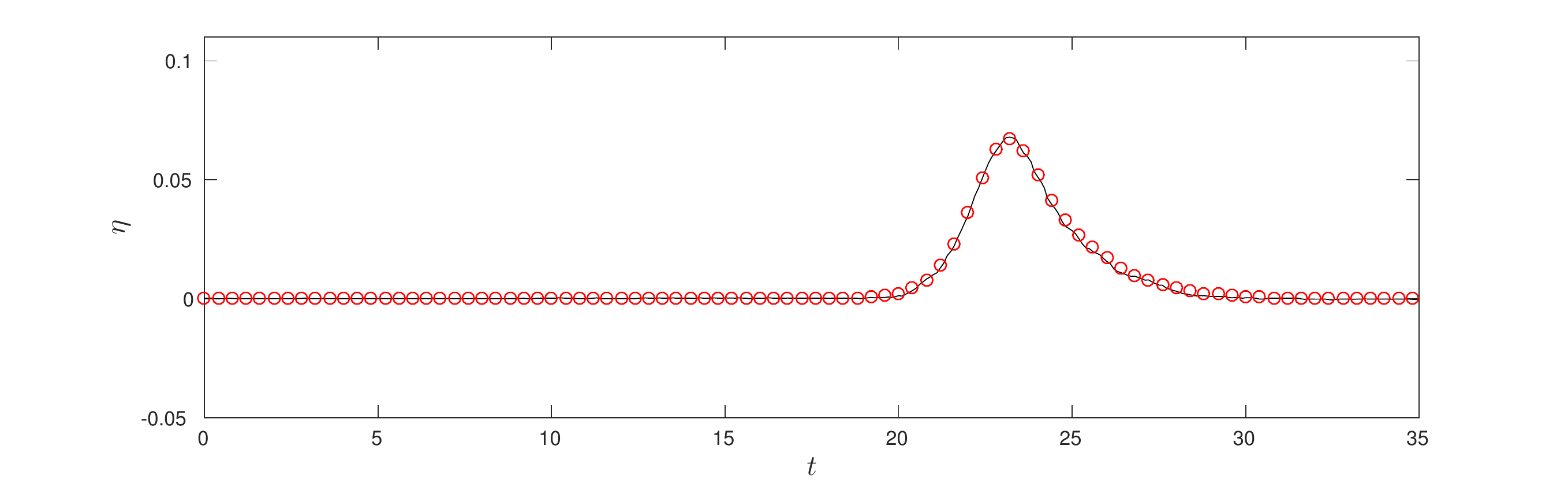}
  \label{fig:HC_24b}}
   \caption{Time histories of the free surface elevation at gauge 1 (a) and gauge 3 (b). Comparison with the numerical VOF results presented in \cite{Lin06}. }
  \label{fig:HC_24}
\end{figure}

A wave flume of $150$ m is sketched in Figure \ref{fig:HC_23a} with the cylinder located at its center ($x=0$ m). In this scenario, positions of gauges 1 and 3 in \cite{Lin06} correspond to $x_{G1}=-31$ m and $x_{G3}=26.5$ m, respectively. This domain is discretised using hybrid meshes with a top layer of curvilinear quadrilaterals. Below, unstructured grid of triangles is employed in the central part ($ x \in [-4.5; 4.5]$ m), fitting the body contour (Triangles size is chosen proportional to their distance to the lower corners of the body). The zones away from the structure $x \in [-75.0; -4.5]$ m $\cup [4.5 ; 75.0]$ m are meshed with regular quadrilaterals of width $1$ m, small enough to guarantee accurate propagation of this wave according to our tests in Section \ref{ssec:5_3_3}. A detail of this mesh corresponding to $x \in [-10.0 ; 10.0]$ m is shown in Figure \ref{fig:HC_23b}.
Similar to previous experiments with solitary waves, the initial condition $(\eta_0,\tilde{\phi}_0)$ is generated using the accurate numerical solution due to \cite{DutCla14} corresponding to a wave peaked at $x = -55$ m. Using the MEL scheme, the time step size is $\Delta t=0.02$ s. The evolution of the free surface elevations at gauge positions are compared with the Lin (2006) results in Figure \ref{fig:HC_24}, and excellent agreement is observed. 
%

\subsection{Solitary Wave Propagation Over a Submerged Semi-circular Cylinder}

\begin{center}
\begin{figure}[!htb]
  \centering
\subfloat[ Mesh ]{
\includegraphics[width=0.98\textwidth]{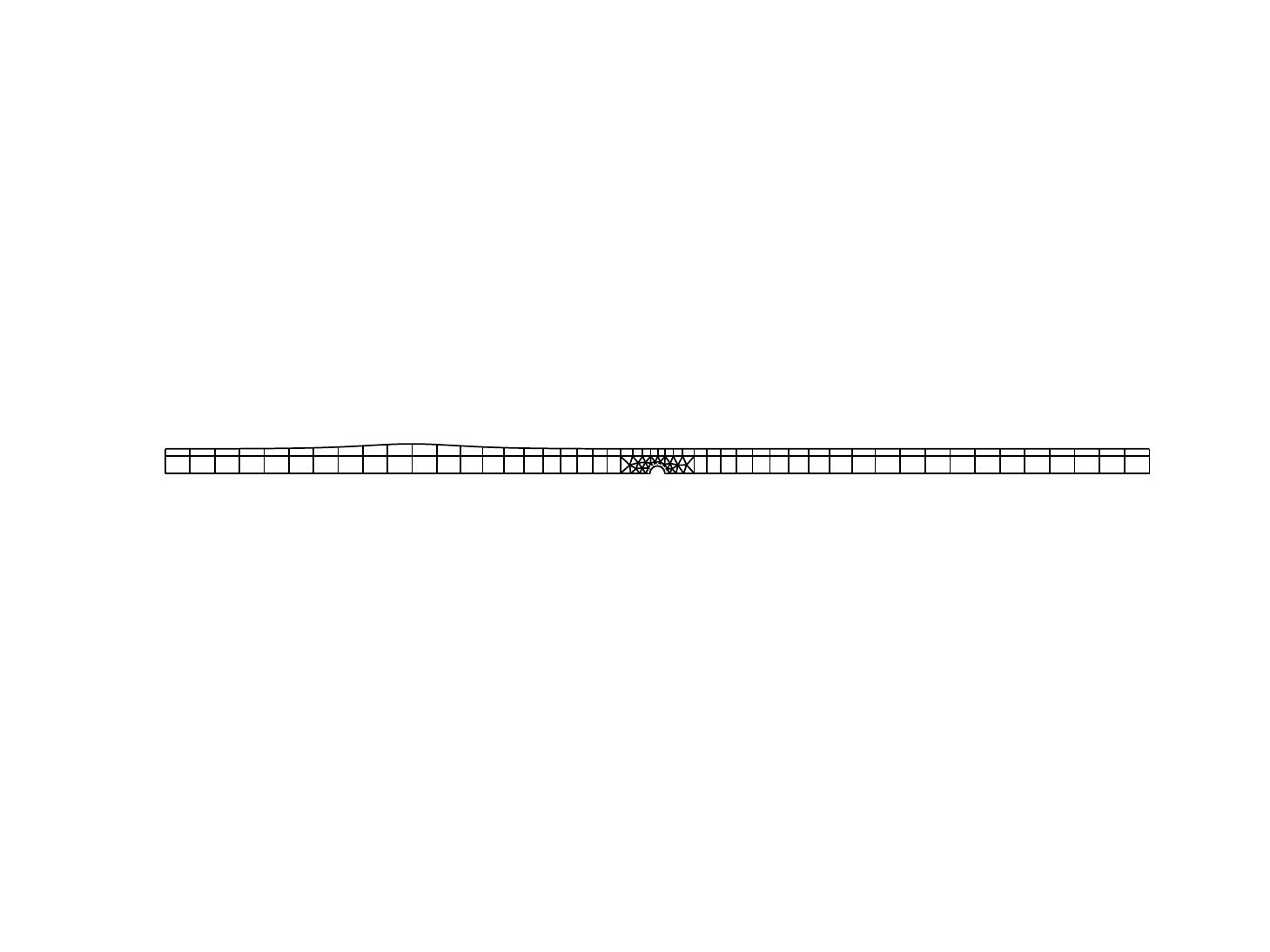}
  }
   \\
\subfloat[ Zoom at center ]{
\includegraphics[width=0.98\textwidth]{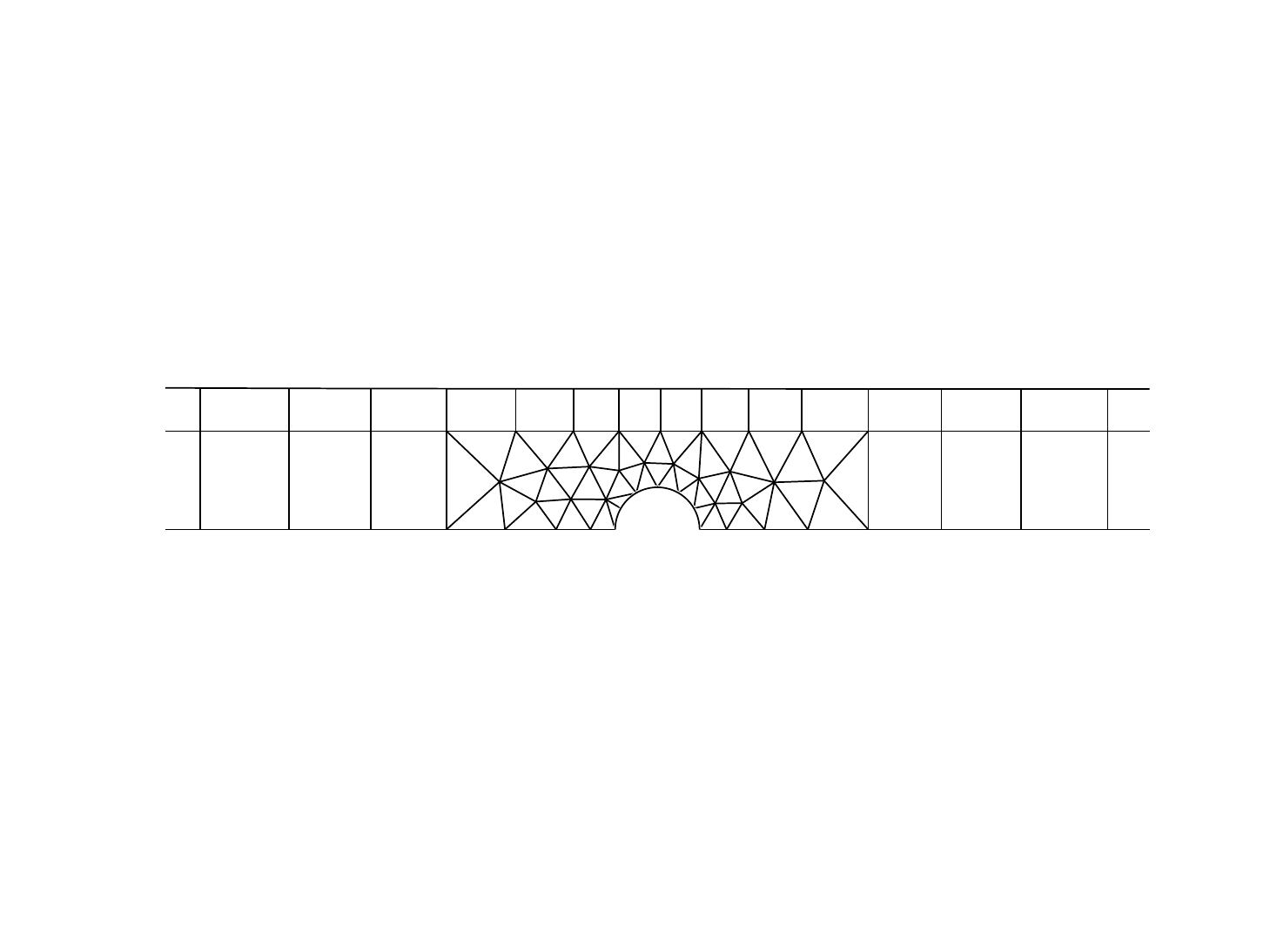}
} 
\caption{Initial mesh for solitary wave propagation over a submerged semi-circular cylinder.}
\label{fig:SCCylMesh}
\end{figure}
\end{center}
\begin{figure}[!htb]
\centering
\subfloat[]{
\includegraphics[width=0.45\textwidth]{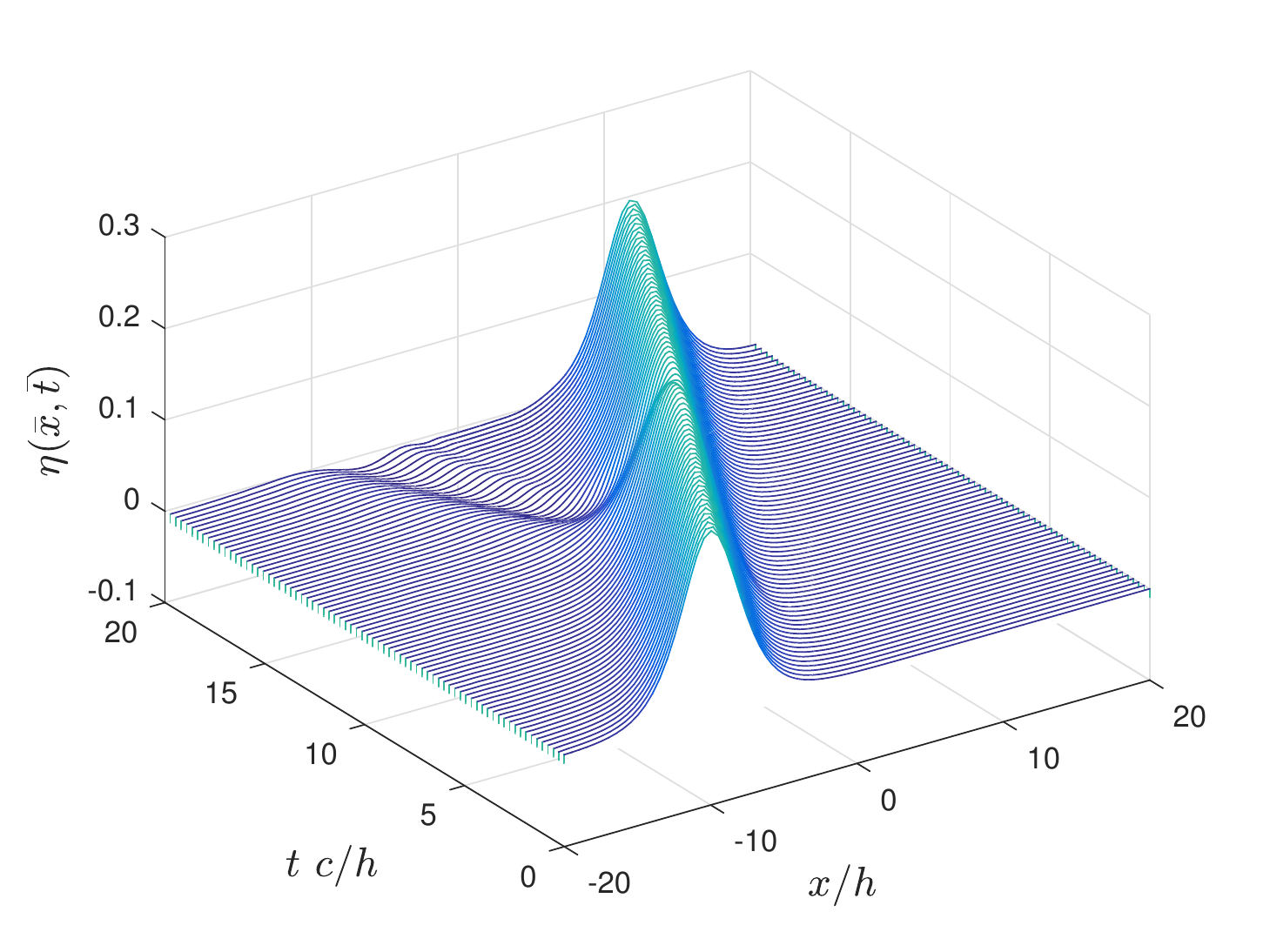}
  \label{fig:HC_a}} \,
\subfloat[]{
\includegraphics[width=0.45\textwidth]{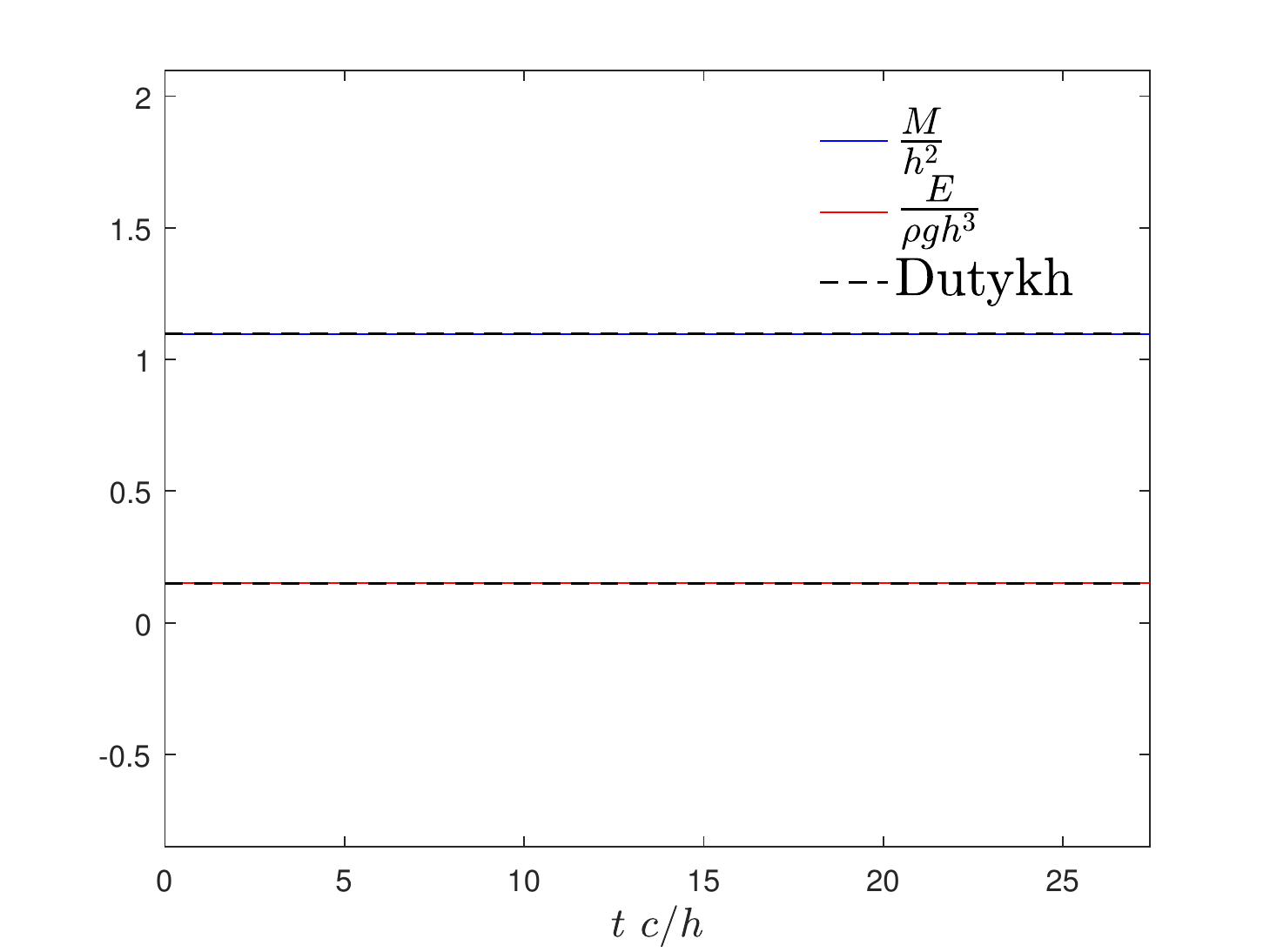}
  \label{fig:HC_b}} \\
\subfloat[ ]{
\includegraphics[width=0.45\textwidth]{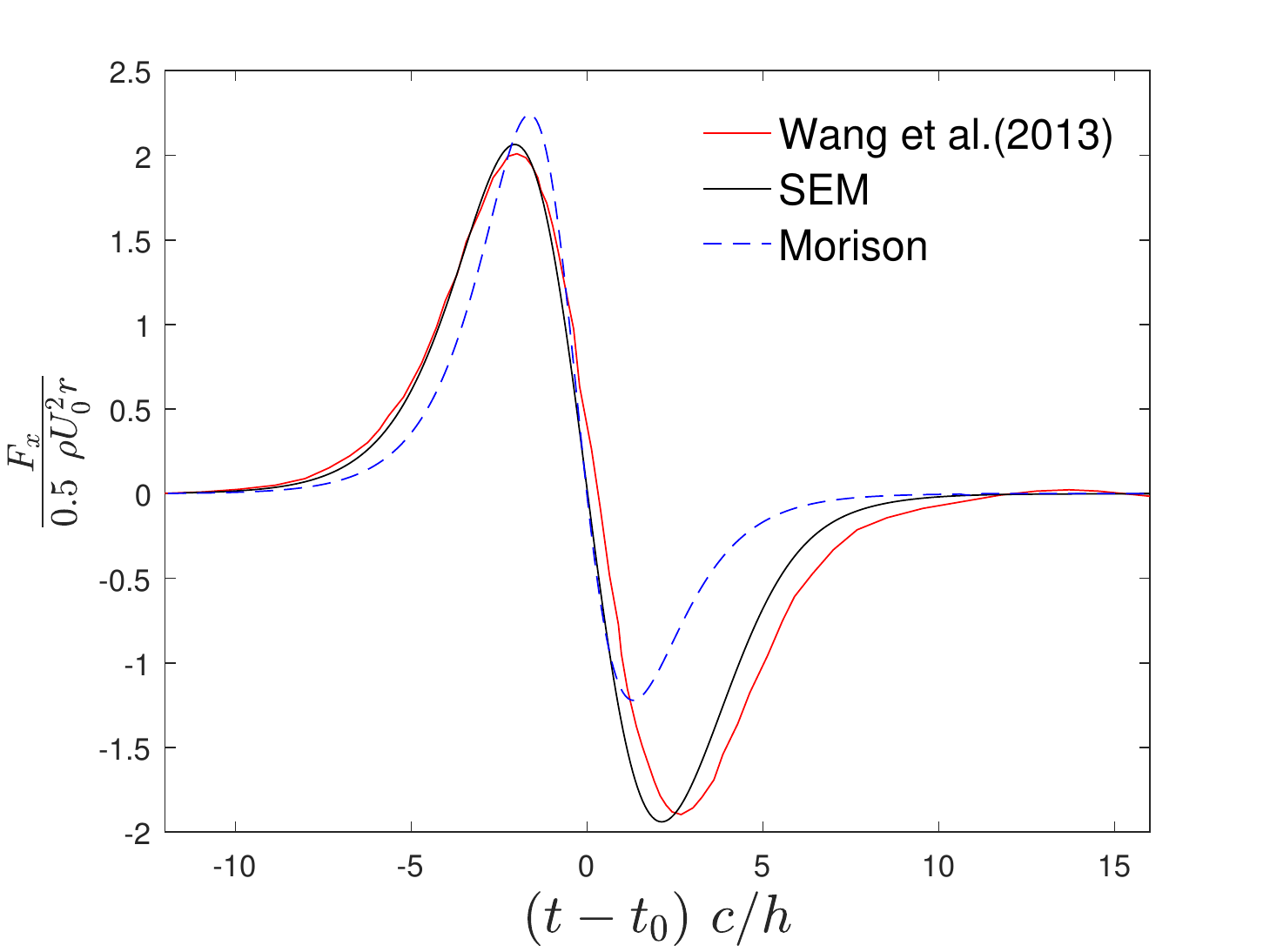}
  \label{fig:HC_c}} \,
\subfloat[ ]{
\includegraphics[width=0.45\textwidth]{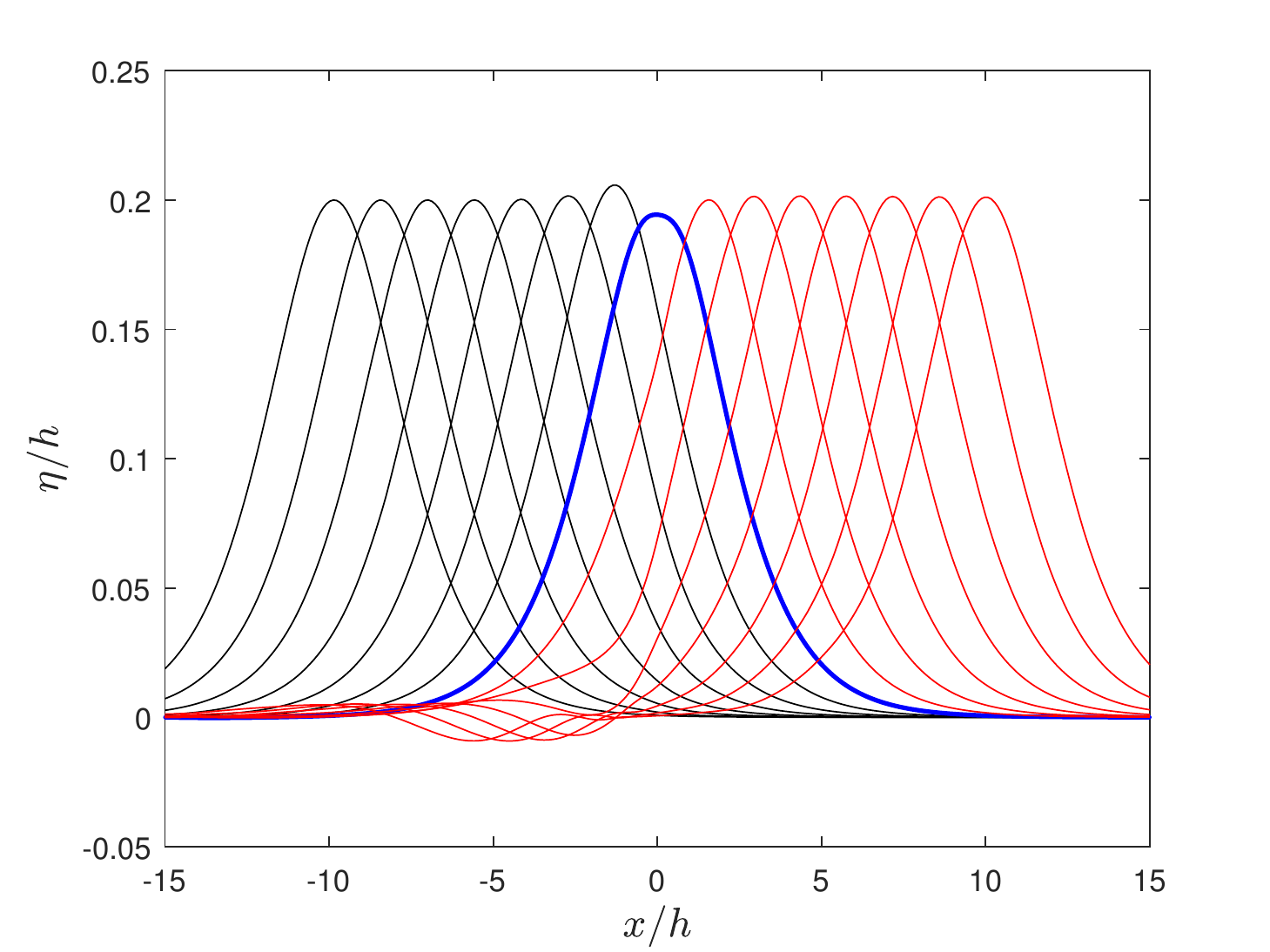}
  \label{fig:HC_c}} \,
\caption{We consider the case with a semi-submerged cylinder of size $R/h = 0.3$. (a) Time evolution of free surface. (b) Computed mass and energy conservation measures. (c) Comparison of the computed horizontal dynamic force on the cylinder with results of Wang et al. (2003) \cite{WangEtAl2013}. (d) Snap shots of the free surface evolution during the initial phase of the interaction with the semi-circular cylinder (solid black) and its transformation in the last phase of interaction (solid red). The wave profile at $t_0$ where the resulting horizontal force component is zero (thickened solid blue). The time spacing used is $\Delta \bar{t} = 1.302$ (0.38 s).}
\label{fig:newexp}
\end{figure}

We consider the numerical experiment described in \cite{WangEtAl2013} on the interaction between a solitary wave of amplitude $a/h=0.2$ (mildly nonlinear) and a submerged semi-circular cylinder with radius $R/h = 0.3$. This experiment serves to validate our force estimation, where the acceleration method due to Tanizawa (1995) is used. The initial mesh is illustrated in Figure \ref{fig:SCCylMesh} and consists of 60 elements in the central part and only 164 elements in total. The computed time evolution of the free surface is given in Figure \ref{fig:newexp} (a). In Figure \ref{fig:newexp} (d) it is seen how the solitary wave propagates undisturbed from the initial condition until it starts the interaction with the semi-circular cylinder resulting in variation in the horizontal dynamic load. During the interaction, a dispersive trail develops after which the solitary wave restore to it's original form. The results are in good agreement with results of Wang et al. The differences in the force curves given in Figure \ref{fig:newexp} (c) is understood to be related to numerical dispersion in the Desingularized Boundary Integral Equation Method (DBIEM) and approximation errors associated with the low-order accurate initial solitary wave. The force curve produced using the SEM solver is for a well-resolved flow and therefore is considered converged to high accuracy. In same figure, the force curve for Morison Equation that takes into account also viscous effects is included and is computed following \cite{KlettnerEames2012,WangEtAl2013}. In figure (b) it is confirmed that the SEM conserves with high accuracy mass and energy during all of the simulation.

\subsection{Solitary wave propagation over a fixed submerged cylinder}

Several studies with both experimental and numerical results can be found regarding the interactions of solitary waves with a submerged circular cylinder \cite{Sib_etal82,Cooker_etal90,ChianErtekin1992,CleMas95}. In the following, we consider two previously reported experiments.

First, we consider the introduction of a fixed submerged cylinder inside the fluid domain with a flat bed following the experiments due to Clement \& Mas (1995) \cite{CleMas95}. Analysis of the hydrodynamic horizontal forces exerted on the cylinder and comparison with experimental results \cite{Sib_etal82} serve as validation in the experiments reported here. 

\begin{figure}[!htb]
  \centering
\subfloat[ Mesh ]{
  \includegraphics[width=1\textwidth]{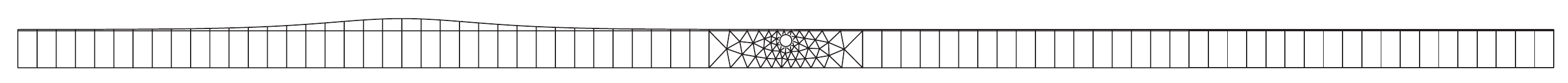} 
  \label{fig:HC_14a}
  }
   \\
\subfloat[ Zoom at center ]{
  \includegraphics[width=0.475\textwidth]{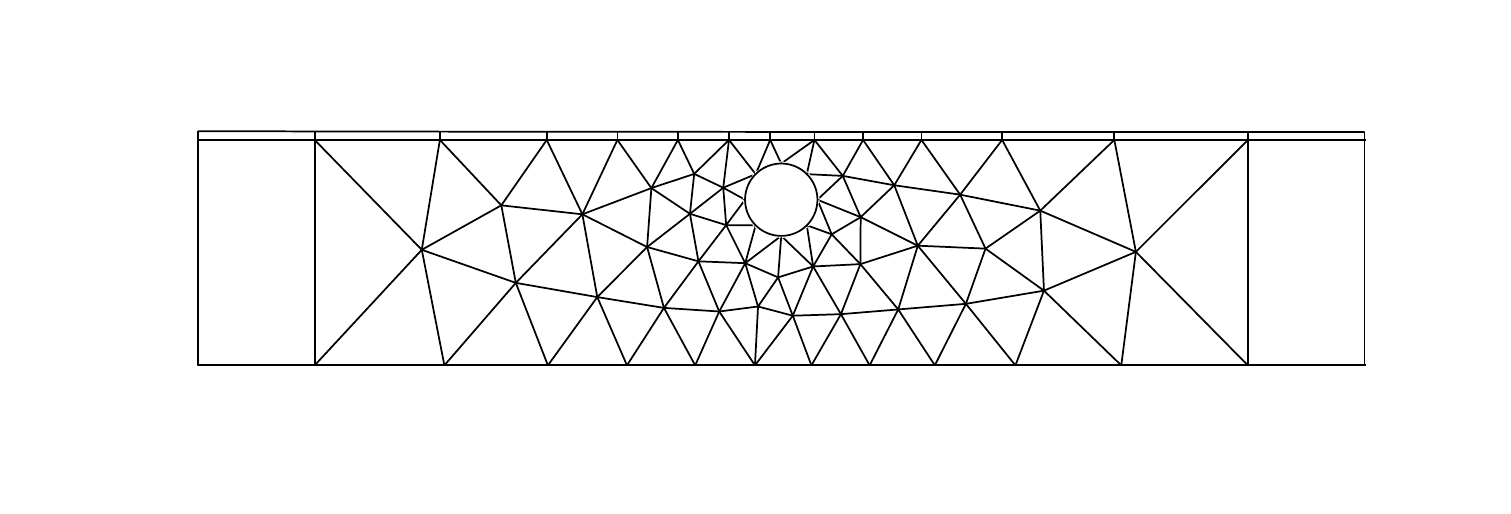} 
  \label{fig:HC_14b}} 
   \caption{Mesh corresponding to case experiment 1 with parameters $D/(2h) = 0.155$, $z_0/h = -0.29$ for the  cylinder and a solitary wave corresponding to $a/h = 0.286$. The total number of elements is 240 in (a), with two layers of 36x2 total elements in each quadrilateral region and 96 elements in the middle triangulated region shown in (b).}
  \label{fig:HC_13}
\end{figure}

Figure \ref{fig:HC_13} shows a typical initial mesh for the experiments. The mesh consist of a small quadrilateral layer near the free surface and contains zones of quadrilateral elements only in first and last part of domain. These zones flank a central part, where a hybrid meshing strategy is employed to adapt using an unstructured triangulation to the submerged cylinder surface in the interior. The initial conditions $\eta_0$, $\phi_0$ for the solitary waves is produced using the method described in \cite{DutCla14}. 

In our first experiment, the solitary wave height is $a/h=0.286$ and corresponds to zone 1 described in \cite{CleMas95} where interaction is said weak. The cylinder has dimensionless radius $D/(2h) = 0.155$ and is positioned with submergence $z_0/h = -0.29$. Figure \ref{fig:HC_14a} shows the resulting time series of the free surface, where we observe a small deformation of the wave during passing the cylinder. This deformation becomes wider at the end and more abrupt in positions just above the cylinder (centered at $x=0$). The computed mass and energy conservation measures given in Figure \ref{fig:HC_14b} confirms stability and accuracy of the simulation. The dimensionless horizontal component is compared in Figure \ref{fig:HC_15a} with experimental results of \cite{Sib_etal82} given in \cite[Figure 2b]{CleMas95}. The results are in good agreement for times before the collision (when horizontal force is zero) and significant discrepancies afterwards where viscous effects are important due to vortex shedding in the experiments \cite{Sib_etal82}. These effects cannot be captured by potential flow models like the present one.

\begin{figure}[!htb]
  \centering
\subfloat[ ]{
  \includegraphics[width=0.425\textwidth]{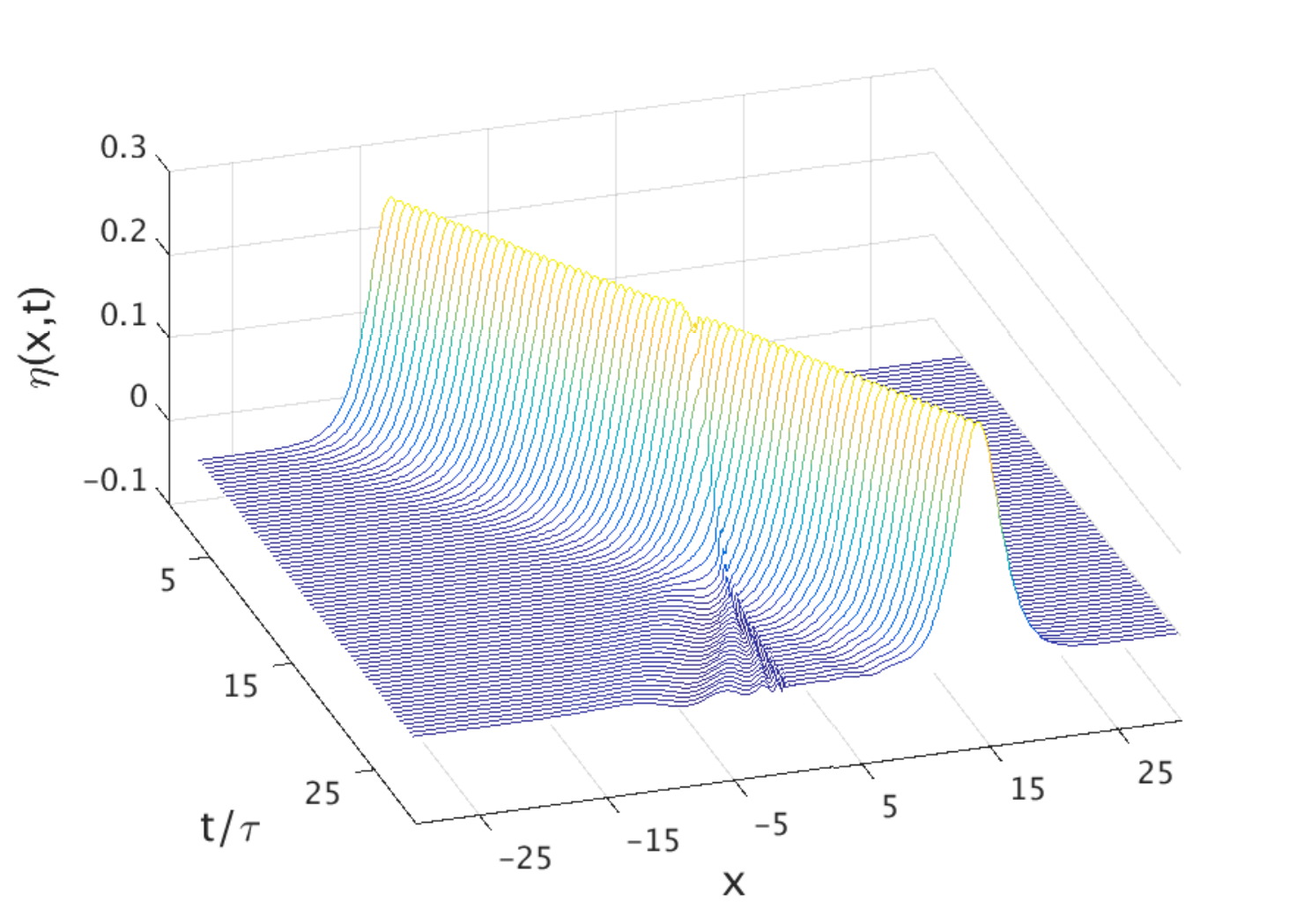}
  \label{fig:HC_14a}} \,
\subfloat[ ]{
  \includegraphics[width=0.425\textwidth]{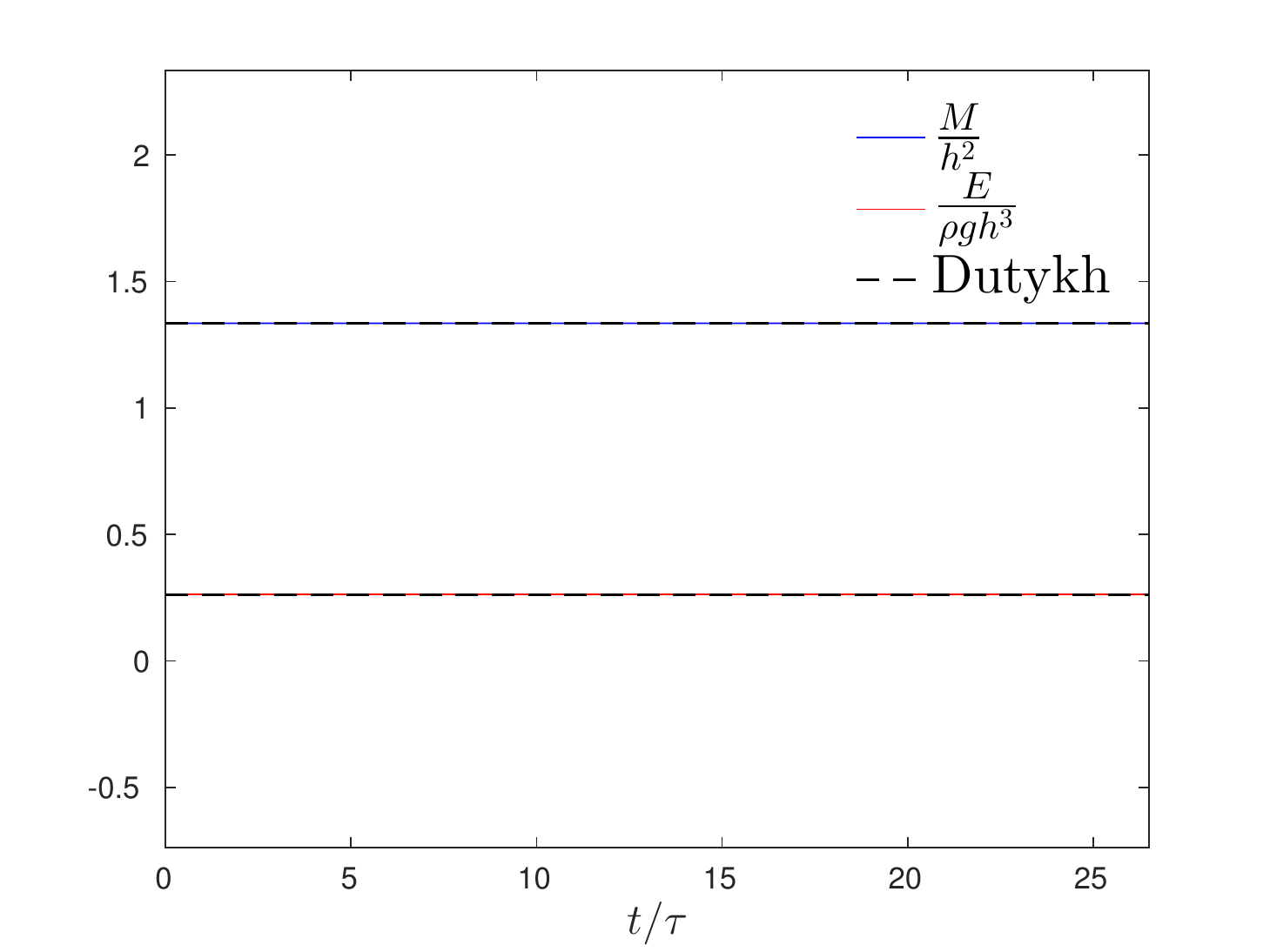} 
  \label{fig:HC_14b}}  \\
\subfloat[ ]{
  \includegraphics[width=0.425\textwidth]{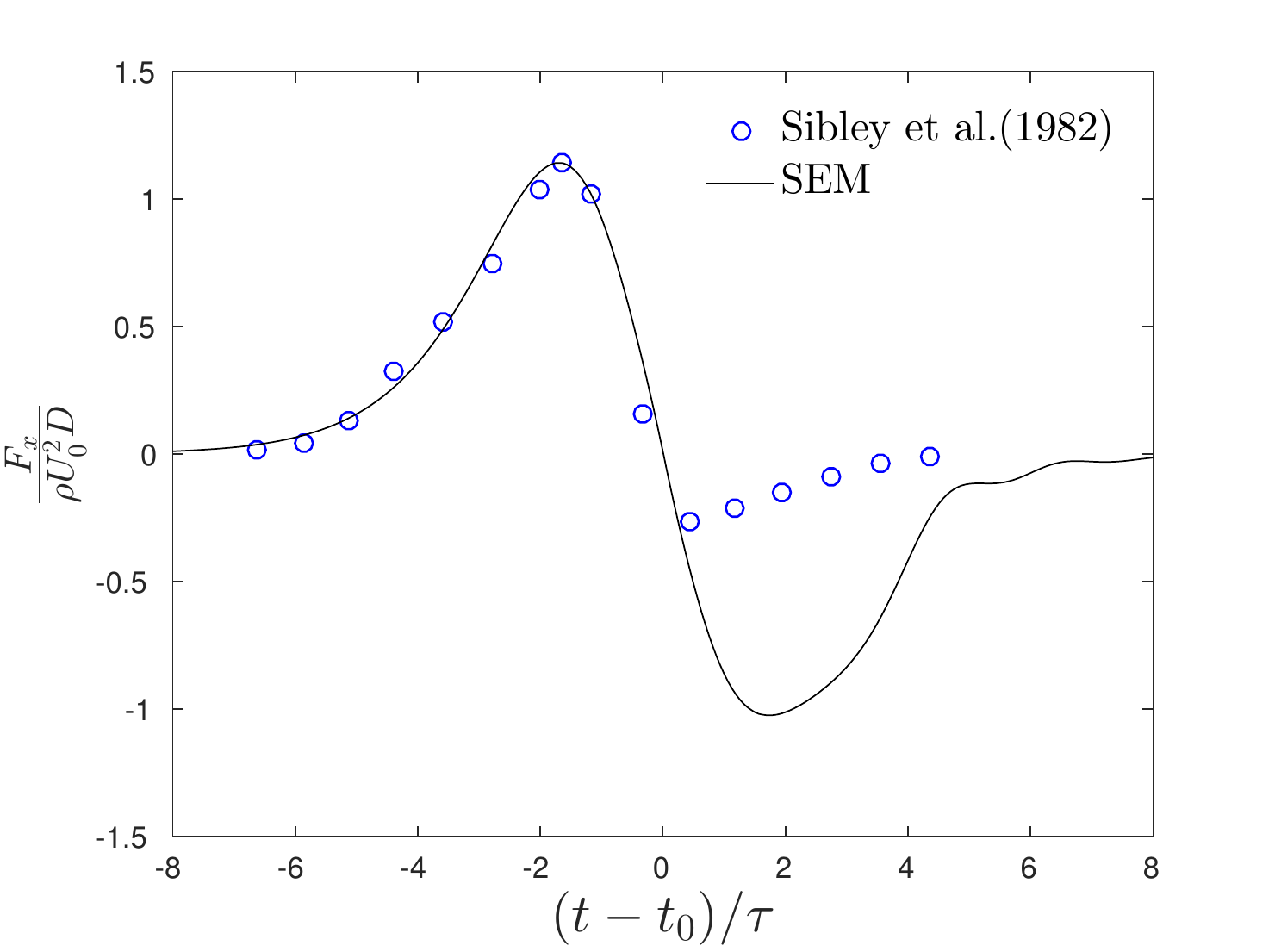} 
  \label{fig:HC_15c}} \,
\subfloat[ ]{
  \includegraphics[width=0.425\textwidth]{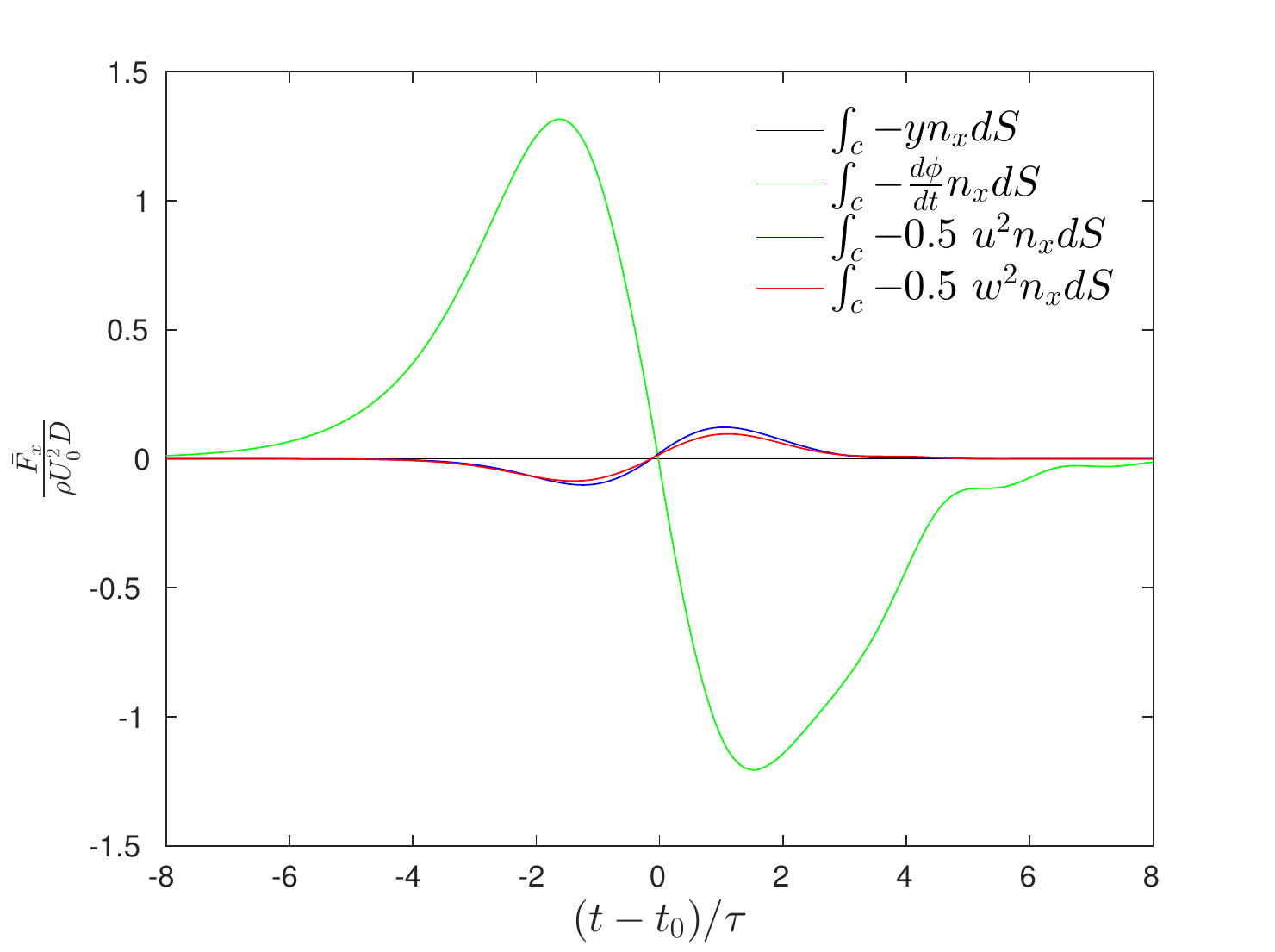} 
  \label{fig:HC_15d}}
   \caption{We consider the case with cylinder size $D/(2h) = 0.155$ and submergence $z_0/h = -0.29$. (a) Time evolution of free surface. (b) Computed mass and energy conservation measures. (c) Horizontal force induced on the cylinder in time for a wave amplitude of $a/h=0.286$ compared with experimental measurements due to Sibley et al. (1982) \cite{Sib_etal82}. (d) Decomposition of the total force in terms of pressure components. 
   }
\label{fig:HC_15}
\end{figure}

\begin{figure}[!htb]
  \centering
\subfloat[ ]{
  \includegraphics[width=0.425\textwidth]{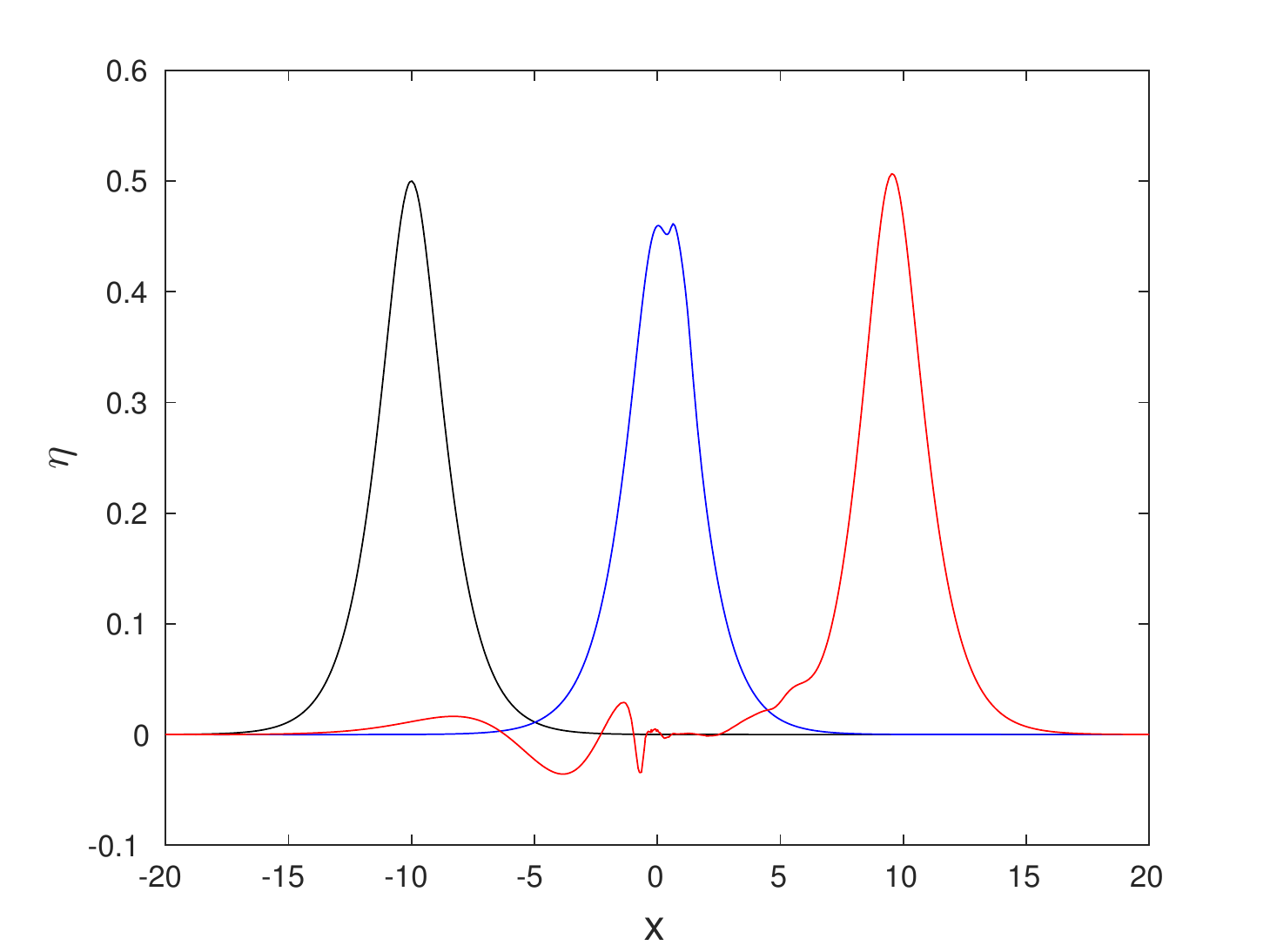}
  \label{fig:HC_16a}} \,
\subfloat[ ]{
  \includegraphics[width=0.425\textwidth]{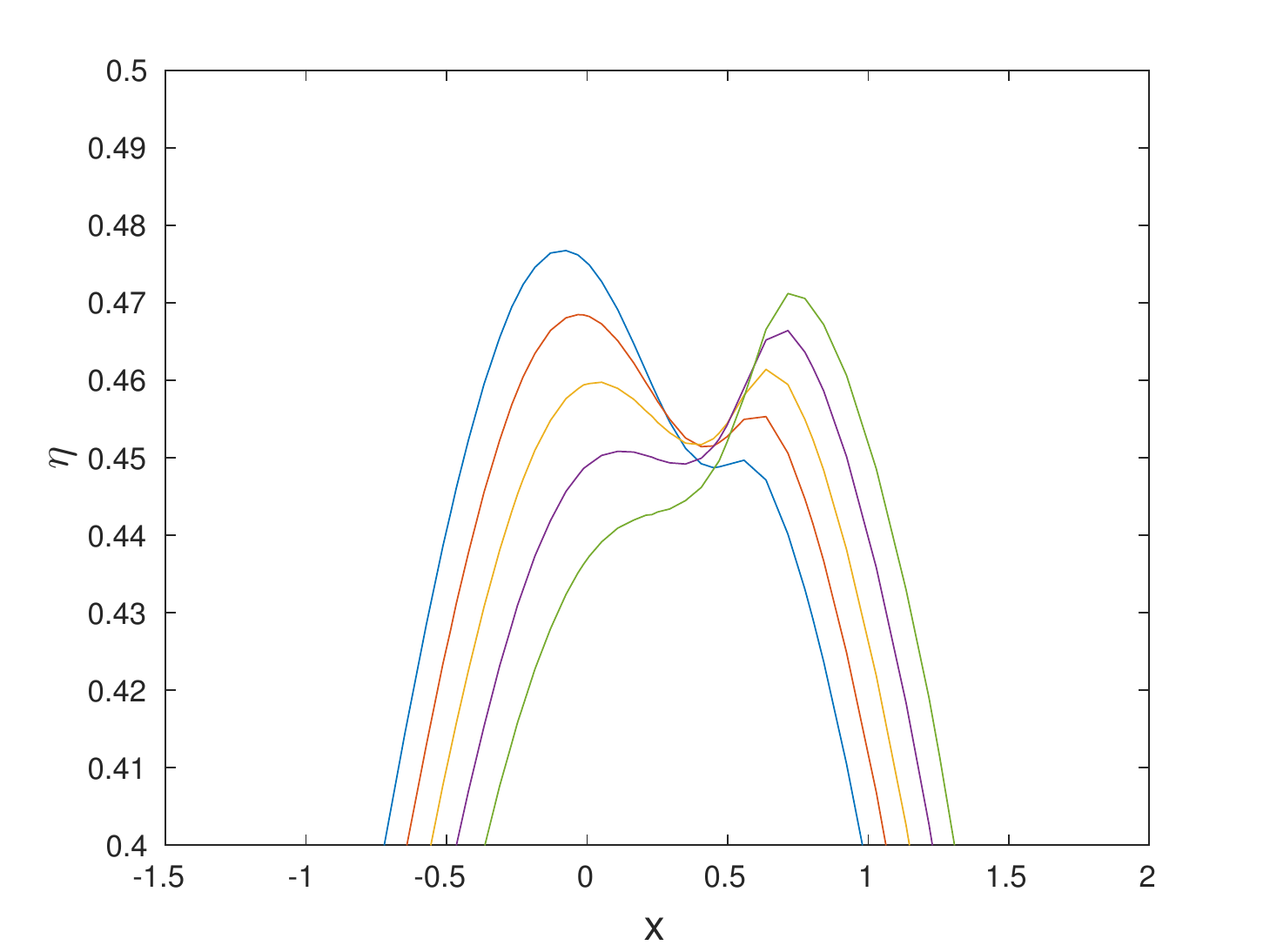}
  \label{fig:HC_16b}} 
   \caption{Case experiment 2: $D/(2h)=0.25$, $z_0/h=-0.5$, $a/h=0.5$, (a) Numerical simulation at initial  position (left), passing the cylinder (middle) and at final time (right). (b) Details of the numerical solution close to cylinder position at instants immediately before and after collision.}
  \label{fig:HC_16}
\end{figure}

\begin{figure}[!htb]
  \centering
\subfloat[ ]{
  \includegraphics[width=0.30\textwidth]{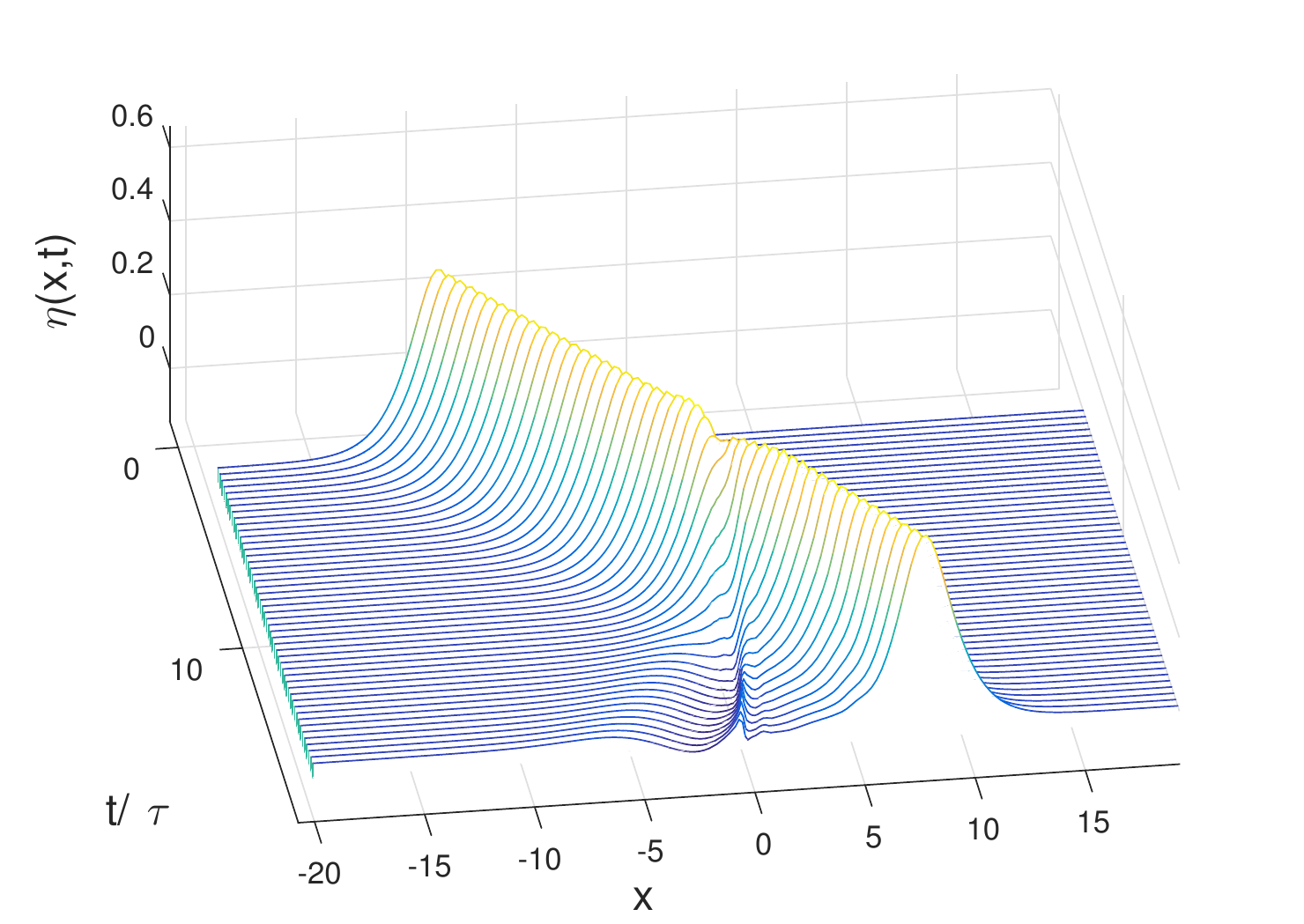}
  \label{fig:HC_17a}} \,
\subfloat[ ]{
  \includegraphics[width=0.30\textwidth]{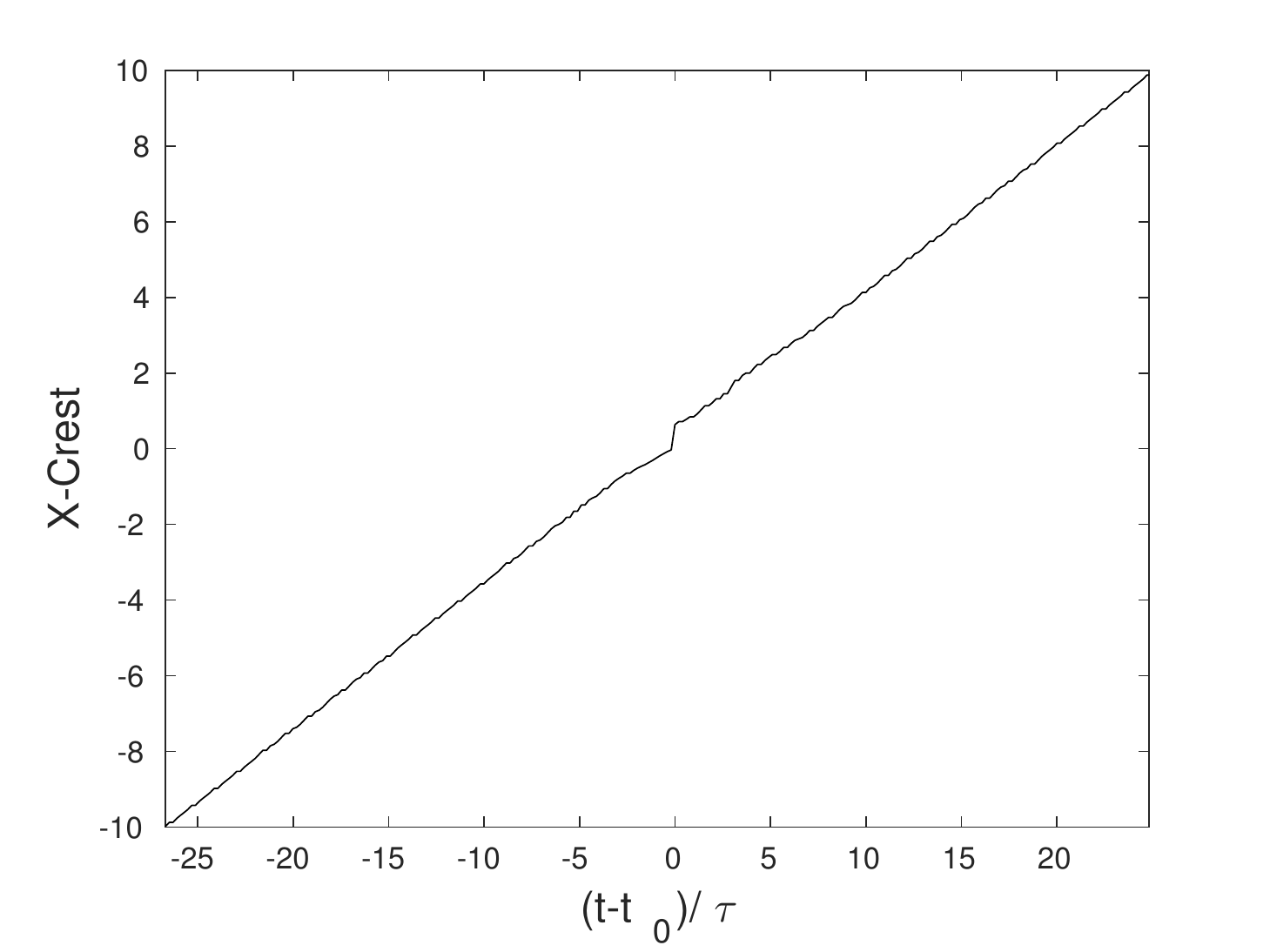}
  \label{fig:HC_18a}} \,
\subfloat[ ] {  
    \includegraphics[width=0.30\textwidth]{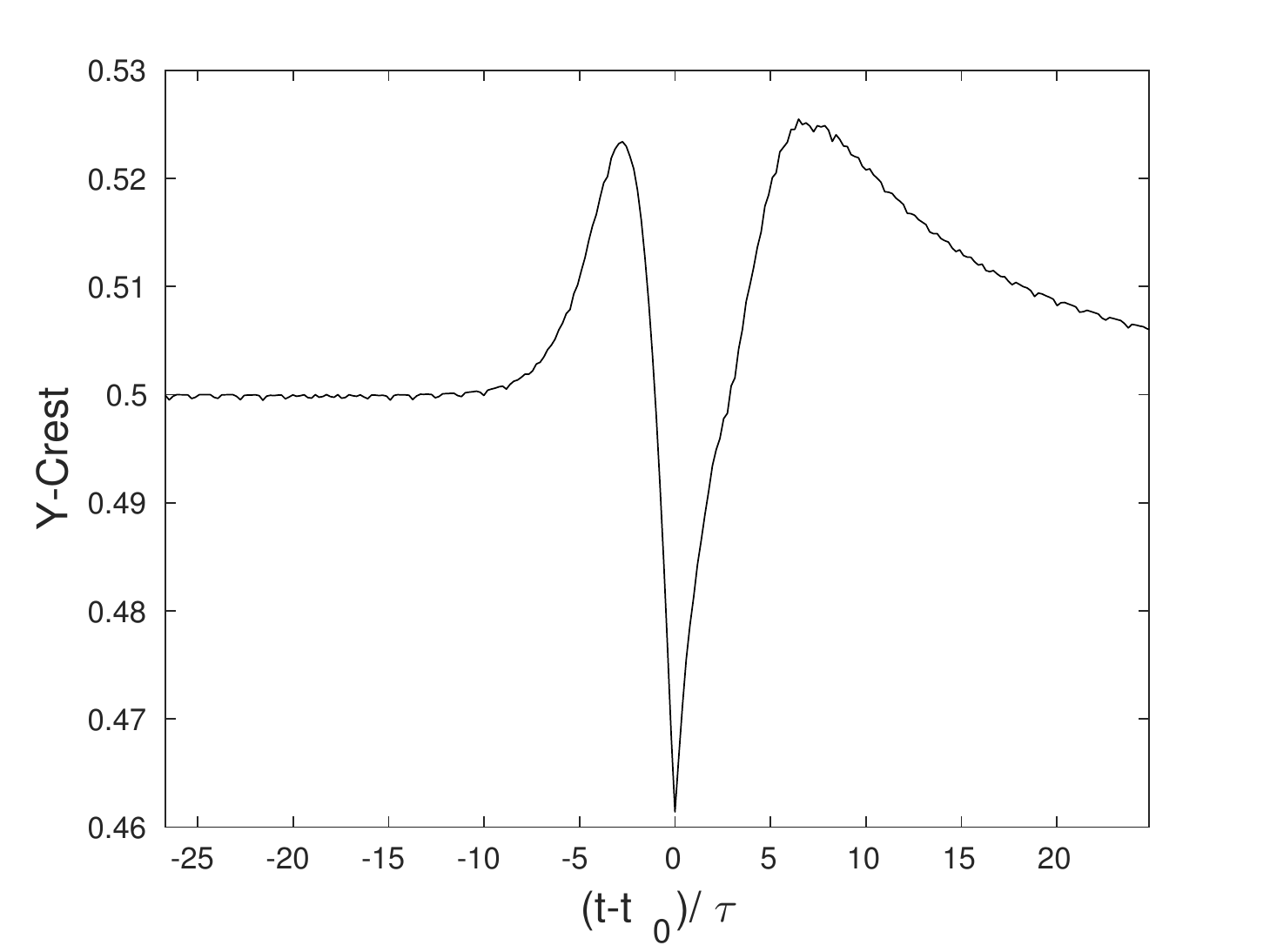}
  \label{fig:HC_18b}} 
   \caption{(a) Computed time series of free surface corresponding to experiment 2 ($D/(2h)=0.25$, $z_0/h=-0.5$, $a/h=0.5$). Free surface node positions ($x$) of the crest during simulation of experiment 2 with dimensionless parameters $D/(2h)=0.25$, $z_0/h=-0.5$, $a/h=0.5$ with (b) $x$-coordinate and (c) $y$-coordinate (height) of the crest illustrated.
   } 
  \label{fig:HC_17}
\end{figure}

\begin{figure}[!htb]
  \centering
\subfloat[Horizontal velocity, $\tfrac{u}{\sqrt{gh}}$]{ \includegraphics[width=0.46\textwidth]{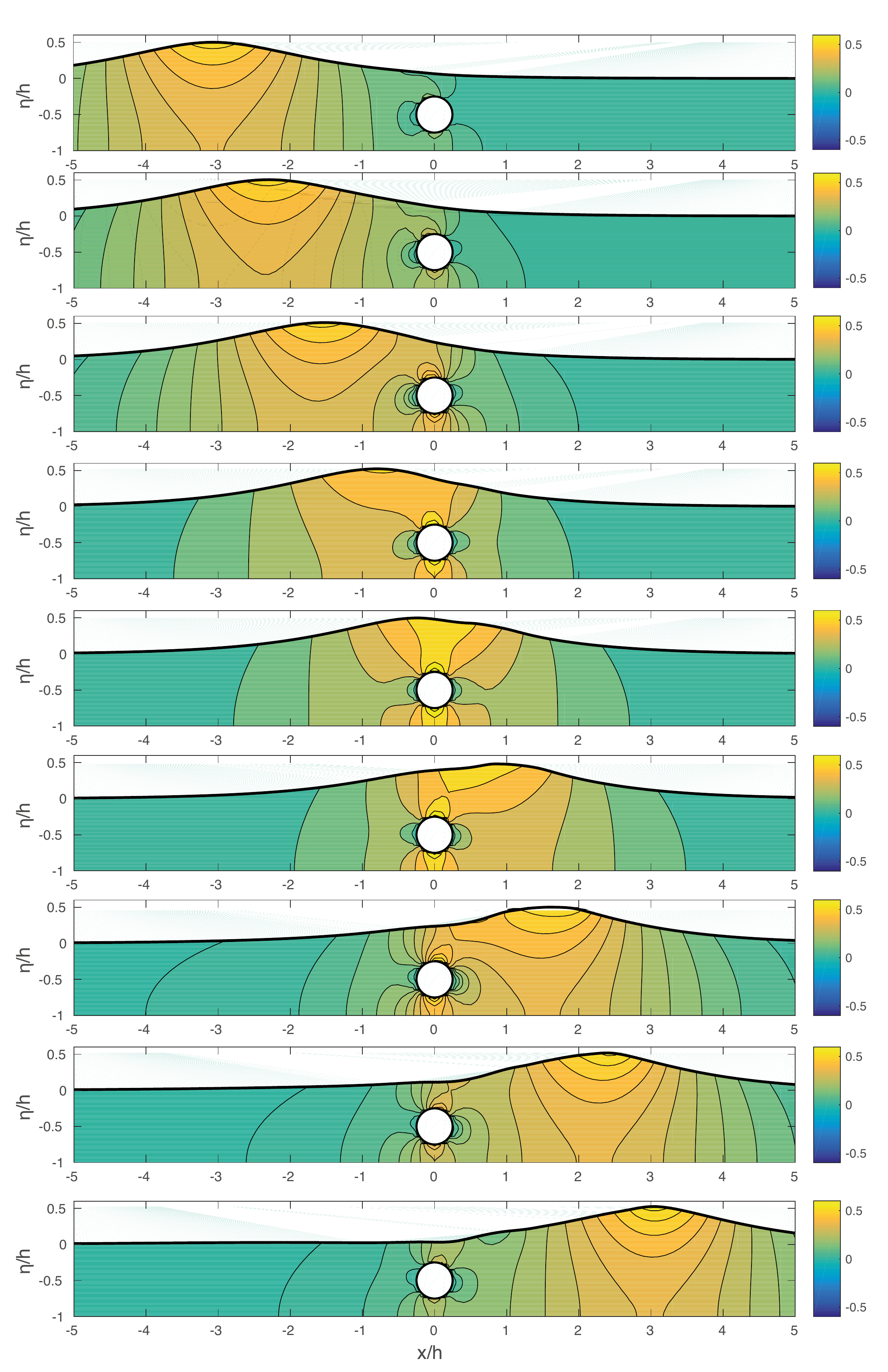}\label{fig:HC_NMu}
}\,
\subfloat[Vertical velocity, $\tfrac{w}{\sqrt{gh}}$]{ \includegraphics[width=0.4625\textwidth]{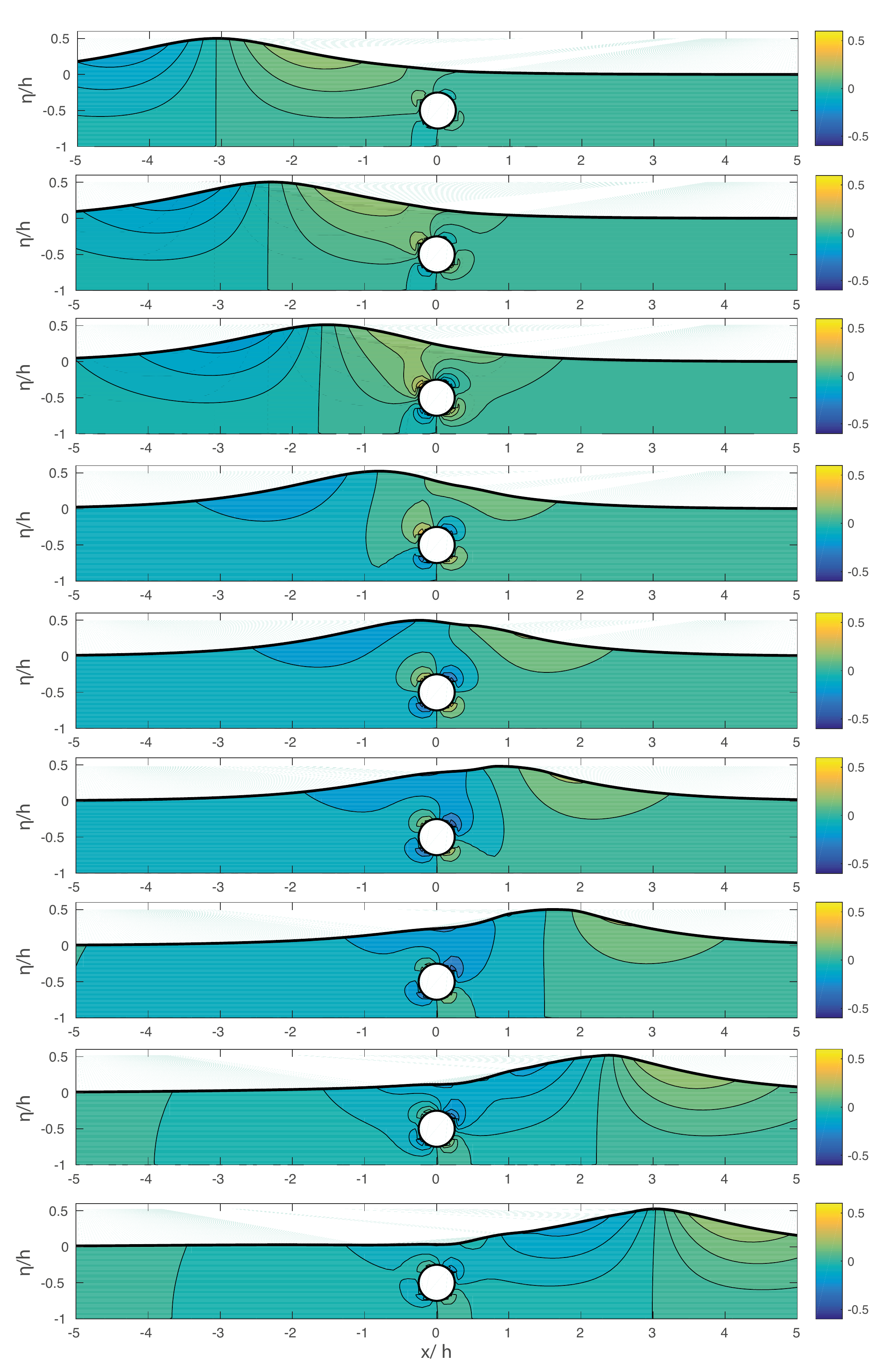}\label{fig:HC_NMw}
}
     \caption{Nonlinear wave-body interaction with fixed cylinder and kinematics contours. Sequences of dimensionless velocity vectors $(u,w)/\sqrt{gh}$ at intervals of $0.02$ s. Solitary wave of $a/h=0.5$, cylinder size of $D/(2h) = 0.25$, submergence of $z_0/h$ = -0.5. Order of the elements is $P=6$. Time increases from top to bottom. 
     }
   \label{fig:HC_NM2}
\end{figure}

In our second experiment, the wave height is $a/h=0.5$ and the cylinder has radius $D/(2h)=0.25$ and is positioned at a submergence $z_0/h =-0.5$. Following \cite{CleMas95} this experiment is located in the crest-crest exchange zone 2. In this zone incident and transmitted waves can be clearly distinguished. The fluid motion in this experiment is studied when the wave passes the cylinder. In Figure \ref{fig:HC_NMu} and Figure \ref{fig:HC_NMw} the horizontal and vertical velocity components in a portion of the fluid domain ($-5\leq x \leq 5$) are illustrated at simultaneous times highlighting the changes in the sub-surface kinematics during the solitary wave interaction with the cylinder. While approaching the position of the cylinder the wave looses height and slows down, cf. Figure \ref{fig:HC_18a} and Figure \ref{fig:HC_18b}. Then the birth of a transmitted wave can be observed, cf. Figure \ref{fig:HC_16b}. The transmitted wave is not shifted and leaves behind a dispersive trail that becomes even more abrupt than observed in compared to wave interaction in our first experiment, cf. Figure \ref{fig:HC_16a}.

\section{Conclusions}

We have presented a new stabilised nodal spectral element model for simulation of fully nonlinear water wave propagation based on a Mixed Eulerian Lagrangian (MEL) formulation. The stability issues associated with mesh asymmetry as reported in \cite{RobertsonSherwin1999} is resolved by using a hybrid mesh with a quadrilateral layer with interfaces aligned with the vertical direction to resolve the free surface and layer just below this level. Our linear stability analysis confirms that the temporal instability associated with triangulated meshes can be fixed, paving the way for considering remaining nonlinear stability issues. By combining a hybrid mesh strategy with the ideas described in \cite{EngsigKarupEtAl2016} on stabilisation of free surface formulation with quartic nonlinear terms, the model is stabilised by using exact quadratures to effectively reduce aliasing errors and mild spectral filtering to add some artificial viscosity to secure robustness for marginally resolved flows. This is combined with a re-meshing strategy to counter element deformations that may lead to numerical ill-conditioning and in the worst cases breakdown if not used. Our numerical analysis confirm this strategy to work well for the steepest nonlinear water waves when using this stabilised nonlinear MEL formulation. In the MEL formulation we avoid the use of a $\sigma$-transform of the vertical coordinate, to introduce bodies of arbitrary geometry that is resolved using high-order curvilinear elements of an unstructured hybrid mesh. By using these elements, only few elements are needed to resolve the kinematics in the water column both with and without complex body surfaces. With this new methodology, we validate the model by revisiting known strenuous benchmarks for fully nonlinear wave models, e.g. solitary wave propagation and reflection, wave-body interaction with a submerged fixed cylinder and wave-body interaction with a fixed surface-piercing structure in the form of a pontoon. The numerical results obtained are excellent compared with other published results and demonstrate the high accuracy that can be achieved with the high-order spectral element method. 

In ongoing work, we are extending the new stabilised spectral element solvers towards advanced and realistic nonlinear hydrodynamics applications in three space dimensions (cf. the model based on an Eulerian formulation in three space dimensions\cite{EngsigKarupEtAl2016ISOPE}) by extending the current computational framework to also handle moving and floating objects of arbitrary body shape. In ongoing work, we consider freely moving bodies and surface-piercing structures with non-vertical boundary at the body-surface intersections and hybrid modelling approaches \cite{PaulsenEtAl2014,VerbruggheEtAl2016}.

\section*{References}
\bibliographystyle{plain}
\bibliography{refs.bib}

\end{document}